\documentstyle[11pt]{book}
\begin{document}


\thispagestyle{empty}

\begin{center}

\vspace{20mm}

{\Huge{Gauged WZNW models and }} \\
\vspace{3mm}
{\Huge{coset constructions }} \\
\vspace{4mm}
{\Huge{in relation to string theories.}}

\vspace{20 mm}
{\huge{Henric Rhedin}} \\
\vspace{10mm}

{\Large Ph.D. Thesis} \\

\vspace{2mm}

{\Large Institute of Theoretical Physics \\
G\"oteborg University and \\
Chalmers University of Technology \\
\vspace{5mm}
G\"oteborg, Sweden \\
\vspace{2mm}
November 1995 \\
}

\end{center}

\newpage

\thispagestyle{empty}

\include{abstract}

\newpage

\renewcommand{\thepage}{\roman{page}}
\setcounter{page}{1}


\newcommand{\al}{\alpha}
\newcommand{\bet}{\beta}
\newcommand{\ga}{\gamma}
\newcommand{\del}{\delta}
\newcommand{\ep}{\epsilon}
\newcommand{\epx}{\varepsilon}
\newcommand{\ze}{\zeta}
\newcommand{\th}{\theta}
\newcommand{\thx}{\vartheta}
\newcommand{\io}{\iota}
\newcommand{\la}{\lambda}
\newcommand{\ka}{\kappa}
\newcommand{\pix}{\varpi}
\newcommand{\rhx}{\varrho}
\newcommand{\si}{\sigma}
\newcommand{\six}{\varsigma}
\newcommand{\yp}{\upsilon}
\newcommand{\om}{\omega}
\newcommand{\phx}{\varphi}
\newcommand{\Ga}{\Gamma}
\newcommand{\De}{\Delta}
\newcommand{\Th}{\Theta}
\newcommand{\La}{\Lambda}
\newcommand{\Si}{\Sigma}
\newcommand{\Yp}{\Upsilon}
\newcommand{\Om}{\Omega}


\newcommand{\be}{\begin{eqnarray}}
\newcommand{\ee}{\end{eqnarray}}
\newcommand{\jt}{\tilde{J}}
\newcommand{\Ra}{\Rightarrow}
\newcommand{\lra}{\longrightarrow}
\newcommand{\pr}{\partial}
\newcommand{\ti}{\tilde}
\newcommand{\ng}{|0\rangle_{gh}}
\newcommand{\pj}{\prod J}
\newcommand{\pjt}{\prod\tilde{J}}
\newcommand{\prb}{\prod b}
\newcommand{\prc}{\prod c}
\newcommand{\bft}{|\tilde{\phi}>}
\newcommand{\bfj}{|\phi>}
\newcommand{\lan}{\langle}
\newcommand{\ran}{\rangle}
\newcommand{\bz}{\bar{z}}
\newcommand{\bw}{\bar{w}}
\newcommand{\bJ}{\bar{J}}
\newcommand{\vacr}{|0\rangle}
\newcommand{\vacl}{\langle 0|}
\newcommand{\IFF}{\Longleftrightarrow}
\newcommand{\phr}{|phys\ran}
\newcommand{\phl}{\lan phys|}
\newcommand{\bT}{\bar{T}(\bz)}
\newcommand{\bD}{\bar{D}}
\newcommand{\bth}{\bar{\th}}
\newcommand{\Gi}{G^{-1}}
\newcommand{\gi}{g^{-1}}
\newcommand{\cG}{{\cal G}}
\newcommand{\cGi}{{\cal G}^{-1}}
\newcommand{\rmg}{{\rm g}}
\newcommand{\rgi}{{\rm g}^{-1}}
\newcommand{\hi}{h^{-1}}
\newcommand{\hJ}{\hat{J}}
\newcommand{\hJt}{\hat{\ti{J}}}
\newcommand{\hg}{\hat{g}}
\newcommand{\hk}{\hat{k}}
\newcommand{\hka}{\hat{\ka}}
\newcommand{\hrho}{\hat{\rho}}
\newcommand{\hal}{\hat{\al}}
\newcommand{\hLa}{\hat{\La}}
\newcommand{\hla}{\hat{\la}}
\newcommand{\hsi}{\hat{\si}}
\newcommand{\hDe}{\hat{\De}}
\newcommand{\hmu}{\hat{\mu}}


\newcommand{\ind}{\indent}
\newcommand{\np}{\newpage}
\newcommand{\hs}{\hspace}
\newcommand{\vs}{\vspace}
\newcommand{\nl}{\newline}
\newcommand{\nn}{\nonumber}
\newcommand{\lef}{\left}
\newcommand{\rig}{\right}
\newcommand{\fra}{\twelvefrakh}
\newcommand{\fraa}{\twelvefraka}
\newcommand{\Bb}{\twelvemsb}
\newcommand{\dagg}{^{\dagger}}
\newcommand{\qd}{\dot{q}}
\newcommand{\cP}{{\cal P}}

\newcommand{\pfu}{partition function}
\newcommand{\hr}{\newline\indent}


\catcode`\@=11
\newfam\msbfam
\font\twelvemsb=msbm10 scaled\magstep1
\font\tenmsb=msbm10 \font\sevenmsb=msbm7 \font\fivemsb=msbm5
\textfont\msbfam=\twelvemsb
\scriptfont\msbfam=\tenmsb \scriptscriptfont\msbfam=\sevenmsb
\def\Bbb{\relax\ifmmode\expandafter\Bbb@\else
 	\expandafter\nonmatherr@\expandafter\Bbb\fi}
\def\Bbb@#1{{\Bbb@@{#1}}}
\def\Bbb@@#1{\fam\msbfam\relax#1}
\catcode`\@=\active


\begin{center}
{\huge Preface}
\end{center}


I will begin by a few words on the layout of this thesis.
At the end of each chapter I give the main sources of information on
the subject dealt with, and at the very end of the first part
there is a more complete set of
references. I do not claim to include all relevant references,
and I apologize if someone has been left out.

This thesis consists of two parts, first an introduction to the subject and
secondly five articles. The introduction should serve two purposes; i)
It should provide with the physical perspective of the articles. ii) It should
also provide with the necessary details of the relevant background, such
that the thesis becomes as self consistent as possible.

We hence start by giving a brief introduction to string theory with
emphasis on the example of the bosonic string. In order to fully
appreciate string theory it is necessary to study the dynamics of the
surface that the string traces out when moving in space-time. This is
described by what is known as conformal field theory, which is the
subject of chapter two. After this digression we return to
string theory to give some insight into how the results of the articles
contained in this thesis could provide with interresting ingredients for
constructions of realistic string theories. This is also continued on
to some extent in chapter seven and eight.

The articles in this thesis are discussing mainly two subjects, gauged
WZNW models and affine branching functions. The background for those
issues is given in chapter seven and eight.

In the articles contained in this thesis we use two major tools. First
affine Lie algebras and its representation theory, and secondly
BRST quantization. In chapter four and five we deal with affine Lie
algebras, and in chapter six we try to illuminate some issues on BRST
quantization. \\

In the second part of this thesis you find the following articles;
\begin{enumerate}
\def\theenumi{\Roman{enumi}}
  \item  S. Hwang and H. Rhedin \\
         {\em The BRST formulation of G/H WZNW models.} \\
         Nucl. Phys B406 (1993) 165

  \item  S. Hwang and H. Rhedin \\
         {\em General branching functions for affine Lie algebras.} \\
         Mod. Phys. Lett. A10 (1995) 823

  \item  H. Rhedin \\
         {\em BRST invariant characters of G/H coset models.} \\
         G\"oteborg ITP 94-16

  \item  S. Hwang and H. Rhedin \\
         {\em Construction of BRST invariant states in $G/H$ WZNW models.} \\
         Phys. Lett. B350 (1995) 38

  \item  H. Rhedin \\
         {\it Gauged supersymmetric WZNW model using the BRST approach.} \\
         G\"oteborg ITP 95-23

\end{enumerate}
henceforth referred to as paper I, II, III, IV and V, respectively.

In paper I is the relation of the gauged WZNW model in the path integral
approach and
the algebraic Goddard-Kent-Olive coset construction discussed. We show,
by use of BRST techniques, that for a restricted set of representations the two
approaches coincide.

In the last section of paper I a branching function for affine Lie algebras
is introduced. This formula for branching functions is further elaborated on
in paper II and III. In paper II we present an independent proof of the
formula's
validity. Paper III deals with explicit general examples which are
calculated in detail.

In paper IV we lift the restriction on the representations introduced in
paper I where it was used to calculate the BRST cohomology. We present a
method to construct states in the cohomology for arbitrary representations,
and find that there exists states at for all ghost numbers in the cohomology.

Paper V deals with the gauged supersymmetric WZNW model. We give a Lagrangian
formulation for the algebraic $N=1$ supersymmetric coset construction. \\

I would like to express my gratitude to my advisor Stephen Hwang. I am greatly
indebted to Stephen for advice, encouragement and support during my years
as a Ph.D. student. Needless to say, he has taught me a great deal, and
always encouraged me to learn more, and hence broaden my knowledge. I have also
had the great pleasure of collaborating with him on several projects.

A word of thanks to all present and former members of the staff of the
institute
for theoretical physics in G\"oteborg. Thank you for providing with
pleasant working conditions.

There are a lot of people that should be acknowledged for support in various
contexts. I will not specify by name in order to make the list as complete as
possible. A thank you to; all former teachers, everybody who has supported my
applications in various contexts, those who has supported me by facilitating
visits to other universities and conferences, all those kind people who has
made visits fruitful, etc.

Finally I would like to thank my family, Anneli and Alexander. Thank you
for encouragement and patience. Anneli is also acknowledged for her
typesetting of large parts of the manuscript.

\vspace{10mm}

\begin{flushright}{ G\"oteborg, November 1995 \\
Henric Rhedin}
\end{flushright}

\tableofcontents
\listoffigures

\newpage

\renewcommand{\thepage}{\arabic{page}}
\setcounter{page}{1}

\pagestyle{headings}


\chapter{String prelude}

At the beginning of this century, two separate theories emerged
which would completely turn old concepts upside down. I am
referring to the theory of general relativity, and quantum mechanics.
Those two, and their implications has kept the
physics society busy for almost a decade. The most difficult task,
the unification of these into a theory for quantum gravity, has however
eluded theoretical physicists so far. (I do not think it is unfair to
only mention theorists at this point since experimental evidence for
quantum gravity seems out of reach at this moment.) There is, however,
room for some optimism, since steady progress towards a theory of
quantum gravity is being made.
In fact the developments in the last few month have made some theoretical
physicists more optimistic than for a long time.

Nature has provided us with four fundamental interactions, strong, weak,
electromagnetic and gravitational. There exists well-behaved quantized
theories for the first three and furthermore, they have been
unified into a common description known as the standard model.
The concepts that are successfully applied to those
interactions, however, fails for gravity, at least to our present
understanding. This failure is related to the pointlike nature of
the particles which gives rise to uncurable divergencies. One naturally
also expects that gravity will be unified.

The standard model, although it must be considered the greatest success
of modern particle physics yet, is far from flawless. There are, for example, a
large number of parameters that must be inserted by hand. The gauge
group structure is rather complicated etc. In a grand unified theory
one would like to put in as little as possible by hand, and also
expects the theory to be beautiful in some framework.

Up until now, there is only one known way of getting rid of the
undesired features that the pointlike nature of particles have,
and at the same time keep consistency, and that is string theory.
The intuitive way to think of a string is in terms of for example a
guitar string. Just as different vibrational modes of the guitar
string gives different sounds different "vibrations" of the
elementary particle string corresponds to different particles.
What is more important is that string theory includes gravity in a
natural way. Some string theories are claimed to
have large enough gauge groups to include
the standard model.

Regarded from the experiment point of view, quantum gravity seems so far
out of reach. The typical energy scale is of the order $10^{19}$GeV which
may be compared with the accelerators of today which are pushing $10^{3}$GeV.
Within reach, in the nearest few decades, is the energy range where
one should be able to put supersymmetry at test. Supersymmetry being a
crucial ingredient of most candidates of realistic string theories might
thus provide with one of perhaps many consistency checks that string
theories must pass before we may actually verify string theories in
direct experiments. This will not be a decisive test in favour of
string theories, because many candidates for theories of everything
incorporates supersymmetry.

In waiting for experiments we may of course discuss phenomenology of
strings, and make predictions and suggestions for experiments.
There is, however, a great deal of issues to be resolved in order
to produce self-consistent and realistic string theories before we discuss
phenomenology. In fact there exist too many candidates for realistic
string theories, so one important issue is to find a criteria which
singles out the candidate.

I will hence in this thesis, when discussing
string theories, never raise issues of phenomenological nature, but
rather discuss string theories for their own sake. Also, I will take the
approach to strings, which emphasizes on what is known as conformal field
theory.  \\

Although string theory appears to be the most natural generalization
of the point particle, this is not the way it appeared
as a candidate
for a unifying theory. On the contrary, as is probably well-known
by now, it was work within the field of strong interactions that
introduced strings to the community of particle physicists.

In the 60's the S-matrix approach to strong interactions dominated
this area of
 particle physics. The basic concept was that only observables should
enter into our
equations, quite the contrary to quantum field theory. Demanding
e.g. Lorentz invariance, analyticity,
duality and unitarity of the scattering matrix we should be given a system of
equations which hopefully were solvable, either perturbatively or in a
self-consistent
manner. These four demands of the S-matrix could heuristically be explained
as follows. Lorentz invariance is obvious.
Consider the scattering of two spinless particles $a+b\longrightarrow
c+d$.
Analyticity means that the S-matrix must only be an analytic function of
Lorentz invariants that describe the scattering process such as the
Mandelstam variables. Duality is assuming
that the same S-matrix describes the "crossed" processes
$a+\bar{c}\longrightarrow
\bar{b}+d$ and $a+\bar{d}\longrightarrow c+\bar{b}$ where "bar"
indicates charge conjugation. The unitarity assumption ensures
conservation of probability. \\
\ind Not much success was met until Veneziano in 1968 wrote down a simple
amplitude which
fulfilled the symmetry of crossing exactly, giving rize to an increased
interest in
those so-called dual models. Later on it was realized by Nambu, Nielsen
and Susskind
that the Veneziano model described an open relativistic string. The
amplitude was also generalized to the closed string, the Virasoro-
Shapiro model. In 1974 Scherk and
Schwarz \cite{Scherk-Schwarz74} suggested that the massless spin-2
particle in the
spectrum could
be interpreted as the graviton. String theories, in
different settings, are now advocated as theories of everything. \\

\section{The Classical Bosonic String.}

The bosonic string is the simplest example of a string theory,
yet it includes most important features, and is thus suitable as an
introductory example. Other examples are various species of strings
with fermionic degrees of freedom, known as superstrings, and the
Heterotic string, named after "heteros" meaning "other", which is a
hybrid, half bosonic string half superstring.

Here we will not use the approach of dual models but rather the more
intuitive view of strings as a natural generalization of the
relativistic point particle. There are two possible topologies for a
one-dimensional compact manifold, closed and open. We will concentrate on the
closed string which has periodic boundary conditions in one coordinate, as
indicated by the name. \\
\ind The massive point particle is described by the action
\be
S=-m\int{\rm
d}\tau\lef(\eta_{\mu\nu}\frac{dx^{\mu}}{d\tau}\frac{dx^{\nu}}{d\tau}\rig)
^{1/2}
\label{sqrtaction}
\ee
where $\eta^{\mu\nu}$ is the Minkowski metric in $D$ dimensions, and $\tau$
parametrizes
the trajectory of the particle. The action is thus proportional to the path
length,
and from the variational principle we should get equations of motion
which extremizes the path. \\
\ind There exists classically an equivalent formulation namely
\be
S=\frac{1}{2}\int{\rm
d}\tau\lef(\frac{1}{e(\tau)}\eta_{\mu\nu}\frac{dx^{\mu}}{d\tau}
\frac{dx^{\nu}}{d\tau}-e(\tau)m^2\rig) .
\ee
Here $e(\tau)$ is present to ensure invariance under reparametrizations of
$\tau$.
One may choose the gauge $e(\tau)=1$, but must of course not forget to impose
the
constraint $\del S/\del e=0$. If we eliminate $e(\tau)$, by using the equations
of
motion in the action, we regain the squareroot form
(\ref{sqrtaction}). \\
\ind The natural generalization for a string is then to take the action
proportional
to the area swept out in space-time as the string moves, the world sheet.
The solutions of the classical equations of motion should be world sheets of
extremal
area. The action thus looks like \cite{Nambu70},\cite{Goto71}
\be
S=T\int{\rm d}\si{\rm d}\tau\lef(\frac{dX^{\mu}}{d\tau}\frac{dX_{\mu}}{d\tau}
\frac{dX^{\nu}}{d\si}
\frac{dX_{\nu}}{d\si}-\lef(\frac{dX^{\mu}}{d\tau}\frac{dX_{\mu}}{d\si}\rig)^2\rig)
^{1/2}\label{namugoto}
\ee
which is a highly inconvenient form because of the squaroot. It can,
however, similarly to the point particle,
be reformulated into the more convenient shape \cite{Brink-Di-Vecchia-Howe76}
\be
S=-\frac{T}{2}\int d^2\si\sqrt{-h}h^{\al\bet}\eta_{\mu\nu}\pr_{\al}X^{\mu}
\pr_{\bet}X^{\nu}\label{stringaction}.
\ee
$h^{\al\bet}$ is the inverse of the metric of the world sheet and $h$ is the
determinant of $h_{\al\bet}$. $T$ is the so-called string tension which is the
fundamental constant of string physics (often one uses the Regge slope
parameter
which is essentially the inverse of the squareroot of T).
Its dimension is [length$]^{-2}$ and since
string theory is supposed to describe gravity its natural scale should
be the Planck length inverted. From now on we will use $T=1/\pi $.

\ind Both (\ref{namugoto}) and (\ref{stringaction}) are invariant under
general coordinate transformations $\tau,\si\lra \tau '(\tau,\si),\si
'(\tau,\si)$,
usually referred to as diffeomorphism invariance. We can use this to eliminate
two
of the three independent components of the world sheet metric $h_{\al\bet}$
and write
it, at least locally, as $h_{\al\bet}=e^{\phi}\eta_{\al\bet}$.
$e^{\phi}$ is an
unknown conformal factor and $\eta_{\al\bet}$ is the flat world
sheet metric.
Inserted into the action, the conformal factor drops out leaving the
free field action. This is of course the other local symmetry of
(\ref{stringaction}), known as Weyl invariance, which is just
invariance under position dependent rescalings of the metric.
Just as for the point particle we must not forget
the constraint equation $\del
S/\del h_{\al\bet}=0$ which is usually formulated in terms of the
vanishing of the energy-momentum tensor
\be
T_{\al\bet}=-\frac{2\pi}{\sqrt{-h}}\frac{\del S}{\del h^{\al\bet}}
\label{emtensor}.
\ee
\ind The equations of motion of the gauge-fixed version of the action
(\ref{stringaction}) for Minkowskian world sheet metric
is the free two-dimensional wave equation. The
general solution, for periodic boundary conditions, can be written as
\be
X^{\mu}(\tau,\si)=q^{\mu}+p^{\mu}\tau+\frac{i}{2}\sum_{n\neq 0}\frac{1}{n}\lef(
\al^{\mu}_ne^{-2in(\tau+\si)}+\ti{\al}^{\mu}_ne^{-2in(\tau-\si)}\rig)
\ee
where $q^{\mu}$ and $p^{\mu}$ may be interpreted as the center of mass
position and the center of mass momentum of the string. The non-zero
equal-time Poisson bracket is given by
\be
[X^{\mu}(\tau,\si),\pr_{\tau}X^{\nu}(\tau,\si ')]_{PB}=\eta^{\mu\nu}\del
(\si-\si ').
\ee

We have here, for simplicity taken the space-time to be Minkowski.
In general we should, however, have $X(\tau,\si)$ to describe
the world sheet of the propagating string in a space-time manifold
${\cal M}$.
Since quantum gravity is expected to be included in string theory ${\cal M}$
should be determined by the dynamics of the string. This would
require a second quantized version which yet lacks a well defined
interpretation. In the first
quantized version we may examine whether the string is propagating in a
consistent way on ${\cal M}$.

\section{Fermionic strings}

In order to incorporate supersymmetry of the space-time
we must have supersymmetry on the world-sheet
\cite{Banks-Dixon-Friedan-Martinec88}. The action of the
fermionic string is the generalization of (\ref{stringaction}) by
local supersymmetry. We must to this end introduce superpartners
for the string coordinate $X^{\mu}$ as well as for the metric
$h^{\al\bet}$. The action is invariant under
supersymmetry transformations as well as reparametrizations
and local Weyl rescaleings.
In analogy to the bosonic
string action we may use invariances of the action to
gauge fix, and this results in the free field action
\be
S=\frac{-1}{2\pi}\int{\rm d}^2\si\lef(\pr^{\al}X^{\mu}\pr_{\al}X_{\mu}-
i\psi^{\mu}\ga^{\al}
\pr_{\al}\psi_{\mu}\rig)
\label{freefermstring}
\ee
Here $X^{\mu}$ is the corresponding field in the bosonic string and
$\psi^{\mu}$ is a Majorana spinor. Note that $\psi$ is a world sheet spinor not
a target space spinor. The supersymmetry transformations of this action are
\be
& &\del X^{\mu}=i\ep_2\psi^{\mu}_1-i\ep_1\psi^{\mu}_2 \nn \\
& &\del\psi^{\mu}_1=(\pr_0-\pr_1)X^{\mu}\ep_2 \hs{15mm}
\del\psi^{\mu}_2=-(\pr_0+\pr_1)X^{\mu}\ep_1
\ee
where  $\psi_1^{\mu}$ and $\psi_2^{\mu}$ are
the upper and lower components of $\psi^{\mu}$ and $\ep_1$ and $\ep_2$
are grassman odd parameters. We have here chosen
the basis for the world sheet gamma matrices
\be
\ga^0=\lef[\begin{array}{rr}
0 & 1 \\ -1 & 0
\end{array}\rig]
\hs{15mm}
\ga^1=\lef[\begin{array}{rr}
0 & 1 \\ 1 & 0
\end{array}\rig].
\ee
Also $\ga_0=-\ga^0$, $\ga_1=\ga^1$ and $\bar{\psi}=\psi^T\ga^0$.

The equations of motion becomes $\pr_+\pr_-X^{\mu}=0$,
$\pr_+\psi_1^{\mu}=0$ and $\pr_-\psi_2^{\mu}=0$ in terms of
$\pr_{\pm}\equiv \pr_0\pm\pr_1$. If we take the parameters $\ep_1,\ \ep_2$ to
be
coordinate dependent we may find the Noeter current of the symmetry.
We find that the current is given by two parts $J_+\propto
\pr_+X_{\mu}\psi^{\mu}_2$
and $J_-\propto \pr_-X_{\mu}\psi^{\mu}_1$. We note that using the equations
of motion $\pr_{\mp}J_{\pm}=0$. Those currents will be one part of
the generators of the superconformal algebra which we will study in general
in the next chapter.

There exists an additional difficulty (or blessing) for the fermionic fields.
When we vary the action in order to obtain Euler-Lagrange equations
of motion, we find that the surface term $\psi_+\del\psi_+-\psi_-\del\psi_-$
is required to vanish. For the of open strings this means that
$\psi_+\del\psi_+-\psi_-\del\psi_-$ is required to vanish at each end of the
string.
Without loss of generality we may chose say $\psi_+(\tau,0)=\psi_-(\tau,0)$.
This leaves us with the possibilities $\psi_+(\tau,2\pi)=\pm\psi_-(\tau,2\pi)$.
The periodic boundary condition is known as Neveu-Schwarz and the anti-periodic
as Ramond boundary condition. In the case of closed strings we may have that
both components vanish separately. This gives four possible combinations
Neveu-Schwarz--Neveu-Schwarz, Neveu-Schwarz--Ramond, Ramond--Neveu-Schwarz and
Ramond--Ramond.
We see from the different boundary conditions that in terms of Laurent
expansions,
Neveu-Schwarz fields will have half-integer modes while Ramond will be
integer moded.

Similarly to the bosonic string we must impose the vanishing of the
stress-energy
tensor. What is less obvious is that we should also require the generator of
the
supersymmetry, i.e. the currents discussed above, to vanish. I will not try to
justify
this but refer to the literature for this point. There is, however, a
convenient
reformulation where this constraint appears naturally. We introduce
what is known as superspace which includes the introduction of two (in two
dimensions) grassman odd coordinates. The theory is extended in such a way
that e.g. the action coincides with the starting point (\ref{freefermstring})
when
the fermionic coordinates have been integrated out. In superspace the
analogy of the stress-energy tensor will contain two parts, the generators
of the supersymmetry discussed above and the stress-tensor of
(\ref{freefermstring}). If we require the vanishing of the superspace
stress-energy tensor we will get the desired constraints. \\



We will return to those models after some general discussion of
world sheet dynamics. This is an example of conformal field theory
or in the fermionic case superconformal field theory.

\begin{itemize}

\item P.D.B. Collins, A.D. Martin and E.J. Squires, {\sc Particle
physics and cosmology}, Wiley interscience 1989

\item M.B. Green, J.H. Schwarz and E. Witten,
{\sc Superstring theory: I, II}, Cambridge University Press 1987

\item B. Greene, {\it Lectures on string theory in four
dimensions}, Trieste summer school on high energy physics
and cosmology 1990

\item J. Polchinski, {\it What is string theory?}, Les Houches summer school
"Fluctuating geometries in statistical mechanics and field theory", 1994

\item J. Schwarz, {\it Superconformal symmetry and superstring
compactification}, Int. J. Mod. Phys. A4 (1989) 2653-2713

\item S. Weinberg, {\it Particle physics: Past and future},
Int. J. Mod. Phys. A1 (1986) 135

\end{itemize}

\chapter{Conformal Field Theory}

The basic concepts of conformal field theories
in two dimensions may be derived in several ways
by analyzing specific examples such as the Ising model
or the world sheet
traced out by a propagating bosonic string. Conformal invariance
is by no means confined to two dimensions, but it is here it becomes
most restrictive in the sense that it provides more information
of the theory than in any higher dimension.

Fortunately two dimensions is, as mentioned above,
the case interesting to string
theories, where conformal field theories provide the theoretical
framework. Conformal invariance constrains the allowed space-time
dimensions or puts constraints on internal degrees of freedom. A
classification of conformally invariant theories could render us useful
information about, for instance, the possible classical solution spaces
for string theories. In two dimensions the demand for
conformal invariance provides much more information about
the theory. This can be
seen to originate from the fact that the conformal algebra here becomes
infinite-dimensional. In higher dimensions, conformal invariance does not
give much more information than scale invariance.
It is, however, enlightening to start in $d$ dimensions and later reduce
to two, and we will thus proceed accordingly.

\section{Conformal Theory In $d$ Dimensions}

We define the conformal group in $d$ dimensions as
the subgroup of coordinate transformations that, up to a space-time
dependent scale factor, leaves the metric invariant. More specificly,
consider a space {\Bb{R}}$^d$ with a flat metric $g_{\mu\nu}$,
having arbitrary signature. The conformal transformation thus induces
\be
g_{\mu\nu}(x)\lra g'_{\mu\nu}(x')=\Om (x)g_{\mu\nu}(x).\label{conftrans}
\ee
One may notice that the Poincar\'{e} group is a subgroup of the conformal
group since it leaves the metric invariant. \\
\ind We determine the infinitesimal generators of the conformal group by
considering infinitesimal coordinate transformations, $x'^{\mu}=x^{\mu}+
\ep^{\mu}$. We find, under this transformation, that the measure
transforms as $ds^2\lra
g'_{\mu\nu}dx^{\mu}dx^{\nu}+(\pr_{\mu}\ep_{\nu}+\pr_{\nu}\ep_{\mu})
dx^{\mu}dx^{\nu}$. Since we demand
that $ds^2$ is invariant under the transformation (\ref{conftrans}) we find
that $g_{\mu\nu}=g'_{\mu\nu}+\pr_{\mu}\ep_{\nu}+\pr_{\nu}\ep_{\mu}
+{\cal O}(\ep^2)$. When we compare this to eq.(\ref{conftrans}) we find
$\pr_{\mu}\ep_{\nu}+\pr_{\nu}\ep_{\mu}=(1-\Om)g_{\mu\nu}$. We now
contract with the inverse of the metric in order to obtain $(1-\Om)=
(2/d)\pr\cdot\ep$. Thus we have
\be
\pr_{\mu}\ep_{\nu}+\pr_{\nu}\ep_{\mu}=\frac{2}{d}\pr\cdot\ep g_{\mu\nu}.
\label{infconf}
\ee
If one acts on this by $\pr^{\nu}$ followed by $\pr^{\mu}$ we will find
that $\pr\cdot\ep$ fulfills the massless Klein-Gordon equation $\Box
\pr\cdot\ep=0$. Acting on (\ref{infconf}) with $\pr^{\nu}$ followed by
$\pr_{\nu}$, symmetrizing in $\mu$ and $\nu$ and then using (\ref{infconf})
again, we finally arrive at
\be
(1-\frac{2}{d})\pr_{\mu}\pr_{\nu}\pr\cdot\ep=0.
\ee
Thus $\ep$ is at the most quadratic in $x^{\mu}$ for $d>2$. We find the
solutions for $d>2$:
\begin{enumerate}
\item $\ep^{\mu}=a^{\mu}$ i.e. ordinary translations, (one part
of the above mentioned Poincar\'{e} group).
\item Rotations $\ep^{\mu}=\om^{\mu}_{\ \nu}x^{\nu}$ where
$\om^{\mu}_{\ \nu}$ is antisymmetric, (the other part of the
Poincar\'{e} group).
\item Dilatations $\ep^{\mu}=\la x^{\mu}$.
\item $\ep^{\mu}=b^{\mu}x^2-2x^{\mu}b\cdot x$ known as
special conformal transformations.
\end{enumerate}

\ind These infinitesimal transformations may be integrated to finite
ones.
The most devious is the special conformal transformation but utilizing a
neat maneuver this may be done in just a few lines. The important step
is to realize that expressed in new coordinates $x^{\mu}/x^2$ the
transformation may be written as
\be
\frac{x'^{\mu}}{x'^2}=\frac{x^{\mu}}{x^2}+b^{\mu}. \label{specconftrans}
\ee
for infinitesimal $b^{\mu}$.
This is nothing but an ordinary translation in those new coordinates and it
will of course integrate to a finite translation. What remains is then to
solve (\ref{specconftrans}) for $x'^{\mu}$ with finite $b^{\mu}$. This
yields the result
\be
x'^{\mu}=\frac{x^{\mu}+b^{\mu}x^2}{1+2b\cdot x+b^2x^2}.
\ee
\ind We can obtain the symmetry generators via Noether's prescription. The
symmetries should be generated by conserved charges which are the
space-integral of the time components of the conserved currents associated
with the symmetries. (Noether's theorem assures the existence of such a
current.) In general, the generators of local coordinate transformations
are constructed from the energy-momentum tensor $T_{\mu\nu}$, which is
symmetric $T_{\mu\nu}=T_{\nu\mu}$ and divergence free $\pr^{\nu}T_{\mu\nu}
=0$. We thus have that conformal transformations are associated with the
current
\be
j_{\mu}=T_{\mu\nu}\ep^{\nu}.
\ee
If we now require that the dilaton current is conserved we find that
$T_{\mu\nu}$ is traceless for conformal invariant theories i.e.
using (\ref{infconf}) $0=\pr
\cdot j^{dil}=1/2T^{\mu}_{\ \mu}\pr\cdot\ep\ \Longleftrightarrow\
T^{\mu}_{\ \mu}=0$. It is now straightforward to verify that the rest of
the currents are conserved without implying any further restrictions on
$T_{\mu\nu}$. We have indeed $\pr\cdot j=\frac{1}{2}T^{\mu\nu}(
\pr_{\mu}\ep_{\nu}+\pr_{\nu}\ep_{\mu})=
T^{\mu\nu}d^{-1}g_{\mu\nu}\pr\cdot\ep=0$ where we first have used eq.
(\ref{infconf}) and then the tracelessness of $T_{\mu\nu}$. \\
\ind We will now establish the restrictions that conformal symmetry
implies on a field theory. Let us first comment on the field content.
We call a field $\phi_i(x)$ "quasi-primary" if it transforms as
\be
\phi'_i(x')=[\Om(x)]^{\De_i/2}\phi_i(x)
\ee
under conformal transformations $x\lra x'$. $\Om$ is the scale factor
and $\De_i$ is known as the scaling dimension.
We will then have the relations
\be
\lan \phi'_1(x'_1)...\phi'_n(x'_n)\ran=\Om(x_1)^{\De_1/2}...
\Om(x_n)^{\De_n/2}\lan \phi_1(x_1)...\phi_n(x_n)\ran. \label{npointtrans}
\ee
for correlation functions.
We can calculate $\Om$ for the different cases of transformations, yielding
$\Om=1,\ \Om=\la^{-2}$ and $\Om=(1+2b\cdot x+b^2x^2)^2$ for Poincar\'{e}
transformations, dilatations, and special conformal transformations,
respectively. Invariance under Poincar\'{e} transformations implies that
correlation functions can only depend on distances $r_{ij}=|x_i-x_j|$. \\
\ind Specializing to the two-point function we find from invariance
under dilatations that $\lan\phi_1(x_1)\phi_2(x_2)\ran=C_{12}r_{12}^{
-\De_1-\De_2}$ by using eq.(\ref{npointtrans}). Finally one may,
utilizing the useful formulae
\be
|x'_1-x'_2|^2=\frac{|x_1-x_2|^2}{(1+2b\cdot x_1+b^2x_1^2)
(1+2b\cdot x_2+b^2x_2^2)}, \label{useful}
\ee
show that unless $C_{12}=0$ one must have $\De_1=\De_2$ i.e.
$\lan\phi_1(x_1)\phi_2(x_2)\ran=C_{12}r_{12}^{-2\De}$. For the
three-point function things look very much similar and the final result is
\be
\lan\phi_1(x_1)\phi_2(x_2)\phi_3(x_3)\ran=C_{123}r_{12}^{-\De_1-\De_2+
\De_3}r_{13}^{-\De_1-\De_3+\De_2}r_{23}^{-\De_2-\De_3+\De_1}.
\label{threepoint}
\ee
In the case of N-point functions for N larger than three, things however
turn to the worse. This is mainly due to the fact that (\ref{useful})
tells us that only cross-ratios $r_{ij}r_{kl}/r_{ik}r_{jl}$ are
invariant under the full conformal group. We will, for N larger than three,
obtain unknown functions of these cross-ratios in the N-point functions,
and the global conformal
invariance will not provide us with any further information. \\

\section{Conformal Theory In Two Dimensions}

Using a flat metric and $d=2$ eq.(\ref{infconf}) reduces to
Cauchy-Riemann's equations
\be
\pr_1\ep_1=\pr_2\ep_2\hs{20mm} \pr_1\ep_2=-\pr_2\ep_1.
\ee
Introducing the complex coordinates $z,\bz=x^1\pm ix^2$ we write $\ep(z)=
\ep^1+i\ep^2 \ \bar{\ep}(\bz)=\ep^1-i\ep^2$. We thus find that conformal
transformations in two dimensions coincide with analytic changes of
variables
\be
z\lra f(z)=z+\ep(z) \hs{20mm} \bz\lra \bar{f}(\bz)=\bz+\bar{\ep}(\bz),
\ee
and we find, using the metric $g_{z\bz}=g_{\bz z}=1/2, \ g_{zz}=g_{\bz\bz}
=0$, $dzd\bz\lra \frac{\pr f}{\pr z}\frac{\pr \bar{f}}{\pr \bz}dzd\bz$,
which means that $\Om^{-1}(z,\bz)=|\pr f/\pr z|^2$. $\bz$ is the complex
conjugate of $z$ but we will treat them as independent coordinates
justified by
the appearance of holomorphic and anti-holomorphic sectors, for example,
in the conformal algebra. \\
\ind We define primary fields of conformal weight $(h,\bar{h})$ to
transform as
\be
\phi(z,\bz)=\lef(\frac{\pr f}{\pr z}\rig)^h\lef(\frac{\pr \bar{f}}{\pr
\bz}\rig)^{\bar h}\phi'\lef(f(z),\bar f(\bz)\rig) \label{primtrans}
\ee
under conformal transformations. N.B. $\bar h$ is not the complex
conjugate of $h$. (\ref{primtrans}) means that the quantity
$\phi(z,\bz)dz^hd\bz^{\bar{h}}$ is conformally invariant. One may also
note the similarity to tensor transformations. Indeed, a tensor in
two-dimensional complex coordinates with $h$ lower $z$-indices and $\bar
h$-lower $\bz$-indices would transform as (\ref{primtrans}). So far, $h$
and $\bar h$ are, in principle, arbitrary real-valued numbers. They may,
for example, be non-integers, which shows the limitation of the tensor
analogy. $h$ and $\bar h$ may, however, be constrained by unitarity as we shall
discover in due time. \\
\ind We now turn to the question of quantum field theory. In a previous
subsection we gave restrictions on the energy-momentum tensor. In these
coordinates its only non-vanishing components are $T_{zz}=T(z)
\ T_{\bz\bz}=\bT$, where we have used conservation of $T$ to obtain
the holomorphic and anti-holomorphic structure. On the complex plane,
time ordering is replaced by radial ordering,
and we define radial order as
\be
R(\phi_1(z)\phi_2(w))=\lef \{ \begin{array}{ll} \phi_1(z)\phi_2(w)
& |z|>|w| \\ \phi_2(w)\phi_1(z) & |z|<|w|
\end{array}
\rig.
\ee
As mentioned above, the conformal transformations are generated by the
charge
\be
Q=\oint\frac{dz}{2\pi i}T(z)\ep(z)+
\oint\frac{d\bz}{2\pi i}\bT\bar{\ep}(\bz).
\ee
Henceforth, we will not bother to write out the anti-holomorphic sector.
The integral is taken around a circle of arbitrary radius centered at
the origin. Variations of fields are given by the commutator of $Q$ and
the fields. The "equal time" concept is then replaced by contour
integration around circles of constant radius. Applying this to a primary
field $\phi$ we get
\be
\del_{\ep}\phi(w)&=&\lef(\oint_{|z|>|w|}-\oint_{|z|<|w|}\rig)\frac{dz}
{2\pi i}\ep(z)R(T(z)\phi(w)) \nn \\
&=&\oint_w\frac{dz}{2\pi i}\ep(z)R(T(z)\phi(w)) \nn \\
&=&h\pr_w\ep(w)\phi(w)+\ep(w)\pr_w\phi(w).
\ee
We have here first deformed the contour in the fashion depicted in
fig.(2.1)
and then used
the infinitesimal version of eq.(\ref{primtrans}). Since we require
this equation to hold, we can read off the residues of the radial ordering
of the energy-momentum tensor with a primary field to be
\be
R(T(z)\phi(w))=\frac{h}{(z-w)^2}+\frac{1}{z-w}\pr_w\phi(w)+{\rm regular\
terms}.
\label{radordt}
\ee

\begin{figure}[H]
\begin{picture}(400,100)(5,5)
\put(100,20){\vector(0,1){70}}
\put(65,55){\vector(1,0){70}}
\put(111,66){\circle*{2}}
\put(104,59){w}
\put(100,55){\circle{50}}
\put(120,55){\vector(0,1){1}}

\put(147,55){\line(1,0){5}}

\put(200,20){\vector(0,1){70}}
\put(165,55){\vector(1,0){70}}
\put(211,66){\circle*{2}}
\put(211,66){w}
\put(200,55){\circle{20}}
\put(210,55){\vector(0,1){1}}

\put(250,52){=}

\put(300,20){\vector(0,1){70}}
\put(265,55){\vector(1,0){70}}
\put(311,66){\circle*{2}}
\put(311,66){w}
\put(311,66){\circle{18}}
\put(321,68){\vector(0,1){1}}


\end{picture}
\caption{Integration contours in the complex plane.}
\end{figure}
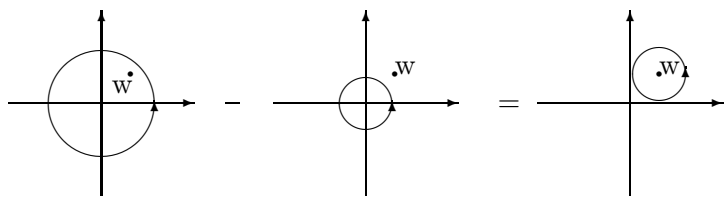
\noindent From now on, we drop the radial order symbol and instead use ordinary
operator product expansions i.e eq.(\ref{radordt}) without $R$.

\ind From (\ref{radordt}) we can derive relations between different
correlation functions, known as conformal Ward identities. One gets
\be
\lan T(z)\phi_1(w_1)....\phi_n(w_n)\ran=\sum^N_{i=1}\lef(\frac{h_i}{(z-
w_i)^2}+\frac{1}{z-w_i}\pr_{w_i}\rig)\lan\phi_1(w_1)....\phi_n(w_n)\ran.
\label{virward}
\ee
This essentially means that we have meromorphic functions with
singularities at the insertion points of the fields. \\
\ind Not all fields satisfy eq.(\ref{primtrans}).
Derivatives of primary fields may, for example,
transform in more complicated ways.
By a secondary field we mean a field that gives higher order poles than
the double pole appearing in (\ref{radordt}), when we take the operator
product expansion with the energy-momentum tensor.
Secondary fields can be obtained by repeatedly taking
operator product expansions of the energy-momentum tensor and a
primary field. Secondary fields are generally denoted descendants. This
means that we can
group all fields into families, each containing one primary field and its
descendants. The number of families may be finite (minimal models) or infinite.
If there exists a larger symmetry algebra then the conformal algebra, we may
classify with respect to this larger algebra. If there are a
finite number of primary fields with respect to the larger
symmetry algebra and we have rational conformal weights, we call the theory
a rational conformal field theory.

The conformal families thus correspond to modules of the two
copies of the Virasoro algebra generated by the holomorphic and
anti-holomorphic parts of the stress-energy tensor.
In order to completely specify the conformal field theory one takes
what is known as the conformal bootstrap approach. It consists of
the assumption that the local fields of the theory $\phx$ for fixed
arguments form an associative algebra over {\Bb C}
\be
\phx_i(z,\bz)\phx_j(w,\bar{w})=
\sum_kC_{ij}^{\ \ k}(z,\bz,w,\bar{w})\phx_k(w,\bar{w}). \label{opalg}
\ee
This is known as the operator algebra of the conformal field theory.
The "structure constants" $C_{ij}^{\ \ k}$ may in principle be
determined by the structure constants of the algebra involving only
primary fields.

The energy-momentum tensor can be regarded as a descendant of
the identity field. One of the simplest ways of convincing oneself of
this fact is displayed below. The operator product expansion of two
stress-energy tensors is
\be
T(z)T(w)=\frac{c/2}{(z-w)^4}+\frac{2}{(z-w)^2}T(w)+\frac{1}{z-w}\pr_wT(w)
+{\rm regular\ terms} \label{opet}
\ee
c is known as the conformal anomaly and
may be restricted by imposing unitarity. Expanding primary fields as
\be
\phi(z)=\sum_{n\in\Bbb Z}\phi_nz^{-n-h} \hs{15mm}
\phi_n=\oint_0\frac{dz}{2\pi i}z^{n+h-1}\phi(z)
\ee
where $h$ is the conformal weight,
and the energy-momentum tensor as
\be
T(z)=\sum_{n\in\Bbb Z}L_nz^{-n-2} \hs{15mm}
L_n=\oint_0\frac{dz}{2\pi i}z^{n+1}T(z)
\ee
we may represent the transformation law (\ref{radordt}) as
\be
[L_n,\phi_m]&=& \oint_0\frac{dw}{2\pi i}\oint_0\frac{dz}{2\pi i}
z^{n+1}w^{m+h-1}(T(z)\phi(w)-\phi(w)T(z)) \nn \\
&=&\oint_0\frac{dw}{2\pi i}\oint_w\frac{dz}{2\pi i}z^{n+1}w^{m+h-1}
\lef(\frac{h}{(z-w)^2}\phi(w)+\frac{1}{z-w}\pr_w\phi(w)\rig) \nn \\
&=&(n(h-1)-m)\phi_{n+m}.
\ee
Redoing this calculation for the stress tensor
we find the well known Virasoro algebra
\be
[L_m,L_n]=(m-n)L_{m+n}+\frac{c}{12}m(m^2-1)\del_{m+n,0}
\label{virasoro}
\ee

To see that the energy-momentum tensor is a descendant we note that
we may extract the descendants from the less singular parts of
operator product expansions of $T(z)$ with a primary field,
\be
T(z)\phi(w,\bar w)\equiv\sum_{n\geq0}(z-w)^{n-2}\hat{L}_{-n}\phi(w,\bar w)
\ee
The descendant fields are as usual given by
\be
\hat{L}_{-n}\phi(w,\bar w)=\oint\frac{dz}{2\pi i}\frac{1}{(z-w)^{n-1}}
T(z)\phi(w,\bar w).
\ee
and are of conformal weight $(h+n,\bar h)$. Applying this to the identity
field we find
\be
\hat L_{-2}1(w)=\oint\frac{dz}{2\pi i}\frac{1}{z-w}T(z)1=T(w)
\ee
confirming that $T(z)$ is indeed a descendant. We note that,
in order to make contact with the
usual Virasoro modes, we have $\hat{L}_{-n}
\phi(z,\bz)\lra L_{-n}\phi(z,\bz)$ in the limit $z,\bz\rightarrow0$.

\ind Introducing the notion of adjoint
\be
\lef(\phi(z,\bz)\rig)^{\dagger}=\phi\lef(\frac{1}{\bz},\frac{1}{z}\rig)
\frac{1}{\bz^{2h}}\frac{1}{z^{2\bar h}} \label{adjprim}
\ee
one sees that the equality for the energy-momentum tensor
\be
T^{\dagger}(z)=\sum L\dagg _m \bz^{(-m-2)} \hs{20mm}
T\lef(\frac{1}{\bz}\rig)\frac{1}{\bz^4}=\sum L_m\bz^{(m-2)}
\ee
results in $L\dagg_m=L_{-m}$. Eq.(\ref{adjprim}) may look a bit funny but
one should bear in mind that the operation of taking the adjoint of
$e^{i(t+\si)}$ in Minkowski space is equivalent to time reversal
$\tau\lra-\tau$ in Euclidian space of $e^{\tau+i\si}$ and is equivalent
to letting $z\lra1/\bz$ on the complex plane $z=e^{i(t+\si)}$. The
factors $\frac{1}{\bz^{2h}}\frac{1}{z^{2\bar h}}$ come from requiring
the correct conformal transformation properties of the adjoint.

We assume the existence of a vacuum state $\vacr$ in the theory.
Requiring that $lim_{z\rightarrow 0}T(z)\vacr$ is well behaved means
$L_m\vacr=0 \ m\geq-1$. The modes $L_{0,\pm1}$ generates a Sl(2,{\Bb{R}})
algebra which means the vacuum should be Sl(2,{\Bb{R}}) invariant. We may
associate a state $|h\ran=\phi(0)\vacr$ to each primary field of
conformal weight $h$. One then finds, using $[L_n,\phi(w)]=h(n+1)w^n\phi(w)
+w^{n+1}\pr_w\phi(w)$, that
\be
L_0|h\ran=h|h\ran \hs{20mm}L_n|h\ran=0 \hs{10mm}n>0. \label{hw}
\ee
We call states that satisfy (\ref{hw}) highest weight states
with respect to the conformal algebra in analogy
to a similar construction in
Lie algebras. Descendant states are given by acting on a highest
weight state with polynomials of $L_{-n}, \
n>0$. Using eq.(\ref{adjprim}) and the definition of $|h\ran$ one can
show that $ \lan h|=|h\ran\dagg$ and $\lan h|L_{n_1}...L_{n_k}, \ n_i>0,
 \ 1\leq i\leq k$ are descendants of the out state. Using the Virasoro
algebra we may show that
\be
\lan h|L_nL_{-n}|h\ran=\lef(2nh+\frac{c}{12}n(n^2-1)\rig)\lan h|h\ran .
\ee
Taking $n$ large we see that $c\geq 0$ and $n=1$ gives $h\geq 0$, if we
want the state-space to have non-negative inner product. \\
\ind The highest weight representations are completely determined by $c$
and $h$. Representations
of the Virasoro algebra are non-unitary unless
\be
c\geq 1 \hs{15mm} h\geq 0
\ee
or
\be
c=1-\frac{6}{m(m+1)} \hs{15mm} m=2,3,... \hs{15mm} \nn \\
h=\frac{((m+1)p-mq)^2-1}{4m(m+1)} \hs{10mm}
1\leq q\leq p\leq m-1.  \label{discr}
\ee
The latter of these conditions were found by Friedan, Qui and Shenker,
\cite{Friedan-Qiu-Shenker84}.
It was shown by Goddard, Kent and Olive \cite{Goddard-Kent-Olive86}
that they also were sufficient and led to unitary
representation. Here explicit realizations in terms of coset constructions
were also found, see chapter 7.3 below for some details. The
representations of (\ref{discr}) contain
null-states, i.e. states that decouple from all other states in the theory
including their adjoint state, yielding zero norm of the state.
We have, for example,
for $m=3 \ h=1/16$ that $(L_{-2}-\frac{4}{3}L^2_{-1})|\frac{1}{16}\ran$
is null. In order to obtain irreducible representations,
we must consider a smaller state space, in which null-states are
excluded. \\

The fields that depend on either the holomorphic or anti-holomorphic
coordinate close under the operator algebra (\ref{opalg}). The holomorphic
fields transform in some representation of some algebra known as the
chiral part of the symmetry algebra. Likewise for the anti-holomorphic
fields which transforms under the anti-chiral algebra. Together the chiral
and anti-chiral parts give the symmetry algebra of the theory.
One part of the symmetry algebra consists of the two copies of the
Virasoro algebra.

In classifying the conformal fields with respect to the Virasoro algebra
one may also find it useful to classify with respect to the full symmetry
algebra. This may render us a finite number of primary fields with
respect to a larger algebra.
In general, the representation theory of the symmetry algebra
is unknown. The most studied exception is when the symmetry algebra
is at least the semidirect sum of the Virasoro algebra  ${\cal V}$
and what is known
as an affine Lie algebra $\hg$ \footnote{Semidirect sum means that the
affine Lie algebra is an ideal of the Virasoro algebra i.e.
$[{\cal V},{\cal V}]\in {\cal V}$, $[{\cal V},\hg]\in \hg$ and
$[\hg,\hg]\in\hg$.}.
If we further require that the stress-tensor
of the theory is bilinear in the affine generators we have what is known
as a WZNW theory. We will in detail discuss those
models in chapter seven.

We will first, however, mention a few words about incorporating supersymmetry
in the conformal field theory.

\section{Superconformal field theory}

We will here briefly mention theories which, apart from
being conformal, also are supersymmetric. We will contain ourselves with
the simplest example namely $N=1$ superconformal theory and stay in
Euclidean superspace.
We begin by a few useful definitions and notations on superspace.
We take $(1,1)$ Euclidean superspace to be parametrized by the coordinates
\be
Z=(z,\th;\bz,\bth)
\ee
and introduce supercovariant derivatives
\be
D=\frac{\pr}{\pr\th}+\th\pr_z \hs{15mm}
\bD=\frac{\pr}{\pr\bth}+\bth\pr_{\bz}
\ee
These will satisfy
\be
D^2=\pr_z \hs{15mm} \bD^2=\pr_{\bz} \hs{15mm} \{D,\bD\}=0.
\ee
The superspace integral is defined from
\be
\int{\rm d}z{\rm d}\bz{\rm d}\th{\rm d}\bth
\equiv\int{\rm d}z{\rm d}\bz D\bD=\int{\rm d}z{\rm d}\bz
\frac{\pr}{\pr\th}\frac{\pr}{\pr\bth}.
\ee
We take the line element to be
\be
{\rm d}s^2={\rm d}S{\rm d}\bar{S} \hs{15mm}
{\rm d}S={\rm d}z-{\rm d}\th\th \hs{10mm}
{\rm d}\bar{S}={\rm d}\bz-{\rm d}\bth\bth
\ee
The superconformal transformations of interest to us are those that
scales the line element and are of the form
\be
{\rm d}S\lra{\rm d}S'=\om{\rm d}S \hs{10mm}
{\rm d}\bar{S}\lra{\rm d}\bar{S}'=\bar{\om}{\rm d}\bar{S} \hs{15mm}
{\rm d}s^2\lra{\rm d}s'^2=\Om{\rm d}s^2 \label{superconftrans}
\ee
where $\om$ and $\bar{\om}$ are functions of $z,\th$ and $\bz,\bth$
respectively. We will from now on only consider the part depending
on $z,\th$ since we have taken the conformal transformation to
decouple in the sense of eq.(\ref{superconftrans}). We will thus
find that we have to copies of the superconformal algebra one
holomorphic and one anti-holomorphic, in analogy to the non-supersymmetric
case.

We now whish to give an alternative derivation of the Virasoro algebra or
rather its
centerless analogy, using infinitesimal generators of the analytic
change of variables. Take as change of variables
\be
f(z)=z+\ep(z)
\ee
where $\ep$ is an infinitesimal function. We define the action of $f$ on a
field by $f:\phi\rightarrow\phi^f$ where $\phi^f(z)\equiv\phi(f^{-1}(z))$. In
terms
of the Laurent expansion $\ep(z)=\sum\ep_nz^{n+1}$ and defining
$\phi^f(z)=(I+\sum\ep_nl_n)\phi(z)$ we find the infinitesimal generators
\be
l_n=-z^{n+1}\pr_z.
\ee
The generators will fulfill a centerless Virasoro algebra
\be
[l_n,l_m]=(n+m)l_{n+m} \label{loopvir}.
\ee

In superspace we must of course include the fermionic coordinate $\th$. It is
not difficult to verify that the generator (I will for convenience use the
same symbol $l_n$ since no confusion should be possible)
\be
l_n=-z^{n+1}\pr_z-\frac{n+1}{2}z^n\th\pr_{\th} \label{supconf1}
\ee
gives the same algebra (\ref{loopvir}). We should also expect a second
generator which corresponds to the supersymmetry mentioned in the
previous chapter. To this end we introduce the infinitesimal generator
\be
g_r=-z^{r+1/2}(\pr_{\th}-\th\pr_z).
\ee
One may check that $l_n$ and $g_r$ obey the algebra
\be
[l_n,g_r]=(\frac{n}{2}-r)g_{r+n}
\hs{15mm} \{g_r,g_s\}=2l_{r+s} \label{supconf2}
\ee
in addition to (\ref{loopvir}). This is the centerless $N=1$ superconformal
algebra. The quantum corrections to this algebra is incorporated by a one
dimensional center which is known as the superconformal anomaly.


This may be described in terms of a generalized stress-tensor generating
transformations in superspace in analogy to the non-supersymmetric case.
We introduce the super stress-energy tensor (super-holomorphic part)
\be
T(Z)=\frac{1}{2}G(z)+\th T(z).
\ee
In terms of the abbreviations $z_{12}=z_1-z_2-\th_1\th_2$ and
$\th_{12}=\th_1-\th_2$ the operator product expansion
of the super stress-energy tensor with itself may be found to be
\be
T(Z_1)T(Z_2)=\frac{1}{4}\frac{\hat{c}}{z_{12}^3}+
\frac{3}{2}\frac{\th_{12}}{z^2_{12}}T(Z_2)+
\frac{1}{2}\frac{D_2T(Z_2)}{z_{12}}+\frac{\th_{12}}{z_{12}}\pr_2T(Z_2)+{\rm
r.t.}
\ee
If we expand both sides of this and identify we find, in addition to the
ordinary stress-energy tensor operator product expansions (\ref{opet}),
also
\be
& &G(z_1)G(z_2)=\frac{\hat{c}}{(z_1-z_2)^3}+
\frac{2T(z_2)}{z_1-z_2}+{\rm r.t.} \nn \\
&
&T(z_1)G(z_2)=\frac{3}{2}\frac{G(z_2)}{(z_1-z_2)^2}+
\frac{\pr_2G(z_2)}{z_1-z_2}+{\rm r.t.}.
\ee
Here $\hat{c}=2c/3$. This corresponds to the $N=1$ superconformal algebra
(\ref{supconf2}) with central extension.

As we saw in the previous chapter we have two types of boundary conditions
Neveu-Schwarz or Ramond. The different
boundary conditions modify the mode expansions of $G(z)$.
For the Neveu-Schwartz case the Laurent coefficients take half-integer values,
while the Ramond case take integer values expressed as functions on the
complex plane. In terms of
commutators we find
\be
& &[L_m,L_n]=(m-n)L_{n+m}+\frac{c}{12}m(m^2-1)\del_{m+n,0} \nn \\
& &\{G_r,G_s\}=2L_{r+s}+\frac{c}{3}(r^2-\frac{1}{4})\del_{r+s,0} \nn \\
& &[L_m,G_r]=(\frac{m}{2}-r)G_{m+r}
\ee
where $r\in{\Bbb Z}+\frac{1}{2}$ for Neveu-Schwartz and $r\in{\Bbb Z}$ for
Ramond.
We note here a difference between the two sectors. In the Neveu-Schwartz sector
$L_0,L_{\pm1},G_{\pm1/2}$ form a centerless subalgebra. For the Ramond sector
the only subalgebra is generated by $L_0,L_{\pm1}$ which is also present
in the purely bosonic case.

The representation theory of the superconformal algebra follows, in
principle, the same steps as the conformal case. In the Hilbert space
picture we apply creation operators $L_{-n}$ and $G_{-r}$ $n,r>0$ on a
highest weight primary to build the Verma modules. In this way we construct
modules
of either the Neveu-Schwartz or the Ramond sector. There exists, however,
fields known as spin fields which interpolates between the sectors. The reason
why we cannot satisfy ourselves by considering either the Neveu-Schwartz or
the Ramond sector will became clear when we discuss the explicit example of
the fermionic string. \\

In analogy to the WZNW model, which is the best studied class of conformal
field theories, one may realize a large class of superconformal field
theories by a supersymmetric extension of the WZNW model. In the
supersymmetric case the superconformal symmetry is accompanied by an
super affine symmetry, giving rise to a super affine Lie algebra. We
will return to those issues in chapter seven.

\begin{itemize}

\item A.A. Belavin, A.M. Polyakov and A.B. Zamolodchikov,
{\it Infinite conformal symmetry in two-dimensional quantum field theory},
Nucl. Phys. B241 (1984) 333

\item P. Ginsparg, {\it Applied conformal field theory}, In: Fields,
strings and critical phenomena, Les Houches, Session XLIX 1988

\item M. Henningsson, {\sc Out of flatland: Applications of
Wess-Zumino-Witten models to string theory}, PhD thesis,
G$\ddot{\rm o}$teborg 1992

\item J.L. Petersen, {\sc Notes on conformal field theory}, University of
Copenhagen, The Niels Bohr institute, Spring 1994

\end{itemize}

\chapter{String interlude}

We will here apply some of the conformal field theory techniques,
described in the last chapter, to string theories. I will in
these examples use the
bosonic string and try to outline the necessary generalizations and
differences to other string theories.

The obvious task is to quantize the string, which at
first sight may seem straightforward for the bosonic case.
There are, however, some subtleties
involved which we will try to expose. \\
\ind Canonical quantization follows in the standard fashion by letting
$[\ ,\ ]_{PB}\lra -i[\ ,\ ]$. This yields the nonvanishing commutator
identities
\be
[\hat{q}^{\mu},\hat{p}^{\nu}]=i\eta^{\mu\nu} \hs{15 mm} [\hat{\al}^{\mu}_m,
\hat{\al}^{\nu}_n]=[\hat{\ti{\al}}^{\mu}_m,\hat{\ti{\al}}^{\nu}_n]
=m\del_{m+n,0}\eta^{\mu\nu}
\ee
where the "hat" indicates operator quantities. (We will not bother to
write
"hats" except for the next few lines.) We also demand that
$\hat{X}^{\mu}(\tau,\si)$ is hermitean which gives the hermiticity properties
\be
(\hat{q}^{\mu})^{\dagger}=\hat{q}^{\mu} \hs{8mm} (\hat{p}^{\mu})^{\dagger}
=\hat{p}^{\mu} \hs{8mm} (\hat{\al}^{\mu}_m)^{\dagger}=\hat{\al}^{\mu}_{-m}
\hs{8mm} (\hat{\ti{\al}}^{\mu}_m)^{\dagger}=\hat{\ti{\al}}^{\mu}_{-m}
\ee
The Fock space is constructed in the usual way by acting with creation
operators
on a vacuum defined by the following properties
\be
\hat{p}|0,0\ran=\hat{\al}_n|0,0\ran=\hat{\ti{\al}}_n|0,0\ran=0 \hs{10mm}
 \forall n>0
\nn \\
\hs{10mm} \lan0,0|0,0\ran=1 \hs{10mm} \lan0,0|=(|0,0\ran)^{\dagger} \hs{15mm}
\ee
A state with momentum $k$ is constructed as $e^{ik\cdot
X(z)}|0,0\ran\equiv|0,k\ran$. This terminology is of course due to
that $p^{\mu}|0,k\ran=k^{\mu}|0,k\ran$.

It is important to note that this state space  contains negatively normed
states. This may be seen from the
example
\be
\lan0,0|\al_1^{\mu}\al_{-1}^{\nu}|0,0\ran=\eta^{\mu\nu}
\ee
which may of course be negative since we have Minkowskian signature on
$\eta^{\mu\nu}$. These states have to be projected out somehow.\\
\ind We now Wick rotate to Euclidean world sheet coordinates i.e.
$ t=-i\tau$ and introduce the complex variables $z=e^{2(\tau+i\si)}\ \
\bar{z}=e^{2(\tau-i\si)}$. This is a mapping from the world sheet cylinder
to the two-dimensional complex plane, where radial ordering substitutes
time ordering among operators. We find
\be
X^{\mu}(z,\bz)=q^{\mu}-\frac{i}{4}ln(z\bz)+\frac{i}{2}
\sum_{n\neq 0}\frac{1}{n}(\al^{\mu}_nz^{-n}+\ti{\al}^{\mu}\bz^{-n})
\ee
If we now split the mode expansion of
$X^{\mu}$ into annihilation $X^{\mu,(+)}$ and creation $X^{\mu,(-)}$
parts we find in these new coordinates the commutation relation
\be
[X^{\mu,(+)}(z,\bz),X^{\nu,(-)}(w,\bw)]=
-\frac{\eta^{\mu\nu}}{4}\lef(ln(z-w)+ln(\bz-\bw)\rig).
\ee

We still have not a well-defined theory due to the negatively normed
states. Also we have not fixed the reparametrization invariance by
just chosing the metric.
What we need is to impose the equations
of motion of the metric, which are lost when fixing the gauge.
Classically this amounts to
\be
T_{\al\bet}\propto\frac{1}{\sqrt{-h}}\frac{\pr S}{\pr h^{\al\bet}}=0.
\ee
We will find that this also projects out the negatively normed states.

For the bosonic string we have that the stress-energy tensor, apart from
being traceless, divides into a holomorphic and an anti-holomorphic part.
We denote the former by $T(z)$ and using the explicit form
$T(z)=-2\!\!:\!\!\pr_zX^{\mu}\pr_zX_{\mu}\!\!:$ we may find the
operator product expansion
\be
T(z)T(w)=\frac{D/2}{(z-w)^4}+\frac{2T(w)}{(z-w)^2}+
\frac{\pr_zT(z)}{z-w}+{\rm regular\ terms} .
\ee
This relation means that the Laurent expansion coefficients of $T(z)$
satisfy a Virasoro algebra with central extension $c=D$.
D may in the flat case be interpreted as the dimension of target space.

If we naively impose the vanishing of the stress-energy tensor we run into
trouble.
Indeed, $L_n|\psi\ran=0$ for all $n$ means $0=[L_n,L_{-n}]|\psi\ran=
(2nL_0+\frac{D}{12}n(n^2-1))|\psi\ran$. We should at the quantum level rather
impose the Gupta-Bleuler like constraint, i.e. that $T(z)$ should
vanish in expectation values between physical states. If we take
\be
(L_0-a)|\psi\ran=0 \hs{15mm} L_n|\psi\ran=0 \ \ n>0 \label{virproj}
\ee
and likewise for the anti-holomorphic sector $\bar{L}$,
we find that $\lan\psi|T(z)|\psi\ran=0$ and likewise for $\bar{T}(\bz)$.
We have here introduced
a constant $a$ for the zero mode which may appear from reordering. In fact
a consistent string theory has in general $a\neq0$.

{}From the projection conditions (\ref{virproj}) we see that any state,
which is a sum of states the
form $L_{-n}|\chi\ran$ $n>0$, for arbitrary $|\chi\ran$,
decouples from every physical state.
If, furthermore, $L_{-n}|\chi\ran$ is a physical state in the
sense of eq.(\ref{virproj}) we have a physical state that decouples from all
other
physical states, and hence must be regarded as zero in the physical state
space.
Those states are known as null-states. The physical state space will be the
physical states defined from (\ref{virproj}) modulo the null-states.

As is well known by now, the ground state of the bosonic string is tachyonic.
This may be seen from the the operator expansion
\be
T(z)\!:\!e^{ik_{\mu}X^{\mu}(z)}\!\!:=
\frac{k^2/8}{(z-w)^2}\!:\!e^{ik_{\mu}X^{\mu}(w)}\!:\!+
\frac{\pr_w\!:\!e^{ik_{\mu}X^{\mu}(z)}\!:\!}{z-w}+{\rm r.t.}
\ee
In the conformal field theory language this means that
$:\!e^{ik_{\mu}X^{\mu}(z)}\!:$ is a conformal primary of weight
$k^2/8$. $:\!e^{ik_{\mu}X^{\mu}(z)}\!:$ is also known as the tachyon vertex
operator.
We then have, using (\ref{hw}) and (\ref{virproj}) that $k^2=8a$. In
what is known as the critical case $a=1$ and we find the mass $M^2=-k^2$ to
be negative.

The next excited level is massless and naturally includes the graviton.
One may proceed to find the conditions for a unitary physical state space
to be $a=1$ and $D=26$ or $a\leq1$ and $D\leq 25$. The critical
case is the one with $a=1$ and $D=26$.
For the critical case the physical degrees of freedom are constrained to the 24
transverse components of $X^{\mu}$.

In the path integral approach to quantization of the bosonic string one must be
careful when chosing the flat metric. Changing variables in this fashion
yields a determinant, that appears from the measure \cite{Alvarez83}.
This may be represented by an action of anti-commuting Fadeev-Popov ghosts
\be
S_{\rm gh}\sim\int{\rm d}^2z(b_{zz}\pr_{\bz}c^z+{\rm c.c.})
\ee
where we have for convenience chosen holomorphic and anti-holomorphic
coordinates.
This action is conformally invariant and the stress-energy tensor is given by
\be
T_{\rm gh}(z)=-2b\pr_zc(z)-\pr_zbc \label{stressgh}
\ee
where $b=b_{zz}$ and $c=c^z$. $b$ and $c$ have conformal weight two and minus
one respectively. Using the operator product expansion $b(z)c(w)=1/(z-w)$
one may easily find the conformal anomaly of (\ref{stressgh}) to be -26.

The critical bosonic string theory is hence anomaly free if $D=26$. This may be
rephrased in
a BRST quantization scheme (see chapter six) to yield $a=1$ as well as $D=26$.
\\

In the case of fermionic strings we must impose that the positive frequency
modes of the superconformal algebra annihilates physical states. We also have,
in analogy to the bosonic string, null states in the Verma module.

We start by examining the ground state $|0,k\ran\equiv {\rm lim}_{z\rightarrow
0}
:\!e^{ik_{\mu}X^{\mu}(z)}\!:|0,0\ran$. We will accordingly with the
analysis for the bosonic case find the mass-shell condition $k^2=8a$.
Consider now the descendant $G_{-1/2}|0,k\ran$. If we require that it is
physical then it must also be a null state. The on-shell condition
$(L_0-a)G_{-1/2}|0,k\ran=0$
reads $k^2/8+1/2=a$ and the zero norm condition
$G_{1/2}G_{-1/2}|0,k\ran=0$ gives $k=0$ and hence $a=1/2$. This
corresponds to the critical fermionic string. Using this value for $a$
for the ground state we find that the tachyon is
present in the spectrum of the Neveu-Schwartz sector.

The undesired tachyon may be excluded from the theory using
what is known as the GSO projection \cite{Gliozzi-Scherk-Olive77}.
This is a truncation of the spectrum to those states which has
$(-1)^{F}=1$ where $F$ is the parity of the state and bosonic
fields $X^{\mu}$ have plus one while fermionic fields $\psi^{\mu}$ has
minus one. The GSO projection yields that we have a local theory in the sense
that in the operator product expansions no branch cuts
will appear in the model. In general the spin field, which we mentioned in
the last chapter, has branch cuts in operator product expansions
with the superconformal generators.

In the path-integral treatment of the
fermionic string we get an additional contribution to the conformal
anomaly from the fermions as well as from the bosonic ghosts, the latter
being superpartners of the fermionic ghosts in the bosonic string. The former
contribute by $D/2$ and the latter by eleven. Adding up we find $3D/2=15$ i.e.
the critical dimension is $D=10$.

In the heterotic string one essentially uses the fermionic string for the
rightgoing modes and the bosonic string for the leftgoing modes.
In the leftgoing sector one keeps ten bosons and adds 32 free fermions
which are not superpartners of the bosons. This exactly cancels the -26
contribution from the fermionic ghosts. \\


We have so far only concerned ourselves with world-sheet supersymmetry.
It turns
out that if we desire space-time supersymmetry we will have $N=2$ supersymmetry
on the world-sheet \cite{Banks-Dixon-Friedan-Martinec88}. The second
supersymmetry
is, however, not a local symmetry of the action. If it were, we would have the
critical dimension reduced to two. This is undesirable since we would
like to interpret the ten dimensions as our four (possibly) flat
space-time dimensions plus internal degrees of freedom.

In space-time the two different boundary conditions have a
remarkable effect. The Neveu-Schwarz sector corresponds to
space-time bosons and the Ramond sector to space-time fermions.
The GSO projection turns out to provide with a pairing in the sense
that for each mass-level we get an equal number of bosons and
fermions.

At this point one may take two different points of views. Either we
insist that we have a geometrical meaning of the redundant dimensions
or we do not. The redundant dimensions in the former case are removed
by compactification.
In the latter case we just substitute the superfluous
degrees of freedom by any appropriate unitary conformal field theory.
We also need to require that our theory is what is known as modular
invariant. We will discuss this issue in chapter eight. This should be
sufficient for obtaining a consistent unitary string theory
propagating in what is no longer necessarily a flat space time. \\

It is thus desirable to analyze conformal field theories and superconformal
field theories in order to
find appropriate candidates for realistic string theories. We will in
chapter seven discuss the WZNW models and their supersymmetric extensions.
In order to fully appreciate the essence of WZNW models we will in the
following two chapters discuss the affine Lie algebra.
This is the algebra of the symmetry which is
present in WZNW models in addition to the conformal symmetry.

\begin{itemize}

\item M.B. Green, J.H. Schwarz and E. Witten,
{\sc Superstring theory: I, II}, Cambridge University Press 1987

\item B. Greene, {\it Lectures on string theory in four
dimensions}, Trieste summer school on high energy physics
and cosmology 1990

\item J. Polchinsky, {\it What is string theory?}, Les Houches summer school
"Fluctuating geometries in statistical mechanics and field theory", 1994

\item J. Schwarz, {\it Superconformal symmetry and superstring
compactification}, Int. J. Mod. Phys. A4 (1989) 2653-2713

\end{itemize}

\chapter{Affine Lie algebras}

As mentioned above, affine Lie algebras arise naturally in the
context of WZNW models. The WZNW model possesses, apart from a conformal
symmetry, also an affine symmetry which has a symmetry algebra of the affine
type. In all the papers contained in this thesis
we make extensive use of affine Lie algebras and
their representation theory. Of special importance is the structure of
Verma modules over affine Lie algebras. This will be discussed separately
in chapter five. \\

We begin with a few lines on Lie algebras in general.
An algebra $a$
is defined as vector spaces $L$ over a field $F$ which are endowed with a
bilinear operation
\be
\diamond:a\times a\lra a \label{algebra}
\ee
A Lie algebra is required to have the additional structure, such that
this map, the Lie bracket, $[,]:g\times g\lra g$ satisfies
\be
& & [x,x]=0 \hs{10mm} \forall x\in g \hs{10mm} ({\rm antisymmetry}) \nn \\
& & [x,[y,z]]+[y,[z,x]]+[z,[x,y]]=0 \hs{5mm} \forall x,y,z\in g
\hs{5mm} ({\rm Jacobi\ identity} )
\ee
We can note that the requirement of antisymmetry, when applied to $x+y$, gives
the
usual antisymmetry relation $[x,y]=-[y,x]$. That $[x,y]=-[y,x]$ implies
$[x,x]=0$ is valid unless the characteristic of the field $F$ is 2 i.e.
that there exists $\xi\in F$ that satesfies $\xi+\xi=0$.

Given an associative algebra $(a,\diamond)$ one may construct a Lie algebra
$g(a)=(a,[,])$ by defining the Lie bracket
\be
[x,y]=x\diamond y-y\diamond x
\ee
where $\diamond $ is the product defined in eq.(\ref{algebra}).

Affine Lie algebras are given by so-called affine Kac-Moody algebras
enlarged by a central extension. Affine Kac-Moody algebras
are a subset of so-called Kac-Moody algebras which are
examples of infinite dimensional Lie algebras. There does not yet
exist a general theory for infinite dimensional Lie algebras and their
representations. More or less detailed studies exist only for four classes
\begin{enumerate}
\item Lie algebras over vector fields and the corresponding groups of
diffeomorphisms.
\item Lie algebras of operators of some Hilbert or Banach space.
\item Lie algebras of smooth mappings of some manifold into a finite
dimensional Lie algebra.
\item Contragradient Lie algebras or Kac-Moody algebras.
\end{enumerate}

We will here only discuss the fourth class. There does, however, exist a
close relation
between affine Lie algebras and one particular example of the third class.
We will elaborate further on this issue below.

The pionering work on Kac-Moody algebras were, as the name suggests,
performed independently by Kac \cite{Kac67,Kac68,Kac68:2} and Moody
\cite{Moody67,Moody68}. Since then there has been a vast number of
papers on generalizations and developments of Kac-Moody algebras and their
representation theory by a number of people, often involving the
ubiquitous Kac. The concepts of Kac-Moody algebras are nicely summarized
in Kac book on this subject \cite{Kac90}.

\section{Preliminaries }

Take an $N\times N$ real matrix of rank $r$. If $A$ meets the following
conditions ($a_{ij}$ $i,j=1,...,N$ are the components of $A$)
\be
& & a_{ii}=2 \hs{3mm} i=1,...,N \nn \\
& & a_{ij}\leq 0 \hs{3mm} i\neq j \nn \\
& & a_{ij}=0 \hs{3mm} \Leftrightarrow \hs{3mm} a_{ji}=0 \nn \\
& & a_{ij} \in {\Bbb Z} \label{kacmoodydef}
\ee
it is known as a generalized Cartan matrix. We may associate with $A$ a Lie
algebra
characterized by $3r$ generators $\{e_i,f_i,h_i|i=1,...,n\}$ obeying the
relations
\be
& & [h_i,h_j]=0 \hs{5mm} [e_i,f_j]=\del_{ij}h_j \hs{5mm} [h_i,e_j]=a_{ij}e_j
\hs{5mm} [h_i,f_j]=-a_{ij}f_j \nn \\
& & (ad_{e_i})^{1-a_{ij}}e_j=0 \hs{5mm} (ad_{f_i})^{1-a_{ij}}f_j=0
\label{definerel}
\ee
where $(ad_{e_i})^me_j$ stands for
\be
\underbrace{ad_{e_i}\circ ad_{e_i}\circ...\circ ad_{e_i}}_{m\ {\rm times}} \nn
\ee
so for example $(ad_{e_i})^2e_j=[e_i,[e_i,e_j]]$. This infinite dimensional Lie
algebra is known as a Kac-Moody algebra. It possesses two well-known subsets,
namely finite dimensional Lie algebras and affine Kac-Moody algebras. In what
follows
we will refer to all Kac-Moody algebras outside those two subsets as indefinite
Lie algebras. We will, in what follows, not pay any attention to the
difference between affine Kac-Moody algebras and affine Lie algebras
except when called for. Hence we will denote the affine algebras by affine Lie
algebras.

Finite dimensional Lie algebras are obtained by imposing the additional
constraint
\be
detA>0 \label{finitedet}
\ee
on the generalized Cartan matrix $A$. We will assume that the reader is
familiar
with the theory and notation of finite dimensional Lie algebras and their
representations.

Affine Lie algebras on the other hand, are described by the system of equations
(\ref{kacmoodydef}) with the restriction that
\be
detA_{\{i\}}>0 \hs{10mm} i=0,...,r \label{affinedet}
\ee
where $A_{\{i\}}$ denotes the matrices obtained from $A$ by deleting the
$i$'th row and column. We have here changed the index $i$ to run from 0 to $r$
for reasons that will be obvious as we proceed. We will in what follows stick
to
this convention when discussing affine Lie algebras.

We will not here deal with the largest and least known subset of Kac-Moody
algebras,
the indefinite Lie algebras. Only a small number of indefinite Lie algebras
has been classified. There exists, however, reasons for string theorists
to be interested in those algebras. The vertex operator algebra of the lowest
lying
mass state of the string obeys an affine Lie algebra. It has been argued that
vertex operators of higher excitations in the string spectrum may obey
indefinite
Lie algebras. There has recently appeared very interesting results pointing in
this direction, see for example \cite{Gebert-Nicolai94} and references therein.

There exists a complimentary way of distinguishing between the three subsets of
Kac-Moody algebras. Without loss of generality we can assume in what follows
that
the generalized Cartan matrix is indecomposable, that is we cannot write $A$ as
a block diagonal matrix by rearrangement of the indices. Take a real
indecomposable
generalized Cartan matrix. We have three mutually exclusive classes provided we
demand the following on both $A$ and $A^T$ the transpose of $A$. Take vectors
$u$ and $v$.
\begin{enumerate}
\def\theenumi{\roman{enumi}}
\item ${\rm det}B\neq 0$: $\exists \ u>0$ such that $Bu>0, \ Bv\geq 0$ implies
$v\geq0$
\item corank$B$=1: $\exists \ u>0$ such that $Bu>0, \ Bv\geq 0$ implies $Bv=0$
\item $\exists \ u>0$ such that $Bu<0, \ Bv\geq 0$ and $v\geq0$ implies $v=0$
\end{enumerate}
for both $B=A$ and $B=A^T$.
$u>0$ means that all entries in $u$ are positive.

It is not to difficult to see that the three cases are disjoint.
The finite and the affine case are obviosly exclusive since they
differ in rank. The finite and the indefinite cases exclude each other
since for the finite case there exists no $v\geq 0$ such that $Av\leq0$
and $Av\neq0$, as is easily verified by substituting $v$ by $-v$ above.
Finally, the affine and the indefinite cases cannot be inclusive
since for the affine case there does not exist a vector $v$ such that
$Av\geq0$ or $Av\leq0$, and hence nor a vector $v<0$ such that $Av<0$ as is
required for the indefinite case.

For the proof that each pair $A$ and $A^T$ are of either finite, affine
or indefinite type we refer to the literature.

It is also important to note that the Kac-Moody algebra $g(A)$ may not
always posses a non-zero invariant bilinear form. If and only if $A$ is
symmetrizable, i.e. one can decompose $A=DB$ for an invertible
diagonal matrix
$D$ and a symmetric matrix $B$, a non-zero invariant bilinear form does
exist. In fact, $A$ is always symmetrizable for the finite and affine
cases.

We can realize the generalized Cartan matrix with a
complex vectorspace {\sc h} and its dual {\sc h}$^\ast$ and
two indexed subsets $\Pi^{\vee}$ of {\sc h} and $\Pi$ of
{\sc h}$^\ast$. They are required to satisfy; i) $\Pi$ and
$\Pi^{\vee}$ are linearly independent. ii) The pairing between the
vectorspaces $\lan\ ,\ \ran$ is such that
$\lan\al^{\vee}_i,\al_j\ran=a_{ij}$ $i,j=1,...,n$ where $\al_i\in\Pi$
and $\al^{\vee}_i\in\Pi^{\vee}$. iii) $n-r=$dim({\sc h})$-n$.
$r$ is the rank of $A$.

Using the well-known terminology for finite Lie algebras we call
$\Pi$ the root basis, $\Pi^{\vee}$ the dual root basis, and their
elements simple roots and simple co-roots, respectively. Also,
{\sc h} is known as the Cartan sub-algebra.

The auxiliary Lie algebra $\ti{g}(A)$, with generators
$e_i$, $f_i$ $i=1,...,n$ and {\sc h}, is defined to obey
\be
& &[e_i,f_j]=\del_{ij}\al^{\vee}_i \nn \\
& & [h,h']=0 \hs{10mm}  \nn \\
& & [h,e_i]=\lan\al_i,h\ran e_i \nn \\
& & [h,f_i]=-\lan\al_i,h\ran f_i
\ee
where $i=1,...,n$ and $h,h'\in${\sc h}. $e_i$ and $f_i$ are known as the
Chevalley generators of $g(A)$.
We take the Kac-Moody algebra
associated with the generalized Cartan matrix $A$ to be the Lie algebra
\be
g(A)=\ti{g}(A)/\tau
\ee
where $\tau$ is the maximal ideal in $\ti{g}(A)$ which intersects
{\sc h} trivially.

We have a rootspace decomposition of $g(A)$ with respect to {\sc h}
\be
g(A)=\bigoplus_{\al}g_{\al}
\ee
for $g_{\al}=\{x\in g(A)|[h,x]=\al(h)x$ for all $h\in${\sc h}$\}$.
Note that the Cartan sub-algebra is $g_0$.
The dimension of $g_{\al}$ is called
the multiplicity of $\al$. From this we define the root system. We take the
root lattice $Q$ to be the
lattice spanned by the simple roots $\al_i$. Define also
\be
Q_+=\sum_{i=1}^n {\Bbb Z}_+\al_i.
\ee
The lattice $Q_+$ is used to define a partial orderring of elements
on {\sc h}$^{\ast}$. We take $\la\geq\mu$ if $\la-\mu\in Q_+$.
An element of the root lattice
$\al$ is called a root if $\al\neq0$ and its multiplicity also is non-zero.
All roots are either positive or negative and we denote the set of
roots, positive roots and negative roots by $\De$, $\De_+$, and
$\De_-$ respectively. We also have that
\be
\De_-=-\De_+
\ee

There exists a triangular decomposition of $g(A)$ such that
$g(A)=${\sc n}$_-\oplus${\sc h}$\oplus${\sc n}$_+$ where
{\sc n}$_+$ and {\sc n}$_-$ are the subalgebras formed by $e_i$
and $f_i$. Obviously $g_{\al}\subset${\sc n}$_+$ if $\al>0$.

We assume from now on that the generalized Cartan matrix $A$
is symmetrizable, that is, it can be written as $A=DB$ for some
diagonal matrix $D$, diag$D=d_1,...,d_n$ all non-zero, and a symmetric
matrix $B$ with entries $b_{ij}$ $i,j=1,...,n$. Fix a
complimentary subspace {\sc h}'' to {\sc h}'$=\sum_i {\Bbb C}$
$\al^{\vee}_i$ in {\sc h}. We define a symmetric bilinear ${\Bbb C}$-valued
form $(.|.)$ on {\sc h} by
\be
(\al^{\vee}_i|h)=\lan\al_i,h\ran d_i \hs{10mm} (h'|h'')=0 \label{bilform}
\ee
for $i=1,...,n$, $h\in${\sc h}, and $h',h''\in${\sc h}''. We have then
for example $(\al^{\vee}_i|\al^{\vee}_j)=b_{ij}d_id_j$ $i,j=1,...,n$.

There exists an isomorphism $\nu:${\sc h}$\rightarrow${\sc h}$^{\ast}$
which we define by
\be
\lan\nu(h),h_1\ran=(h|h_1)
\ee
for $h,h_1\in${\sc h}. We thus get $\nu(\al^{\vee}_i)=d_i\al_i$ and
it follows from $(\al^{\vee}_i|\al^{\vee}_j)=b_{ij}d_id_j $ that
$(\al_i|\al_j)=
b_{ij}=a_{ij}/d_i$.

Next we extend the bilinear form to the whole of $g(A)$ which has the
properties
\begin{enumerate}
\def\theenumi{\alph{enumi}}
\item $(.|.)$ is invariant i.e. $([x,y]|z)=(x|[y,z])$ for $x,y,z\in g.$
\item The restriction of $(.|.)$ to the Cartan sub-algebra is defined as above
(\ref{bilform}) and is non-degenerate.
\item $(g_{\al}|g_{\bet})=0$ if $\al+\bet\neq0$
\item $(.|.)|_{g_{\al}+g_{-\al}}$ is non-degenerate for $\al\neq 0$.
\item $[x,y]=(x|y)\nu^{-1}(\al)$ for $x\in g_{\al}$, $y\in g_{-\al}$ and
$\al\in\De$.
\end{enumerate}

We now take a decomposition $A=DB$ where $d_i$ $i=1,...,n$ are positive
rational numbers, and assume $A$ to be indecomposable. We then have that
$(\al_i|\al_i)>0$, $(\al_i|\al_j)\leq0$ for $i\neq j$, and using $a_{ii}=2$,
$\al^{\vee}_i=
\frac{2}{(\al_i|\al_i)}\nu^{-1}(\al_i)$ and we hence find the
usual expression for the elements of the Cartan matrix
\be
a_{ij}=\frac{2(\al_i|\al_j)}{(\al_i|\al_i)}.
\ee
The bilinear form $(.|.)$ is known as the standard bilinear form.


\section{Classification of affine Lie algebras }

We again list the properties of an affine Cartan matrix
$A$ necessary for a complete classification of affine Lie algebras.
\begin{enumerate}
\item $a_{ii}=2$.
\item $a_{ij}=0$ $\Longleftrightarrow$ $a_{ji}=0$.
\item $a_{ij}\leq0$ for $i\neq j$.
\item $a_{ij}\in{\Bbb Z}$.
\item $A$ is indecomposable.
\item det$A=0$ and det$A_{\{i\}}>0$ for all $i=1,...,n$.
\end{enumerate}

\noindent Again $A_{\{i\}}$, which is known as a principal minor, denotes the
matrix obtained from $A$ by deleting row and column number $i$.

Obviously the affine Lie algebras of rank $r$ contain finite Lie algebras
obtained at rank $r+1$ since det$A>0$ and the conditions the two cases
have in common implies det$A_{\{i\}}>0$ see (\ref{kacmoodydef}),
(\ref{finitedet})
and (\ref{affinedet}). We can thus use the classification of finite Lie
algebras to classify the affine ones. We hence demand that the submatrices
$A_{\{i\}}$ $i=1,...,n$ of an affine Cartan matrix $A$ are
Cartan matrixes of finite Lie algebras or direct sums there of.

We will give a few examples and start at affine rank one. Using our set
of rules we have the general form
\be
\lef[\begin{array}{rr}
2 & -p \\ -q & 2
\end{array}\rig].
\ee
Demanding the determinant to vanish we have the solutions $p=q=2$ or
$p=4$ and $q=2$. Those correspond to $A^{(1)}_1$ and $A^{(2)}_1$. The former
as we will explain below the affine extension of the finite Lie algebra
$su(2)$ and the latter is an example of a so called twisted affine Lie
algebra. We will comment on the twisted case in a subsequent section.

For rank two we choose the example where by deleting row one and column one
of the affine Cartan matrix we obtain the Catran matrix corresponding to the
finite Lie algebra $C_2$ that is we take
\be
\lef[\begin{array}{rrr}
2 & -p & -q  \\ -r & 2 & -1 \\ -s & -2 & 2
\end{array}\rig].
\ee
det$A$=0 yields now $0=4-p(s+2r)-2q(r+s)$. The only possible solutions
consistent with the Cartan matrix rules are either $r=p=0$, $q=1$ and $s=2$
or $q=s=0$, $p=1$ and $r=2$. They correspond to the affine Lie algebras
$C^{(2)}_2$ and$C^{(1)}_2$ respectively.

Proceeding along those lines one finds seven infinite series $A^{(1)}_r$
$r\geq2$, $B_r^{(1)}$ $r\geq3$, $C_r^{(1)}$ $r\geq2$, $D_r^{(1)}$ $r\geq2$,
$B_r^{(2)}$ $r\geq3$, $C_r^{(2)}$ $r\geq2$, $\widetilde{B}_r^{(2)}$ $r\geq3$,
and nine exceptional affine algebras $A^{(1)}_1$, $E^{(1)}_6$, $E^{(1)}_7$,
$E^{(1)}_8$, $F^{(1)}_4$, $G^{(1)}_2$, $A^{(2)}_1$, $F^{(2)}_4$, $G^{(3)}_2$.

In analogy to finite Lie algebras we may introduce Dynkin diagrams and we have
listed them in fig.(4.3). We see that if we delete a node of an affine
Dynkin diagram we recover a finite Lie algebra Dynkin diagram or sums thereof.
This of course corresponds to deleting rows and columns in the affine Cartan
matrix. Taking the example given above with the choice $q=s=0$ we have
the Cartan matrix
\be
\lef[\begin{array}{rrr}
2 & -1 & 0  \\ -2 & 2 & -1 \\ 0 & -2 & 2
\end{array}\rig] \label{affG}
\ee
which upon deleting rows and columns each in term produces the principal minors
\be
\lef[\begin{array}{rr}
2 & -1 \\ -2 & 2
\end{array}\rig] \hs{10mm}
\lef[\begin{array}{rr}
2 & 0 \\ 0 & 2
\end{array}\rig] \hs{10mm}
\lef[\begin{array}{rr}
2 & -1 \\ -2 & 2
\end{array}\rig] \label{principmin}
\ee
The affine Cartan matrix (\ref{affG}) corresponds to the  affine
Dynkin diagram $C^{(1)}_2$ found in fig.(4.1).
\begin{figure}[H]
\begin{picture}(400,30)(0,0)

\put(169,20){\circle{10}}
\put(174,23){\line(1,0){22}}
\put(174,17){\line(1,0){22}}
\put(187,20){\line(-1,1){10}}
\put(177,9){\line(1,1){10}}
\put(200,20){\circle{10}}
\put(205,23){\line(1,0){22}}
\put(205,17){\line(1,0){22}}
\put(212,20){\line(1,1){10}}
\put(222,9){\line(-1,1){10}}
\put(231,20){\circle{10}}

\end{picture}
\caption{$C^{(1)}_2$}
\end{figure}
\noindent The three finite Dynkin diagrams corresponding to the principal
minors
(\ref{principmin}) corresponds to $C_2$, $A_1\oplus A_1$ and $C_2$
and are depicted
in fig.(4.2).
\begin{figure}[H]
\begin{picture}(400,30)(0,0)

\put(50,20){\circle{10}}
\put(55,23){\line(1,0){22}}
\put(55,17){\line(1,0){22}}
\put(62,20){\line(1,1){10}}
\put(72,9){\line(-1,1){10}}
\put(81,20){\circle{10}}

\put(180,20){\circle{10}}

\put(200,20){\circle{10}}

\put(299,20){\circle{10}}
\put(304,23){\line(1,0){22}}
\put(304,17){\line(1,0){22}}
\put(317,20){\line(-1,1){10}}
\put(307,9){\line(1,1){10}}
\put(330,20){\circle{10}}

\end{picture}
\caption{$C_2$, $A_1\otimes A_1$ and $C_2$}
\end{figure}
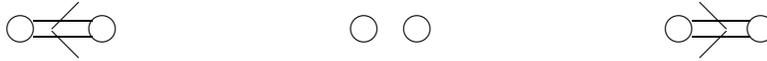

We have in fig.(4.3) depicted the Dynkin diagrams of the affine Lie
algebras as well as the twisted affine Lie algebras. The twisted case
corresponds to those with superscript different from one. The naming of the
affine Lie algebras is chosen such

\begin{figure}[H]
\begin{picture}(420,400)(0,0)

\put(0,0){\framebox(200,400)}
\put(200,0){\framebox(220,400)}

\put(5,375){$A_1^{(1)}$}
\put(90,380){\circle{10}}
\put(120,380){\circle{10}}
\put(95,383){\line(1,0){21}}
\put(95,377){\line(1,0){21}}
\put(88,366){{\bf 1}}
\put(118,366){1}

\put(5,330){$A_r^{(1)}$}
\put(110,345){\circle{10}}
\put(104,345){\line(-2,-1){46}}
\put(115,345){\line(2,-1){46}}
\put(55,317){\circle{10}}
\put(165,317){\circle{10}}
\put(60,317){\line(1,0){20}}
\put(85,317){\circle{10}}
\put(90,317){\line(1,0){10}}
\put(105,315){...}
\put(120,317){\line(1,0){10}}
\put(135,317){\circle{10}}
\put(140,317){\line(1,0){20}}
\put(115,348){{\bf 1}}
\put(52,303){1}
\put(82,303){1}
\put(163,303){1}
\put(132,303){1}

\put(5,260){$B_r^{(1)}$}
\put(43,248){\circle{10}}
\put(43,282){\circle{10}}
\put(46,251){\line(1,1){10}}
\put(46,279){\line(1,-1){10}}
\put(60,265){\circle{10}}
\put(65,265){\line(1,0){20}}
\put(90,265){\circle{10}}
\put(95,265){\line(1,0){10}}
\put(107,263){...}
\put(119,265){\line(1,0){10}}
\put(134,265){\circle{10}}
\put(138,268){\line(1,0){22}}
\put(138,262){\line(1,0){22}}
\put(153,265){\line(-1,1){10}}
\put(143,254){\line(1,1){10}}
\put(164,265){\circle{10}}
\put(48,240){1}
\put(48,282){{\bf 1}}
\put(58,251){2}
\put(88,251){2}
\put(132,251){2}
\put(162,251){1}
\put(162,243){{\sl 2}}

\put(5,205){$C_r^{(1)}$}
\put(55,210){\circle{10}}
\put(60,213){\line(1,0){22}}
\put(60,207){\line(1,0){22}}
\put(73,210){\line(-1,1){10}}
\put(63,199){\line(1,1){10}}
\put(86,210){\circle{10}}
\put(91,210){\line(1,0){10}}
\put(103,208){...}
\put(115,210){\line(1,0){10}}
\put(130,210){\circle{10}}
\put(135,213){\line(1,0){22}}
\put(135,207){\line(1,0){22}}
\put(142,210){\line(1,1){10}}
\put(152,199){\line(-1,1){10}}
\put(161,210){\circle{10}}
\put(84,196){1}
\put(84,188){{\sl 2}}
\put(128,196){1}
\put(128,188){{\sl 2}}
\put(53,196){{\bf 1}}
\put(159,196){1}

\put(5,155){$D_r^{(1)}$}
\put(43,143){\circle{10}}
\put(43,177){\circle{10}}
\put(46,146){\line(1,1){10}}
\put(46,174){\line(1,-1){10}}
\put(60,160){\circle{10}}
\put(65,160){\line(1,0){20}}
\put(90,160){\circle{10}}
\put(95,160){\line(1,0){10}}
\put(107,158){...}
\put(119,160){\line(1,0){10}}
\put(134,160){\circle{10}}
\put(139,160){\line(1,0){20}}
\put(164,160){\circle{10}}
\put(167,163){\line(1,1){10}}
\put(178,146){\line(-1,1){10}}
\put(181,143){\circle{10}}
\put(181,177){\circle{10}}
\put(48,135){1}
\put(48,177){{\bf 1}}
\put(58,146){2}
\put(88,146){2}
\put(132,146){2}
\put(162,146){2}
\put(186,135){1}
\put(186,177){1}

\put(5,85){$E_6^{(1)}$}
\put(108,123){\circle{10}}
\put(108,103){\line(0,1){15}}
\put(108,98){\circle{10}}
\put(108,78){\line(0,1){15}}
\put(108,73){\circle{10}}
\put(113,73){\line(1,0){20}}
\put(138,73){\circle{10}}
\put(143,73){\line(1,0){20}}
\put(168,73){\circle{10}}
\put(83,73){\line(1,0){20}}
\put(78,73){\circle{10}}
\put(53,73){\line(1,0){20}}
\put(48,73){\circle{10}}
\put(116,120){{\bf 1}}
\put(116,95){2}
\put(106,59){3}
\put(136,59){2}
\put(166,59){1}
\put(76,59){2}
\put(46,59){1}

\put(5,20){$E_7^{(1)}$}
\put(108,40){\circle{10}}
\put(108,20){\line(0,1){15}}
\put(108,15){\circle{10}}
\put(113,15){\line(1,0){15}}
\put(133,15){\circle{10}}
\put(138,15){\line(1,0){15}}
\put(158,15){\circle{10}}
\put(163,15){\line(1,0){15}}
\put(183,15){\circle{10}}
\put(88,15){\line(1,0){15}}
\put(83,15){\circle{10}}
\put(63,15){\line(1,0){15}}
\put(58,15){\circle{10}}
\put(38,15){\line(1,0){15}}
\put(33,15){\circle{10}}
\put(116,37){2}
\put(106,1){4}
\put(131,1){3}
\put(156,1){2}
\put(181,1){1}
\put(81,1){3}
\put(56,1){2}
\put(31,1){{\bf 1}}

\put(205,370){$E_8^{(1)}$}
\put(358,390){\circle{10}}
\put(358,370){\line(0,1){15}}
\put(308,365){\circle{10}}
\put(313,365){\line(1,0){15}}
\put(333,365){\circle{10}}
\put(338,365){\line(1,0){15}}
\put(358,365){\circle{10}}
\put(363,365){\line(1,0){15}}
\put(383,365){\circle{10}}
\put(388,365){\line(1,0){15}}
\put(408,365){\circle{10}}
\put(288,365){\line(1,0){15}}
\put(283,365){\circle{10}}
\put(263,365){\line(1,0){15}}
\put(258,365){\circle{10}}
\put(238,365){\line(1,0){15}}
\put(233,365){\circle{10}}
\put(366,387){3}
\put(356,351){6}
\put(381,351){4}
\put(406,351){2}
\put(331,351){5}
\put(306,351){4}
\put(281,351){3}
\put(256,351){2}
\put(231,351){{\bf 1}}

\put(205,325){$F_4^{(1)}$}
\put(258,330){\circle{10}}
\put(263,330){\line(1,0){20}}
\put(288,330){\circle{10}}
\put(293,330){\line(1,0){20}}
\put(318,330){\circle{10}}
\put(323,333){\line(1,0){22}}
\put(323,327){\line(1,0){22}}
\put(336,330){\line(-1,1){10}}
\put(326,319){\line(1,1){10}}
\put(349,330){\circle{10}}
\put(354,330){\line(1,0){20}}
\put(379,330){\circle{10}}
\put(256,316){\bf{1}}
\put(286,316){2}
\put(316,316){3}
\put(347,316){2}
\put(347,308){\sl{4}}
\put(377,316){1}
\put(377,308){\sl{2}}

\put(205,288){$G_2^{(1)}$}
\put(288,293){\circle{10}}
\put(293,293){\line(1,0){20}}
\put(318,293){\circle{10}}
\put(323,296){\line(1,0){22}}
\put(323,293){\line(1,0){20}}
\put(323,290){\line(1,0){22}}
\put(336,293){\line(-1,1){10}}
\put(326,282){\line(1,1){10}}
\put(349,293){\circle{10}}
\put(286,279){\bf{1}}
\put(316,279){2}
\put(347,279){1}
\put(347,271){\sl{3}}

\put(200,268){\framebox(220,1)}

\put(205,250){$A_1^{(2)}$}
\put(300,255){\circle{10}}
\put(330,255){\circle{10}}
\put(305,258){\line(1,0){22}}
\put(305,252){\line(1,0){22}}
\put(300,260){\line(1,0){30}}
\put(300,250){\line(1,0){30}}
\put(317,255){\line(-1,1){10}}
\put(307,244){\line(1,1){10}}
\put(298,241){2}
\put(298,233){\bf{1}}
\put(328,241){1}
\put(328,233){\sl{2}}

\put(205,210){$B_r^{(2)}$}
\put(270,215){\circle{10}}
\put(275,218){\line(1,0){22}}
\put(275,212){\line(1,0){22}}
\put(282,215){\line(1,1){10}}
\put(292,204){\line(-1,1){10}}
\put(301,215){\circle{10}}
\put(306,215){\line(1,0){10}}
\put(318,213){...}
\put(330,215){\line(1,0){10}}
\put(345,215){\circle{10}}
\put(350,218){\line(1,0){22}}
\put(350,212){\line(1,0){22}}
\put(363,215){\line(-1,1){10}}
\put(353,204){\line(1,1){10}}
\put(376,215){\circle{10}}
\put(268,201){\bf{1}}
\put(299,201){2}
\put(299,193){\sl{1}}
\put(343,201){2}
\put(343,193){\sl{1}}
\put(374,201){1}

\put(205,170){$\widetilde{B}_r^{(2)}$}
\put(270,175){\circle{10}}
\put(275,178){\line(1,0){22}}
\put(275,172){\line(1,0){22}}
\put(288,175){\line(-1,1){10}}
\put(278,164){\line(1,1){10}}
\put(301,175){\circle{10}}
\put(306,175){\line(1,0){10}}
\put(318,173){...}
\put(330,175){\line(1,0){10}}
\put(345,175){\circle{10}}
\put(350,178){\line(1,0){22}}
\put(350,172){\line(1,0){22}}
\put(363,175){\line(-1,1){10}}
\put(353,164){\line(1,1){10}}
\put(376,175){\circle{10}}
\put(268,161){2}
\put(268,153){\bf{1}}
\put(299,161){2}
\put(343,161){2}
\put(374,161){2}
\put(374,153){\sl{1}}

\put(205,120){$C_r^{(2)}$}
\put(254,108){\circle{10}}
\put(254,142){\circle{10}}
\put(257,111){\line(1,1){10}}
\put(257,139){\line(1,-1){10}}
\put(271,125){\circle{10}}
\put(276,125){\line(1,0){20}}
\put(301,125){\circle{10}}
\put(306,125){\line(1,0){10}}
\put(318,123){...}
\put(330,125){\line(1,0){10}}
\put(345,125){\circle{10}}
\put(350,128){\line(1,0){22}}
\put(350,122){\line(1,0){22}}
\put(363,125){\line(-1,1){10}}
\put(353,114){\line(1,1){10}}
\put(376,125){\circle{10}}
\put(259,100){1}
\put(259,142){\bf{1}}
\put(271,111){2}
\put(299,111){2}
\put(343,111){2}
\put(374,111){2}
\put(374,103){\sl{1}}

\put(205,65){$F_4^{(2)}$}
\put(258,70){\circle{10}}
\put(263,70){\line(1,0){20}}
\put(288,70){\circle{10}}
\put(293,73){\line(1,0){22}}
\put(293,67){\line(1,0){22}}
\put(306,70){\line(-1,1){10}}
\put(296,59){\line(1,1){10}}
\put(318,70){\circle{10}}
\put(323,70){\line(1,0){20}}
\put(349,70){\circle{10}}
\put(354,70){\line(1,0){20}}
\put(379,70){\circle{10}}
\put(256,56){2}
\put(258,48){\sl{1}}
\put(286,56){4}
\put(286,48){\sl{2}}
\put(316,56){3}
\put(347,56){2}
\put(377,56){\bf{1}}

\put(205,18){$G_2^{(3)}$}
\put(288,23){\circle{10}}
\put(293,26){\line(1,0){22}}
\put(293,23){\line(1,0){20}}
\put(293,20){\line(1,0){22}}
\put(306,23){\line(-1,1){10}}
\put(296,12){\line(1,1){10}}
\put(318,23){\circle{10}}
\put(323,23){\line(1,0){20}}
\put(349,23){\circle{10}}
\put(289,9){3}
\put(289,1){\sl{1}}
\put(316,9){2}
\put(347,9){\bf{1}}

\end{picture}
\caption{Affine Dynkin diagrams}
\end{figure}
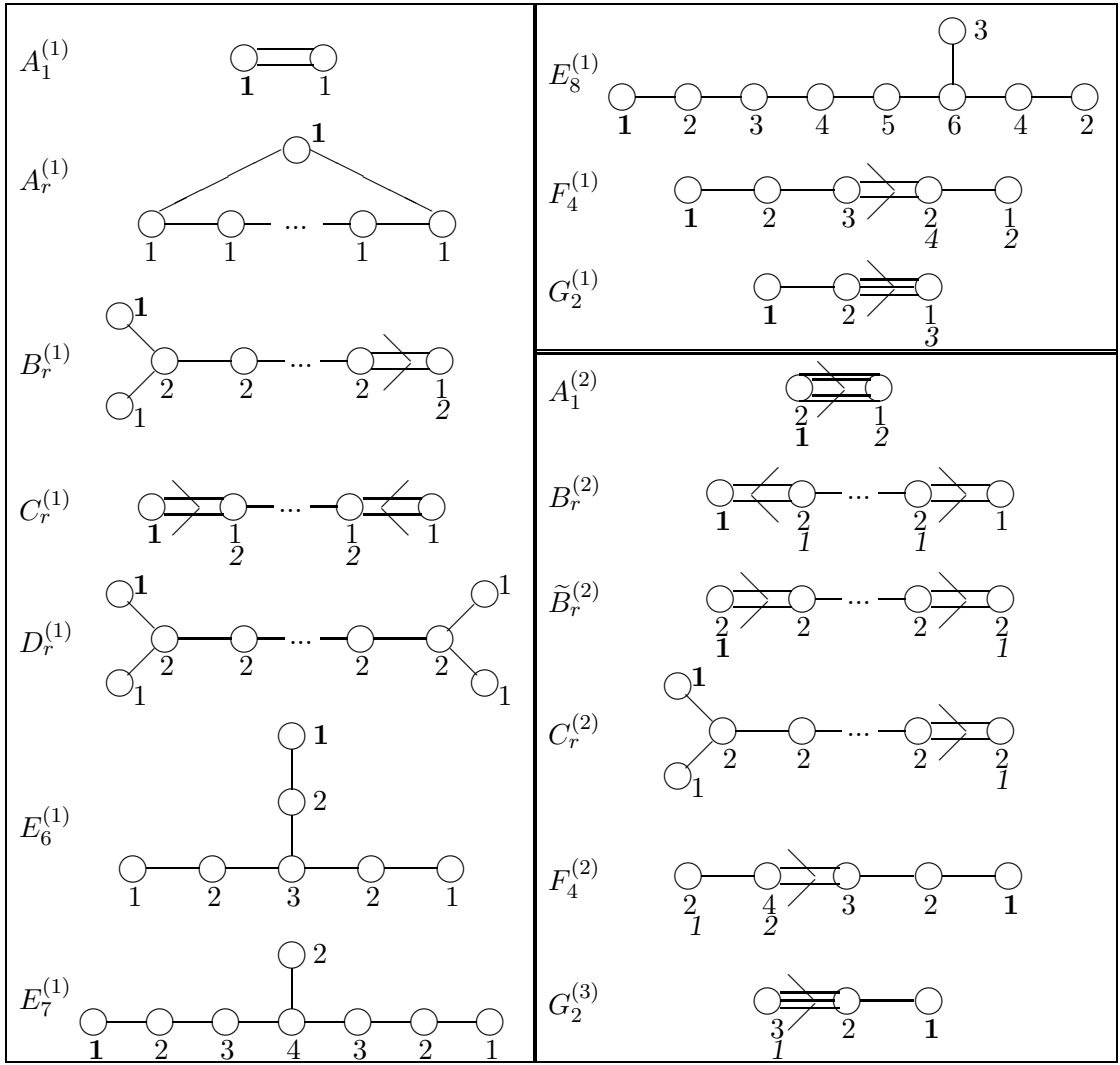

\noindent
that removing the boldface node of a Dynkin
diagram $X^{(1)}_r$ will produce the simple Lie algebra $X_r$ where
$X$ is one of $A,B,C,D,F,E$ or $G$.

The number attached to the nodes are the Coxeter $a_i$ and dual Coxeter
$a^{\vee}_i$ labels.
The dual Coxeter labels are displayed in italics below the Coxeter labels
whenever they differ. The Coxeter labels and the dual Coxeter labels
are the entries of the left and right eigenvectors to the affine Cartan
matrix corresponding to the zero eigenvalue i.e.
\be
\sum_{i=0}^ra_ia_{ij}=0=\sum_{i=0}^ra_{ij}a^{\vee}_j.
\ee
The sum of those numbers, the Coxeter number g and the dual Coxeter
number g$^{\vee}$, appear in different contexts in representation theory
of affine Lie algebras.

\section{Loop algebras in connection to affine Lie algebras}

We pointed out above that there exists a close relation
between affine Lie algebras and one representative of the class of
infinite dimensional Lie algebras appearing from smooth maps from a
manifold onto a finite Lie algebra. More precisely we can realize an
affine Lie algebra $\hat{g}$ without center by considerring maps from
the unit circle onto a finite Lie algebra $g$. These are known as loop
algebras.

We take the generators of the finite Lie algebra to be $t^a$ $ a=1,...,d$.
$S^1$ is represented by the unit circle on the complex plane parametrized by
the coordinate $z$.
Consider the generators
\be
t^a_n\equiv t^a\otimes z^n
\ee
where $\otimes$ is some formal multiplication. Using the Lie bracket of the
finite Lie algebra $[t^a,t^b]=f^{ab}_{\ \ c}t^c$,
we can define the bracket operation of the objects $t^a_n$ as
\be
[t^a_n,t^b_m]\equiv[t^a,t^b]\otimes(z^nz^m)=f^{ab}_{\ \ c}t^c_{n+m}.
\ee
Here $f^{ab}_{\ \ c}$ are the structure constants of the finite Lie algebra.
Recapitulating the criteria for the finite Lie algebra, it is easy
to see that both anti-symmetry and the Jacobi identity are met. This
infinite dimensional Lie algebra is known as the loop algebra
$g_{\rm loop}$ over $g$.
It is worth noting that subalgebra generated by the zeromodes $t^a_0$ of the
loop algebra is isomorphic to the finite Lie algebra $g$.

With hindsight of what is to come, we now investigate the possibilities
for central extensions of the loop algebra. A central extension or
central element is, as the name suggests, related to the center of the algebra.
We know that affine Lie algebras have a one dimensional center, and if
loop algebras are the centerless part of the affine Lie algebras then they must
allow for a central extension.

Indeed, using the fact that the dual Coxeter labels is an eigenvector
of the Cartan matrix with zero eigenvalue, it is easy to construct the central
extension of an affine Lie algebra
\be
K\equiv\sum_i aa^{\vee}_ih_i.
\ee
for $h^i$ being the elements of the Cartan subalgebra.
We have using (\ref{definerel}) trivially $[K,h^i]=0$ and also
\be
[K,e_i]=aa^{\vee}_ja_{ji}e_i=0 \hs{15mm}
[K,f_i]=-aa^{\vee}_ja_{ji}f_i=0
\ee
Since the affine Cartan matrix has only one zero eigenvalue this extension
is unique up to the multiplicative constant $a$.

We start by performing the same investigation for finite Lie algebras.
Take a simple finite Lie algebra with generators $t^a$. Add now $s$ generators
$K^i$ $i=1,...,s$ defined to be in the center of the algebra that is
\be
[t^a,K^i]=0 \hs{15mm} [K^i,K^j]=0 \hs{10mm} \forall a,i,j.
\ee
The original Lie bracket is generalized to
\be
[t^a,t^b]=f^{ab}_{\ \ c}t^c+f^{ab}_iK^i
\ee
The new structure constants $f^{ab}_i$ are constrained by the Jacobi
identity. Note that the trivial choice $f^{ab}_i=0$ $\forall i$ is
of course allowed, yielding a direct sum of $g$ and $s$ $u(1)$ factors.

There exists, however, more general solutions of this type. We can
change basis in $g$ and take $\ti{t}^a=t^a+u^a_iK^i$ for some
constants $u^a_i$. From the Lie bracket it now follows that the new structure
constants in this basis are
\be
\ti{f}^{ab}_i=f^{ab}_i-f^{ab}_{\ \ c}u^c_i.
\ee

Consider the basis for $g$ where $f^{abc}$ are totally antisymmetric
and $f^{ab}_{\ \ c}f^{cd}_{\ \ a}=c_g\del^{bd}$ where $c_g$ ia the quadratic
Casimir in the adjoint representation. The Jacobi identity requires
\be
f^{ab}_i=\frac{2}{c_g}f_{ec}^{\ \ b}f^{ea}_{\ \ d}f^{cd}_i
\ee
and we can thus choose
\be
u^a_i=\frac{-1}{c_g}f^{a}_{\ bc}f^{bc}_i
\ee
which gives $\ti{f}^{ab}_i=0$.

It is therefore always possible for simple finite Lie algebras (and
more generally semi-simple as well) to find a basis such that the
extension decouples and we have the direct sum of $g$ and $u(1)$'s.
This is not always the case, the
Poincar$\acute{\rm e}$ algebra, for example, allows for non trivial central
extension. This is in order since the Poincar$\acute{\rm e}$ algebra is not
semi-simple.

Remaking this for loop algebras we take the extended bracket as
\be
[t^a_m,t^b_n]=f^{ab}_{\ \ c}t^c_{m+n}+(f^{ab}_i)_{mn}K^i.
\ee
For $m=n=0$ we know that there exists a basis such that $f^{ab}_i=0$.
This is also possible for $m\neq 0$ $n=0$. It does not
work for both $m$ and $n$ non-zero, as can be seen from the fact that the
algebra is graded by the modenumbers $m$ and $n$. We hence have that for
fixed $n$, $t^a_n$ transforms in the adjoint representation of the
finite Lie algebra. It also follows that $(f^{ab}_i)_{mn}$ is an invariant
tensor with respect to the indices $a,b$ under the adjoint
representation of $g$. From the theory for finite Lie algebras we learn that
up to a normalization this tensor is unique and is known as the Killing form.
The Killing form $\ka^{ab}$ is defined as the map
\be
\ka : \begin{array}{l} g\times g\lra F \\
(x,y)\lra \ka(x,y)\equiv {\rm tr}({\rm ad}_x\circ {\rm ad}_y)
\end{array}
\ee
where $F$ is the base field, compare to the discussion on general algebras at
the very beginning of this chapter.
Thus $(f^{ab}_i)_{mn}=\ka^{ab}f_{mn}$ where we have excluded the index $i$
since the Killing form is unique in the sense stated above. Furthermore,
the Killing form is symmetric which means that $f_{mn}=-f_{nm}$ in
order to satisfy the anti-symmetry condition for the Lie bracket.

The Jacobi identity now reads
\be
f_{l,m+n}+f_{n,l+m}+f_{m,n+l}=0.
\ee
Note that putting $l=0$ yields the anti-symmetry condition $f_{m,n}+f_{n,m}=0$
where we use $f_{m,0}=0$. If we choose $l=-(x+1)m$ and $n=xm$ we end up with
\be
f_{m,-m}+f_{xm,-xm}+f_{-(x+1)m,(x+1)m}=0
\ee
for $x$ an arbitrary integer.
If we iterate this equation in $x$ we find that $f_{xm,-xm}=xf_{m,-m}$
which implies that $f_{m,n}$ scales with $m$. Take now
$f_{m,n}=m\ti{f}_{m,n}$. Anti-symmetry with $n=-m$ gives
$m(\ti{f}_{m,-m}-\ti{f}_{-m,m})=0$ which means that
$\ti{f}_{m,n}=\ti{f}_{n,m}$.
We may now use this to find $(m+n)\ti{f}_{m,n}=0$ and hence
$\ti{f}_{m,n}\propto\del_{m+n,0}$. The tensor $\ka^{ab}m\del_{m+n,0}$
fulfills all Lie bracket requirements, so we are free to choose a
convenient normalization, which is conventionally chosen to $1/2$.

We have not yet proven that this algebra is an affine Lie algebra
but this will be shown to be case in the next section.

\section{The root system of affine Lie algebras}

One way of finding the roots of affine Lie algebras is to
work iteratively with the defining relations (\ref{definerel}). We will
satisfy ourselves by displaying one simple example namely $A_1^{(1)}$.
In the Chevalley-Serre basis we have two Cartan generators $h_0$ and
$h_1$ and the "root" generators $e_0$, $e_1$, $f_0$ and $f_1$ such that
\be
& & [h_i,h_j]=0 \hs{10mm} [e_i,f_j]=\del_{ij}h_i \hs{10mm}
[h_i,e_j]=(-1)^{i+j}2e_j \hs{10mm} [h_i,f_j]=(-1)^{i+j+1}2f_j \nn \\
& & [e_i,[e_i,[e_i,e_j]]]=0 \hs{15mm} [f_i,[f_i,[f_i,f_j]]]=0 \hs{10mm}
i\neq j
\ee
We denote the simple roots $\al_0$ and $\al_1$ corresponding to the
root generators with the corresponding index.

We now start to inspect our relations. We have for example from
$[e_0,f_1]=0$ that $\al_0-\al_1$ is not a root. $({\rm ad}_{e_0}e_1)^3=0$
means that $\al_0+\al_1$ and $2\al_0+\al_1$ are roots but $3\al_0+\al_1$
is not. Using the Serre relation $({\rm ad}_{e_1}e_0)^3=0$ yields the
new root $\al_0+2\al_1$ while $\al_0+3\al_1$ is not a root. Furthermore
$[e_1,[e_0,[e_0,e_1]]]$ is not constrained to vanish which gives the new
root $2\al_0+2\al_1$. One may find more relations that are unconstraind
and iterate this procedure the desired number of times but since we
know the result we introduce the parametrization $\al_0=\del-\th$ and
$\al_1=\th$. The roots found above becomes $\del,\ \th+2\del,\ \th+\del,
\ 2\del$ and the forbidden combinations $-2\th+\del,\ -2\th+3\del,\
2\th+\del$. After a few more steps it is not hard to convince
oneself that the root system in this parametrization is spanned by
$\pm\th+n\del$ $n\in{\Bbb Z}$ and $n\del$ $n\in{\Bbb Z}\backslash 0$.

It is, however, more convenient to study the root system in the
Cartan-Weyl basis using the realization of affine Lie algebras as centrally
extended loop algebras. In the previous section we were given the
commutation relations, (in a physicists notation)
\be
& & [J^a_m,J^b_n]=if^{ab}_{\ \ c}J^c_{m+n}+\ka^{ab}\frac{K}{2}m\del_{m+n,0}
\nn \\
& & [J^a_m,K]=0
\ee
Proceeding in a similar manner as for the case of finite Lie algebras
we choose a maximal set of commuting generators, known as the Cartan
sub-algebra. Since the zero modes of the affine Lie algebra should
coincide with a finite dimensional Lie algebra, the Cartan sub-algebra of the
finite dimensional algebra should be a part of the Cartan sub-algebra of the
affine
case. It is also clear that $K$ should belong here, since it is in the
center. We thus take $J_0^i$ $i=1,...,r$ and $K$ to be in the affine
Cartan subalgebra.

In a standard way we now add the step operators corresponding to the
roots $J^{\pm\al}_m$, and demand that
\be
[J_0^i,J^{\pm\al}_m]=\pm\al^iJ_m^{\pm\al}
\ee
Using this we may consider $[J^{\al}_m,J^{\bet}_n]$. From the Jacobi
identity we find that
\be
[J_0^i,[J^{\al}_m,J_n^{\bet}]]=(\al^i+\bet^i)[J^{\al}_m,J_n^{\bet}].
\ee
We have the following three possibilities;
\begin{enumerate}
\item $\al+\bet$ is not a root and not zero
\item $\al+\bet$ is a root
\item $\al+\bet=0$.
\end{enumerate}
The first case just means that we must take $[J^{\al}_m,J^{\bet}_n]=0$. For the
second
case we have that $[J^{\al}_m,J^{\bet}_n]$ is proportional to the
generator $J^{\al+\bet}_{m+n}$ ($m+n$ since we can introduce a gradation
in the mode number), let the constant of proportionality be $\ep(\al,\bet)$.
In the final case we will use the invariance of the bilinear form
$([x,y]|z)=(x|[y,z])$. First note that $[J_0^i,[J^{\al}_m,J_n^{-\al}]]=0$
tells us that $[J^{\al}_m,J_n^{-\al}]$ is commuting with the Cartan subalgebra
(we will see below that it is in general not in the Cartan subalgebra).
Under the scalar product, using its invariance property, we then find
$([J^{\al}_m,J_n^{-\al}]|\xi_iJ_0^i)=
(J^{\al}_m|\xi_i\al^iJ^{-\al}_n)$ and we know that $(g_{\al}|g_{\bet})
\neq0$ if and only if $\al+\bet=0$. This yields that
$[J^{\al}_m,J_n^{-\al}]$ should be proportional to $\al_iJ^i_{m+n}$.

Adding the central extension, when appropriate, we find the algebra
\be
[J_m^i,J_n^j]&=&\frac{K}{2} m\delta^{ij}\delta_{m,-n}
\nonumber \\
{}~[J_m^i, J_n^\alpha ]&=& \alpha ^iJ_{m+n}^\alpha
\nonumber\\
 ~[J_m^\al,J_n^\beta] &=& \left\{\begin{array}{ll} \epsilon(\al,\beta)J_{m+n}
^{\al+\beta}\hs{30mm} &
\mbox{if $\alpha +\beta$ is a root}
\nn \\
\frac{2}{\al^2}(\al_iJ^i_{m+n}+\frac{K}{2}m\del_{m+n,0}) \hs{7mm} & \mbox{ if
 $\al=-\beta$ } \\
 0 \hs{48mm} & \mbox{otherwise}\label{cwbasis}
\end{array}
\right. \\
& &[K,J_m^a]=0 \hs{10mm} \forall a.
\ee

This is, however, not the whole story. Taken as above, the Cartan
sub-algebra is not maximal since $[J^i_0,J^j_m]=0$ and $[K,J^i_m]=0$
$\forall i,j$ and $m$. We must thus add at least one element to the Cartan
sub-algebra which removes this degeneracy. It is also clear that this
element should be connected to the roots of $J^i_m$. One candidate that
meets those requirements is minus the zero mode of the
Virasoro algebra connected to the affine Lie algebra through the
Sugawara construction. Adding the generator $d=-L_0$ we find the full
centrally extended loop algebra in the Cartan-Weyl basis as (\ref{cwbasis})
and in addition
\be
[d,J_m^a]=mJ_m^a \hs{15mm} [d,K]=0 \hs{15mm} [d,d]=0.
\ee

We may now look a bit closer on the roots. We have three different types of
Cartan sub-algebra generators $J_0^i$, $K$ and $d$. They probe separate
parts of the roots or their corresponding step operators which may be
sorted into two groups: $J^{\al}_m$ where $\al$ is a root of the finite
Lie algebra, and $J^i_m$ $m\neq 0$. The different Cartan sub-algebra generators
probe $\al$, the level $k$ and the mode number respectively. We thus take a
generic affine root to have three entries corresponding to these three
types of generators. $J^{\al}_m$ corresponds to the root $\hat{\al}=
(\al,0,m)$ and the generator $J_m^i$ to the root $m\del=(0,0,m)$
i.e. we take $\del=(0,0,1)$. 

We derive the affine Killing form by comparing to the finite
dimensional Lie algebra
and imposing the generalized restrictions that holds in the finite case
namely symmetry, bilinearity and associativity. Generalizing the finite Lie
algebra definition $\ka(x,y)\equiv {\rm Tr}({\rm ad}_x\circ {\rm ad}_y)$
would not work since the affine trace becomes infinite. Associativity means
that
\be
\hat{\ka}([x,y],z)=\hat{\ka}(x,[y,z]).
\ee
Choosing specific set of generators $x,y,z$ and using, in addition to
associativity,  symmetry and bilinearity
will give us $\hat{\ka}$. \\
\begin{tabular}{ccccc}
 & x & y & z & restriction on $\hat{\ka}$ \\
1 & $J^a_m$ & $d$ & $J^b_n$ & $(m+n)\hka(J^a_m,J^b_n)=0$ \\
2 & $J^a_m$ & $J_0^b$ & $J_{-m}^c$ & $f^{ab}_{\ \ d}\hka(J^d_m,J^c_{-m})=
f^{bc}_{\ \ d}\hka(J^a_m,J^d_{-m})$ \\
3 & $J^a_m$ & $J_1^b$ & $J_{-m-1}^c$ & $f^{ab}_{\ \
d}\hka(J^d_{m+1},J^c_{-m-1})=
f^{bc}_{\ \ d}\hka(J^a_m,J^d_{-m})$ \\
4 & $J^a_m$ & $J_n^b$ & $K$ & $\hka(J^a_m,K)=0$ \\
5 & $J^a_m$ & $J_{-m}^b$ & $K$ & $f^{ab}_{\ \ c}\hka(J^c_0,K)+\frac{1}{2}
\del^{ab}\hka(K,K)=0$ \\
6 & $J^a_m$ & $J_n^b$ & $d$ &
$f^{ab}_{\ \ c}\hka(J_{m+n}^c,d)+m\ka^{ab}\del_{m+n,0}\hka(k,d)=
-n\hka(J^a_m,J^b_n)$ \\
\end{tabular} \\

\noindent 1 and 2 implies together that
$\hka(J^a_m,J^b_n)\propto\del_{m+n,0}\ka^{ab}$.
3 shows that the constant is $m$ independent and we can therefor choose
normalization
\be
\hka(J^a_m,J^b_n)=\del_{m+n,0}\ka^{ab}.
\ee
4 provides only the information listed above. 5 gives, using 4
\be
\hka(K,K)=0
\ee
Using 6 and symmetry of $\hka$
we get
\be
\hka(J^a_m,d)=0 \hs{15mm} \hka(k,d)=1.
\ee
We cannot constrain $\hka(d,d)$ in this way but we may define
$\hka(d,d)=\eta$. If we introduce $d'=d-\frac{1}{2}\eta K$, which obeys
the same algebra as $d$, we get $\hka(d',d')=0$. We may therefore choose
\be
\hka(d,d)=0.
\ee
In summary we have thus
\be
\hka^{ab}=
\lef[\begin{array}{ccc}
\ka^{ab}\del_{m+n,0} &\ 0\  &\ 0\ \\
0 & 0 & 1 \\
0 & 1 & 0
\end{array}\rig].
\ee

The scalar product of two arbitrary elements $\hat{\la}=(\la,k,n)$ and
$\hat{\la}'=(\la',k',n')$ thus becomes $(\hat{\la}|\hat{\la}')=
(\la|\la')+kn'+n'k$. Note that the roots $n\del$ has zero scalar product
with all roots including itself and we will refer to those as lightlike
roots in what follows. The non-lightlike roots, referred to as real roots
in what follows, have positive inner product.

We define the set of positive roots
\be
\hat{\De}^+=\{\hat{\al}=(\al,0,n)\in\hat{\De};n>0\ {\rm or}\ n=0\
{\rm and}\ \al\in\De^+\}
\ee
where $\hat{\De}$ is the set of affine roots, and $\De^+$ is the set of
positive roots of the finite dimensional Lie algebra. The set of negative roots
are $\hat{\De}^-=-\hat{\De}^+$ and $\hat{\De}=\hat{\De}^++\hat{\De}^-$,
which is a disjoint union. The set of simple affine roots are then given as
\be
\hat{\De}_s=\{\hat{\al}_{(i)}=(\al_{(i)},0,0) \ i=1,...,r\ {\rm and}\
\hat{\th}=\hat{\al}_{(0)}=(-\th,0,1)\}
\ee
where $\al_{(i)}\in\De_s$ are the simple roots and $\th$ the highest
root of the finite dimensional Lie algebra.

We should also define the affine analogy of the sum of positive roots in the
finite case. There are, however, infinitely many positive affine roots. We
rather take the definition of the affine sum of positive roots $\hat{\rho}$
from one of the properties of the finite algebra sum of positive roots
$\rho$ namely
\be
(\hat{\rho}|\hat{\al}_{(i)})\equiv(\hat{\al}_{(i)}|\hat{\al}_{(i)}).
\ee
{}From this it follows that $\hat{\rho}=(\rho,c_g,n)$ where $c_g$
is the quadratic Casimir in the adjoint representation. Without loss of
generality we may choose $n=0$.
There also exists a non-ambiguous definition of $\hat{\rho}$ in terms of
fundamental weights which yields this result.

It is not difficult to see that the Cartan matrix that follows from the
roots in this realization of the affine Lie algebra coincides with the
Cartan matrix given with the Chevalley-Serre realization. The easiest way
of convincing oneself is to compare the Dynkin diagrams of the two
realization. This finally proves the isomorphism of the
defining relations (\ref{definerel}) and the realization of affine Lie
algebras as centrally extended loop algebras.

\section{Twisted affine Lie algebras}

Twisted affine Lie algebras can be obtained in much the same
way as the untwisted ones and they are, for example, realized as slightly
generalized centrally extended loop algebras.
Recall that the loop algebras were obtained when considering analytic maps
from $S^1$ onto some finite Lie algebra $g$. We have the general loop
algebra element realized as $x\otimes{\cal P}(z)$ where $x\in g$ and
${\cal P}(z)$ is a Laurent polynomial in the coordinate $z$ of $S^1$.

The generalization we can allow for is to take the twisted boundary
condition
\be
x\otimes{\cal P}(e^{2\pi i}z)=\om(x){\cal P}(z) \label{twistedbc}
\ee
where $\om$ is a finite automorphism of $g$. Finiteness means that there
exists a finite integer $l$ such that $\om ^l=1$. This induces a gradation
of $g$ in the way that defining $x\in g_{[j]}\ \IFF\ \om(x)=e^{2\pi ij/l}x,\
j=0,...,l-1$ we have
\be
g=\bigoplus_{j=0}^{l-1}g_{[j]}
\ee
This shows that c.f. (\ref{twistedbc}) we have to take
\be
{\cal P}(e^{2\pi i}z)=e^{2\pi ij/l}{\cal P}(z)
\ee
for the polynomial multiplying $x\in g_{[j]}$. Proceeding exactly as for
untwisted
case we find the natural bracket structure
\be
[J^a_{m+j/l},J^b_{n+j'/l}]=f^{ab}_{\ \ c}J^c_{m+n+(j+j')/l}+
(m+\frac{j}{l})\frac{k}{2}\ka^{ab}\del_{m+n+(j+j')/l,0}.
\ee

Not all twisted algebras obtained in this way are mathematically distinct
from the untwisted algebra. If the automorphism is what is known as an
inner automorphism then we may regain the untwisted algebra by a
redefinition of the generators.

Inner automorphisms are of the form
\be
\om(x)=\ga x\ga^{-1}
\ee
fore some groupelement $\ga\in G$ with Lie algebra $g$. We may now take
\be
\ga=e^{2\pi i\chi\cdot H}
\ee
for some vector $\chi$ and $H$ denoting the Cartan sub-algebra of $g$.
$\chi$ is
constrained by $\om^l=1$ which transforms into
\be
l\chi\cdot\al\in{\Bbb Z} \hs{10mm} \forall\al\in\De.
\ee
It is not difficult to check that $\om$ leaves the Cartan sub-algebra
invariant while
\be
\om(E^{\al})=e^{2\pi i\chi\cdot\al}E^{\al}.
\ee
for a generic root generator $E^{\al}$ of the affine Lie algebra $\hg$.
This implies that if we introduce the new generators
\be
\ti{J}^{\al}_n=J^{\al}_{n+\chi\cdot\al} \hs{8mm}
\jt_n^i=J^i_n+\chi^i\del_{n,0}K \hs{3mm} n\in{\Bbb Z} \hs{8mm}
\widetilde{K}=K \hs{8mm} \ti{d}=d-\chi_iJ^i_0
\ee
then they satisfy an untwisted affine Lie algebra.

For inner automorphisms there thus exists an isomorphism between the
twisted and untwisted algebras. Physically there may, however, be a
difference between the two ways of representing the algebra. In particular,
the vacuum of the untwisted algebra, i.e. the state of lowest $\ti{d}$
eigenvalue will transform in some representation of $g$. The vacuum
state of the twisted algebra, corresponding to lowest $d$ eigenvalue, on the
other hand, transforms in some representation of $g_{[0]}\subset g$ the
subalgebra which commutes with $\chi_iJ^i_0$. Choosing the twisted way of
representing the algebra thus breaks the symmetry from $g$ to $g_{[0]}$.

The outer automorphisms gives a non-isomorphic algebra and corresponds to
the Dynkin  diagrams presented in fig.(4.3) with superscripts different
from one. For construction of root systems, a non-trivial matter, we
refer to the literature quoted at the end of this chapter.

It is important to note that there exists two different notations for
twisted affine Lie algebras. Referring to their respective subalgebras
$g_{[0]}$  we have here denoted the possible twisted cases
$A^{(2)}_1,F^{(2)}_4,B^{(2)}_r,\ti{B}^{(2)}_r,C^{(2)}_r$ and $G^{(3)}_2$.
The upper index is the order of the automorphism denoted $l$ above.
More common in the literature is to denote the twisted cases by letters
referring to $g$ the finite Lie algebra that underlies the particular
realization in terms of centrally extended loop algebras. Here we then
have $A^{(2)}_2,E^{(2)}_6,D^{(2)}_{r+1},A^{(2)}_{2r},A^{(2)}_{2r-1}$ and
$D_4^{(3)}$ respectively.

\section{Generalizations of affine Lie algebras}

We will here briefly mention two possible generalizations
of the affine Lie algebra structure, both introducing fermionic
degrees of freedom. The two generalizations will in what follows
be called affine Lie superalgebras and superaffine Lie algebras.
Affine Lie superalgebras are related to Lie superalgebras in much the
same way as affine Lie algebras are related to finite Lie algebras.
Superaffine Lie algebras, on the other hand, is the supersymmetric
generalization of affine Lie algebras.

The affine Lie superalgebra has the following bracket structure
\be
& & [J_m^A,J_n^B]=f^{AB}_{\ \ \ \ C}J^C_{m+n}+
\frac{k}{2}m\del_{m+n,0}\Om^{AB} \nn \\
& &\{\psi^a_m,\psi^b_n\}=g^{ab}_AJ^A_{m+n}+
\frac{k}{2}m\del_{m+n,0}\om^{ab} \nn \\
& &[\psi_m^a,J^A_n]=h^{aA}_b\psi_{n+m}^b
\ee
where $J$ are even elements and $\psi$ are odd elements with respect
to a ${\Bbb Z}_2$ gradation. $f^{AB}_{\ \ \ \ C},g^{ab}_A$ and
$h^{aA}_b$ are structure constants and $\Om^{AB}$ and $\om^{ab}$ are
bilinear forms. Under certain restrictions it is possible to obtain a
Virasoro algebra via a Sugawara construction. In this context the algebra
is the symmetry algebra of a supergroup valued WZNW model.

The bracket structure of superaffine Lie algebras is
\be
& & [J_m^a,J_n^b]=f^{ab}_{\ \ c}J^c_{m+n}+
\frac{k}{2}m\del_{m+n,0}\ka^{ab} \nn \\
& &\{\psi^a_r,\psi^b_s\}=
\frac{k}{2}m\del_{r+s,0}\ka^{ab} \nn \\
& &[\psi_r^a,J^b_m]=f^{ab}_{\ \ c}\psi_{r+m}^c. \label{superaffine}
\ee
Here $J_m^a$ are generators of an affine Lie algebra, $\psi^a_r$ is a
fermionic generator. $f^{ab}_{\ \ c}$ are structure constants and $\ka^{ab}$
the Killing form of the underlying finite dimensional Lie algebra.

This is the symmetry algebra of the supersymmetric WZNW model c.f. chapter
seven
which is $N=1$ superconformal. Indeed by introducing the shifted
affine current
\be
\hJ^a_n\equiv J^a_n+\frac{i}{k}f^a_{\ bc}\sum_r\psi_{n-r}^b\psi^c_r
\ee
which obeys an affine Lie algebra of level $k-c_g$, $c_g$ being the
quadratic Casimir in the adjoint representation of $g$, we can realize the
$N=1$ superconformal algebra generators $L_n$ and $G_r$ as
\be
& & L_n=\frac{1}{k}g_{ab}\lef(\sum_l:\hJ^a_{n-l}\hJ^b_l:+
\sum_r(r-\frac{n}{2}):\psi^a_{n-r}\psi^b_r:+a\del_{n,0}\rig) \nn \\
& & G_r=\frac{2}{k}g_{ab}\sum_s\hJ^a_{r-s}\psi^b_s-
\frac{2i}{3k^2}f_{abc}\sum_{s,t}:\psi^a_{r-s-t}\psi^b_s\psi^c_t:
\ee
where $a=3/16$ for Ramond and $a=0$ for Neveu-Schwartz boundary conditions.

\begin{itemize}

\item J.F. Cornwell, {\sc Group theory in physics: Volume III
Supersymmetries and infinitedimensional algebras}, Academic Press 1989

\item J. Fuchs, {\sc Affine Lie algebras and quantum groups},
Cambridge University Press 1992

\item P. Goddard and D. Olive, {\it Kac-Moody and
Virasoro algebras in relation to quantum physics}
Int. J. Mod. Phys. 1 (1986) 303-414

\item J. Humphreys, {\sc Introduction to Lie algebras and
representation theory}, Springer-Verlag 1980

\item V.G. Kac, {\sc Infinite dimensional Lie algebras},
Cambridge University Press 3'd ed. 1990

\end{itemize}

\chapter{Verma modules over affine Lie algebras}

In the models which we discuss in the papers
contained in this thesis the affine Lie
algebra appears as the constraint algebra, and the state space consists of
what is known as Verma modules over the affine Lie algebra. In the Verma
module there appears null-vectors in much the same way as for the case of
Virasoro algebras. The structure and presence of null-vectors is of
considerable
importance when studying the BRST cohomology over complexes including
Verma modules. Due to their presence there may arise new states in the
BRST cohomology of non-zero ghost number. I will discuss those aspects
in more detail at the end of chapter seven when the necessary ingrediences
have been introduced.

We will here discuss modules which are diagonal with
respect to the Cartan sub-algebra $h$. If a module $V$ over an affine Lie
algebra $\hg(A)$ is diagonalizable with respect to the Cartan sub-algebra
it allows for a weight space decomposition
\be
V=\bigoplus_{\la}V_{\la}
\ee
for weights $\la\in${\sc h}$^{\ast}$. More precisely
\be
V_{\la}=\{v\in V;h(v)=\la(h)v\}
\ee
for $h\in${\sc h}. $V_{\la}$ is called the weight space, $\la$ is a weight
if $V_{\la}\neq0$, and dim$V_{\la}$ is known as the multiplicity of $\la$.

\section{Highest weight modules}

One important sub-class of weight modules are so-called
highest weight modules.
We call a $g(A)$ module $V$, highest weight module if there exists an element
$v_{\La}\in V$
such that $v_{\La}$ is annihilated by all generators of the algebra
corresponding to positive roots, $h(v_{\La})=\La(h)v_{\La}$ $h\in${\sc h},
and all elements of the module may be obtained by applying generators of the
algebra, corresponding to negative roots on $v_{\La}$. It follows from the
algebra that highest weight modules allow for a weight space decomposition.

The eigenvalues of the Cartan sub algebra on the highest weight vectors are
known as highest weights. We denote them
by $\hat{\La}=(\La,k/2,n)$, where $\La$
is the eigenvalue of the Cartan sub algebra of the underlying finite Lie
algebra, $k$ is the eigenvalue of $K$ the central element, and $n$ the
eigenvalue of the derivation $d$. We can always shift $n$ to zero and we
take this as our convention.

{}From the fact that $K$ is in the center we see that all vectors of a
module $V$ have the same eigenvalue $k$ of $K$, as the highest weight vector.
Furthermore, by representing a vector $v_{\la}$ as
\be
v_{\la}=\prod_jJ^{\al_j}_{-n_j}\prod_kJ^{i_k}_{-n_k}v_{\La}
\hs{10mm} n_j\geq0,n_k>0, \al_j\in\De
\ee
we see from
\be[d,J^a_{-n}]=-nJ^a_{-n} \hs{15mm} \Longrightarrow \hs{15mm}
dv_{\la}=-(\sum_jn_j+\sum_kn_k)v_{\la}.
\ee
$k$ is called the level of the weight and $N=\sum_jn_j+\sum_kn_k$
is known as the grade.

Of particular interest are the highest weights, known as dominant weights, and
their corresponding modules, known as integrable highest weight modules.
Dominant weights are defined to obey
\be
\hat{\La}\cdot\hat{\al}^{(i)\vee}\in {\Bbb Z}_{\geq0} \hs{10mm} \forall
\hat{\al}^{(i)}\in\hat{\De}_s \label{dominan}
\ee
which, expressed in terms of quantities related to the finite Lie algebra,
means
\be
\La\cdot\al^{(i)}\geq0 \hs{10mm} \forall \al^{(i)}\in\De_s
\hs{10mm} {\rm and } \hs{10mm} \La\cdot\th\leq k/2
\ee
where $\th$ is the highest root of the finite dimensional algebra $g$.

These conditions can be obtained by demanding that the sub-modules of the
irreducible module of the $sl(2)$ sub-algebra generated by
$E^{\hat{\al}}=J^{\al}_n$,
$E^{-\hat{\al}}=J^{-\al}_{-n}$ and
$H^{\hat{\al}}=[E^{\hat{\al}},E^{-\hat{\al}}]$
are finite dimensional. Using the fact that finite-dimensional representations
of $sl(2)$ have integer weights, and in addition, that for a
highest weight this integer is
non-negative, we find
\be
\hat{\la}\cdot\hat{\al}^{\vee}\in{\Bbb Z} \hs{15mm}
\hat{\La}\cdot\hat{\al}^{\vee}\in{\Bbb Z}_{\geq0} \hs{15mm}
\hs{10mm} \forall \hat{\al}\in\hat{\De}_R^+ \label{cond1}
\ee
Here $\hla$ is any weight of the representation.
Only a finite number of these conditions are independent corresponding to the
simple roots $\hat{\al}^{(i)}\in\hat{\De}_s$. Furthermore,
the first condition in (\ref{cond1}) follows from the second which means that
this is enclosed in
eq.(\ref{dominan}).

The condition for a weight to be dominant is equivalent to what is known as
local
nilpotency, which is what must be required in order to be able to
exponentiate a
representation of the algebra to a representation of the group. Hence the name
integrable modules or integrable representation.

Fundamental weights $\La_{(i)}$ $i=0,...,l$ are for the case of semi-simple
affine Lie algebras, defined to be dual to the simple co-roots.
\be
\hat{\La}_{(i)}\cdot\hat{\al}^{(j)\vee}=\del_i^{\ j}
\ee
{}From this we deduce, using the known vectors $\hat{\al}^{\vee}$, that
\be
\hat{\La}_{(0)}=(0,\frac{1}{2}\th^2,0) \hs{15mm}
\hat{\La}_{(i)}=({\La}_{(i)},m_i,0)
\ee
where $m_i=a^{\vee}_i\th^2/2$. This can be used to define the analogue of the
quadratic Casimir for affine Lie algebras. We take the Weyl vector $\hrho$ as
twice the sum of the fundamental weights
\be
\hat{\rho}=2\sum_{i=0}^r\hat{\La}_{(i)}=(\rho,c_g,0)
\ee
where $c_g$ is the quadratic Casimir of the adjoint representation
of the finite Lie algebra $g$. Note that this coincides with our previous
definition. As remarked above, the usual definition of the quadratic
Casimir in terms of operators does not make sense, since it
includes an infinite sum
over positive roots.
We take the eigenvalue definition using the Weyl vector. We then have
\be
(\hat{\La},\hat{\La}+\hat{\rho})=(\La,\La+\rho)=C_{\La}
\ee
which is independent of the level of the module.

\section{Affine Weyl group}

We define the Weyl group $\hat{W}$ to be generated by the
identity element and the reflections
\be
\hat{\si}(\hat{\la})=\hat{\la}-\hat{\la}\cdot\hat{\al}^{\vee}\hat{\al}
\hs{15mm}
\hat{\al}\in\hat{\De}_R.
\ee
Note that for light-like roots this definition would not make sense since we
would divide by zero.
To justify the
notion reflection we note that, as for the finite case,
\be
\hat{\si}_{\hat{\al}}(\hat{\la})\cdot\hat{\al}^{\vee}=-\hat{\la}\cdot\hat{\al}^{\vee}
\ee
The Weyl group is generated by the simple reflections
$\hat{\si}_{\hat{\al}^{(i)}}$
for simple roots $\hat{\al}^{(i)}$ $i=0,...,r$. The positive roots other than
$\hat{\al}^{(i)}$ are permuted by these simple reflections. We also have, as
usual,
that $\hat{\si}_{\hat{\al}}(\hat{\si}_{\hat{\al}}(\hat{\la}))=\hat{\la}$. What
is
new compared to the finite case is the action on light-like roots which are
invariant under the Weyl reflections since $\hat{\al}\cdot\hat{\del}=0$ for any
root $\hat{\al}$ and
$\hat{\del}\in\hat{\De}_I$ .

Take an arbitrary weight $\hat{\la}=(\la,k,m)$. Its Weyl reflection with
respect
to a root $\hat{\al}=(\al,0,n)$ becomes
\be
& &\hat{\si}_{\hat{\al}}(\hat{\la})=\hat{\la}-
\frac{2\hat{\la}\cdot\hat{\al}}{\hat{\al}\cdot\hat{\al}}\hat{\al}\nn \\
& &=(\la-\la\cdot\al^{\vee}\al-kn\al^{\vee},k,m-n\la\cdot\al^{\vee}
-kn^2\frac{2}{\al\cdot\al})
\ee
Introducing the translation
\be
t_{\bet}:\hat{\la}=(\la,k,m)\lra
\lef(\la+k\bet,k,m+\frac{1}{2k}\lef(\la\cdot\la-
(\la+k\bet)\cdot(\la+k\bet)\rig)\rig)
\ee
we se that we can write $\hat{\si}_{\hat{\al}}$ as $n$ translations followed by
a
finite Weyl transformation of the first entry
\be
\hat{\si}_{\hat{\al}}(\hat{\la})=\lef(\si_{\al}(\la+nk\al^{\vee}),k,
m+\frac{1}{2k}\lef(\la\cdot\la-(\la+nk\al^{\vee})\cdot(\la+nk\al^{\vee})\rig)\rig)
\ee
which we may represent as
\be
\hat{\si}_{\hat{\al}}=\si_{\al}\circ(t_{\al^{\vee}})^n
\ee
where we define $\si_{\al}$ to act as a finite Weyl transformation on the
component $\la$ of $\hat{\la}$ and as the identity on the two other components.

Defining the co-root lattice as the lattice spanned by the co-roots we see
that it generates translations on this lattice and obviously
\be
t_{\al}\circ t_{\bet}=t_{\al+\bet}
\ee
It is also possible to show that
\be
t_{\si(\al)}=\si\circ t_{\al}\circ\si^{-1} \hs{15mm} \forall \si\in W
\ee
so the translation group is a subgroup of $\hat{W}$. Thus $\hat{W}$ is the
semidirect product of the finite Weyl group $W$ and the group of translations
on the co-root lattice.

All
affine weights are generated by the set of dominant weights $P_+$ and the
Weyl group, where the dominant weights are defined from
\be
P_+=\{\hla \ |\ (\hla+\hrho/2)\cdot\hal^{(i)}\geq0\ \ \forall
\ \hal^{(i)}\in\hDe_s\}.\label{dominant}
\ee
Note that this definition is slightly different to the usual one
(\ref{dominan}).
Furthermore, the Weyl group is generated by the elements corresponding
to simple roots. This will be useful when we consider the structure of
singular vector in Vema modules over affine Lie algebras.

\section{Verma modules}

Take a $g(A)$ module $M_{\La}$ of highest weight $\La$. We call
it a Verma module if every highest weight $g(A)$ module over $\La$ is a
quotient of $M_{\La}$.

Some important properties of Verma modules $M_{\La}$ are i) $M_{\La}$ is
unique up to isomorphisms for every $\La$. ii) $M_{\La}$ contains a unique
proper maximal submodule $M'_{\La}$. iii) There exists a unique irreducible
submodule $L_{\La}=M_{\La}/M'_{\La}$. Obviously iii) follows from ii). In ii)
proper means that $M'_{\La}$ cannot be identical to $M_{\La}$, but the trivial
submodule without any elements is allowed.

An irreducible module is a module that contains exactly one proper maximal
submodule namely the trivial submodule. A fully reducible module is a module
which is a direct sum of irreducible modules. To get
the irreducible submodule $L_{\La}$  one must divide out any elements of the
form
\be
v\in M_{\la}, \hs{10mm} v\in\!\!\!\!\!/U, \hs{10mm} n_+(v)\subset U
\ee
where $M_{\la}$ is defined from the weights space decomposition of $M_{\La}$
$M_{\La}=\bigoplus_{\la\leq\La}M_{\la}$ and
$U$ is a submodule of $M_{\La}$. $v$ is called a primitive vector and $\la$ a
primitive weight. Note that the vectors $v$ of the type $\ n_+(v)=0$ are
primitive. We denote
this set of primitive vectors singular vectors. The singular vectors
and descendants of singular vectors
will be denoted null-vectors. The reason for this is that null-vectors decouple
from all vectors in the Verma module including its
conjugate. In the irreducible module they must, therefore, be identified with
the zero element, (although they have a non-zero weight).

The singular vectors $v_{\la}$ satisfy the condition of a highest weight
vector, and thus the Verma module $M_{\La}$ contains Verma modules $M_{\la}$
as submodules. Verma showed, for finite dimensional Lie algebras, that there
is exactly one Verma submodule for
each distinct singular vector \cite{Verma68}.
In general, a Verma module is not generated by its Verma submodules due to
prescence of primitive vectors that are not singular, and this fact will
obscure the structure of Verma modules. This was first discovered in
\cite{Bernshtein-Gel'fand-Gel'fand71}.

For finite Lie algebras a necessary condition for existence of primitive
vectors that are not singular were given by Conze and Dixmier
\cite{Conze-Dixmier72}. If $M_{\La}$
contains a sub-module which is not engraded by its Verma sub-modules then
there exists three elements $\la_1,\la_2,\la_3$ of $W(\La)$, the set of Weyl
reflected weights, all different such that
\be
\La\succ\la_1\succ\la_3 \hs{15mm} \La\succ\la_2\succ\la_3 \hs{15mm}
\la_2\geq\la_1
\ee
The notation here is; $\la_1\succ\la_3$ means that one may obtain $\la_3$ from
Weyl reflections of $\la_1$ while $\la_2\geq\la_1$ means that $\la_2-\la_1$
can be written as a sum of positive roots with positive integers.

Choose a series of sub-modules $M_0,M_1,M_2,...$ such that $M_{\la}=M_0\supset
M_1
\supset M_2\supset ...$. If all quotients of the type $M_i/M_{i+1}$ are
irreducible we call this a Jordan-H\"older series.
We define the multiplicity
$[M:L(\la)]$ of a irreducible submodule in a Verma module as the number of
times $L(\la)$ appears in the Jordan-H\"older series.

The multiplicity is in general different from one for at least one submodule
if the rank is larger than two. Jantzen see for example \cite{Jantzen80}
showed that the multiplicity for simple
finite dimensional Lie algebras is one for rank $\leq2$ and this was
generalized by
Rocha-Carridi and Wallach to affine Lie algebras
\cite{Rocha-Caridi-Wallach83:2}. For the finite dimensional case one may,
in fact, state that if $g$ contains k simple factors of
$A_n$ $n\geq3$, $C_n$ $n\geq3$, $D_n$ $n\geq4$ $E_n$ $n=6,7,8$ or $F_4$
then for a regular dominant weight $\La$, $M_{\La}$ contains an irreducible
sub-module $L_{\la}$ with multiplicity $\geq2^k$ \cite{Deodhar-Lepowsky77}.

We illustrate the existence of primitive non-singular vectors in a simple
example for finite Lie algebras. We choose $sl(4,{\Bbb C})$ as our algebra,
with Dynkin diagram as given in fig.(5.1).
\begin{figure}[H]
\begin{picture}(400,30)(0,0)

\put(180,20){\circle{10}}
\put(185,20){\line(1,0){20}}
\put(210,20){\circle{10}}
\put(215,20){\line(1,0){20}}
\put(240,20){\circle{10}}

\put(177,5){$\al_1$}
\put(207,5){$\al_2$}
\put(237,5){$\al_3$}

\end{picture}
\caption{sl(4,C)}
\end{figure}

With a suitable choice of numbering the positive roots are
$\{\al_1, \al_2, \al_3, \al_1+\al_2, \al_2+\al_3, \al_1+\al_2+\al_3\}$.
We choose a highest weight $\la$ such that $\la\cdot\al_2=0$,
$\la\cdot\al_1=\la\cdot\al_3=-1/2$.
The singular vectors of this module are $\la-\al_2,\ \la-\al_2-\al_1,\
\la-\al_2-\al_3,\ \la-\al_2-\al_1-\al_3,\ \la-2\al_2-\al_1-\al_3$
which gives rise to sub-modules of Verma type in $M_{\la}$, and we will denote
these $M_{\la_2},\ M_{\la_{21}},\ M_{\la_{23}},\ M_{\la_{213}},\
M_{\la_{2132}}$,
respectively. This exhausts the possible singular vectors. The singular vectors
of $M_{\la}$ are given by the elements
$E^{-\al_2}v_{\la},\ E^{-\al_1}E^{-\al_2}v_{\la},\ E^{-\al_3}E^{-\al_2}v_{\la},
\ E^{-\al_3}E^{-\al_1}E^{-\al_2}v_{\la},\   $
and $E^{-\al_2}E^{-\al_3}E^{-\al_1}E^{-\al_2}v_{\la},\ $ where $v_{\la}$ is the
highest weight vector, and $E^{\al}$ are generators corresponding to simple
roots.
Take the element
\be
z=(E^{-\al_1}E^{-\al_2}E^{-\al_3}-E^{-\al_3}E^{-\al_2}E^{-\al_1})v_{\la}
\ee
of weight $\la-\al_1-\al_2-\al_3$. Obviously $z\in M_{\la}$ but
$z\in\!\!\!\!\!/ M_{\la_2}$ and all singular vectors of $M_{\la}$ except
$E^{-\al_2}v_{\la}$ are also singular vectors i.e. contained in $M_{\la_2}$.
A simple calculation now shows that
\be
E^{\al_1}z\propto E^{-\al_3}E^{-\al_2}v_{\la} \hs{15mm}
E^{\al_3}z\propto E^{-\al_1}E^{-\al_2}v_{\la} \hs{15mm}
E^{\al_2}z=0
\ee
and we thus find that
\be
n_+(z)\in M_{\la_2}
\ee
Thus $z$ cannot be in irreducible sub-module $L_{\la}$ and the sub-module
generated by $n_-(z)$ is not a Verma module.

\section{Structure of singular vectors in Verma modules}

We would here like to establish when singular vectors
appear in the Verma module and their generic form. The ocurrence of
singular vectors in Verma modules $M_{\La}$
is given by the Kac-Kazhdan determinant \cite{Kac-Kazhdan79},
which is proportional to
\be
\prod_{\hat{\al}>0}\prod_{n=1}^{\infty}\lef((\hat{\La}+\frac{\hat{\rho}}{2})
\cdot\hat{\al}-
\frac{n}{2}\hat{\al}\cdot\hat{\al}\rig)
^{P(\hat{\eta}-n\hat{\al})}.
\ee
Here $\hat{\rho}$ is as usual the sum of fundamental weights, $\hat{\eta}$ is a
grading of the module with respect to positive roots, and $P(\hat{\la})$ is
the Kostant partition function which is the number of distinct
sets of positive integers
$\{m_i\}$ such that $\hat{\la}=\sum_i m_i\hat{\al}$ for the positive roots
$\hat{\al}$.
We thus have the gradation
\be
M_{\hat{\La}}=\bigoplus_{\hat{\eta}\in\hat{\De}^+}(V_{\hat{\La}})_{\hat{\eta}}
\ee
where elements in $(V_{\hat{\La}})_{\hat{\eta}}$ differ from the highest weight
$\hat{\La}$ by $\hat{\eta}$. Intuitively we may think of the Kac-Kazhdan
determinant
as the determinant of the matrix of all inner products of states at a
certain weight $\hat{\La}-\hat{\eta}$.

The Verma module $V_{\hat{\La}}$ will be irreducible if and only if
$2(\hat{\La}+\hat{\rho}/2)\cdot\hat{\al}\neq n\hat{\al}\cdot\hat{\al}$ for any
$\hat{\al}\in\hat{\De}^+$ and positive integer $n$. If, on the other hand,
$2(\hat{\La}+\hat{\rho}/2)\cdot\hat{\al}= n\hat{\al}\cdot\hat{\al}$ for any
$\hat{\al}\in\hat{\De}^+$ and positive integer $n$ there exists a singular
vector at weight $\hLa-n\hal$ in the Verma module $M_{\La}$.

We note that if we
restrict ourselves to real roots i.e. that $(\hat{\al},\hat{\al})\neq0$
then we may put the equation $2(\hat{\La}+\hat{\rho}/2)\cdot\hat{\al}=
n\hat{\al}\cdot\hat{\al}$ in the form
\be
\hsi^{\hrho}_{\hal}(\hLa)-\hLa=-n\hal
\ee
where $\hsi^{\hrho}_{\hal}$ is the $\hrho$-centered Weylreflection defined
as $\hsi^{\hrho}_{\hal}(\hLa)=\hsi(\hLa+\hrho/2)-\hrho/2$. For light-like roots
the conditions for singular vectors all reduce to $k+c_g=0$.

We may represent the $\hrho$-centered Weyl reflection of a weight
$\hla_0$ with respect to the simple root $\hal^{(i)}$ by
\be
v_{\hla_0}\lra(f_i)^{\ga_i}v_{\hla_0} \label{hwtonulltrans}
\ee
where the power $\ga_i$ is defined from $\hsi^{\hrho}_{\al^{(i)}}
(\hla_0)-\hla_0\equiv-\ga_i\hal^{(i)}$, and $f_i$ is the affine algebra
generator corresponding to $\hal^{(i)}$.

We may now convince
ourselves that $(f_i)^{\ga_i}v_{\hla_0}$ is a singular-vector
simply by acting on it by $n_+$. It is sufficient to check the generators
corresponding to simple roots i.e. to check with $e_i$. The generators of $e_i$
are given by $J^{-\th}_1$ and $J^{\al^{(i)}}_0$ corresponding to $\hal^{(0)}$
and $\hal^{(i)}$ for $i=1,...,r$. The creators $f_i$ are given by switching
the sign of the affine roots. If we choose $i=0$ in (\ref{hwtonulltrans})
the non-trivial check will correspond to $J^{-\th}_1$ and becomes
\be
J^{-\th}_1(J^{\th}_{-1})^{\ga_0}v_{\hla_0}=
\ga_0\frac{2}{\th^2}(-\th\cdot\la_0+\frac{k}{2}-\th^2\frac{\ga_0-1}{2})
(J^{\th}_{-1})^{\ga_0-1}v_{\hla_0}.
\ee
We have thus a singular vector if $\ga_0=0$ or $\ga_0=
(k-2\th\cdot\la_0)/\th^2+1$, and the latter condition exactly corresponds
to the definition of $\ga_0=2(\hla_0+\hrho/2)\cdot\hal^{(0)}/\th^2$
from the Weyl reflection. It is not difficult
to find the corresponding
result for the choice of another simple root.

Following this prescription we may thus represent a singular vector in the
module $M_{\hla_0}$ as
$v_{\hla_{i_n...i_1}}=(f_{i_n})^{\ga_{i_n}}...(f_{i_1})^{\ga_{i_1}}v_{\hla_0}$.
If we do this for two singular vectors $v_{\hla_{i_n...i_1}}$ and
say $v_{\hla_{j_m...j_1}}$ we may eliminate $v_{\hla_0}$ to find
\be
v_{\hla_{i_n...i_1}}=(f_{i_n})^{\ga_{i_n}}...(f_{i_1})^{\ga_{i_1}}
(f_{j_1})^{-\ga_{j_1}}...(f_{j_m})^{-\ga_{j_m}}v_{\hla_{j_m...j_1}}.
\label{singvect}
\ee
All such expressions like (\ref{singvect}) do not make sense, since
the total power of each $f_i$
for arbitrary choices of $v_{\hla_{i_n...i_1}}$ and
$v_{\hla_{j_m...j_1}}$ may be negative. It may be shown
\cite{Malikov-Feigin-Fuks86} that
$v_{\hla_{i_n...i_1}}$ is a singular vector in the module
$M_{\hla_{j_m...j_1}}$ if one may obtain the sequence $j_1...j_m$
from the sequence $i_1...i_n$ by deleting $m-n$ $i$'s. We may then
of course rewrite (\ref{singvect}) in a form with only positive powers
using analytically continued commutation relations. \\

We will take the example of $\widehat{su}(2)$ to shed some light
on this construction. For convenience we take $\th^2=1$ for the rest
of this chapter.
In this case we have the fundamental weights
$\hmu_0=(0,1,0)$ and $\hmu_1=(1,1,0)$. We may parametrize an arbitrary
weight by the spin $j$ and level $k$ as $\hla=(k/2-j)\hmu_0+j\hmu_1$.
The simple roots are $\hal_0=(-1,0,1)$ and $\hal_1=(1,0,0)$ and $\hrho=
(1,c_g,0)$ where $c_g=2$.
We parametrize an arbitrary positive
real root as
$\hal=n'\hal_0+(n'-1)\hal_1$ $n'>0$ and $\hal=n'\hal_0+(n'+1)\hal_1$
$n'\geq0$.

We then have that for the two cases that if there exists solutions
to the equations
\be
& & 2j+1=-n+n'(k+c_g) \ \ \ n\geq1 \ \ n'\geq1 \nn \\
& & 2j+1=n-n'(k+c_g) \ \ \ n\geq1 \ \ n'\geq0
\ee
we have a singular vector of weight $j_s=j+n$ and $j_s=j-n$ respectively
and total mode number $N=nn'$ in the module $M_j$.

The structure of embedding of Verma modules can be visualized as in
fig.(5.2).

\begin{figure}[H]
\begin{picture}(400,40)(0,0)

\put(100,20){\circle*{2}}
\put(117,2){\vector(-1,1){15}}
\put(117,38){\vector(-1,-1){15}}

\put(119,1){\circle*{2}}
\put(119,39){\circle*{2}}
\put(155,2){\vector(-1,1){34}}
\put(155,38){\vector(-1,-1){34}}
\put(125,1){\vector(-1,0){4}}
\put(128,1){\line(1,0){4}}
\put(135,1){\line(1,0){4}}
\put(142,1){\line(1,0){4}}
\put(149,1){\line(1,0){4}}

\put(125,39){\vector(-1,0){4}}
\put(128,39){\line(1,0){4}}
\put(135,39){\line(1,0){4}}
\put(142,39){\line(1,0){4}}
\put(149,39){\line(1,0){4}}

\put(157,1){\circle*{2}}
\put(157,39){\circle*{2}}
\put(195,2){\vector(-1,1){34}}
\put(165,1){\vector(-1,0){4}}
\put(168,1){\line(1,0){4}}
\put(175,1){\line(1,0){4}}
\put(182,1){\line(1,0){4}}
\put(189,1){\line(1,0){4}}

\put(195,38){\vector(-1,-1){34}}
\put(165,39){\vector(-1,0){4}}
\put(168,39){\line(1,0){4}}
\put(175,39){\line(1,0){4}}
\put(182,39){\line(1,0){4}}
\put(189,39){\line(1,0){4}}

\put(197,1){\circle*{2}}
\put(197,39){\circle*{2}}
\put(235,2){\vector(-1,1){34}}
\put(205,1){\vector(-1,0){4}}
\put(208,1){\line(1,0){4}}
\put(215,1){\line(1,0){4}}
\put(222,1){\line(1,0){4}}
\put(229,1){\line(1,0){4}}

\put(235,38){\vector(-1,-1){34}}
\put(205,39){\vector(-1,0){4}}
\put(208,39){\line(1,0){4}}
\put(215,39){\line(1,0){4}}
\put(222,39){\line(1,0){4}}
\put(229,39){\line(1,0){4}}

\put(237,1){\circle*{2}}
\put(237,39){\circle*{2}}

\put(250,20){.....}

\end{picture}
\caption{Embedding of Vermamodules for $\widehat{su}(2)$ I}
\end{figure}

The dots correspond to Verma modules and the arrows point
towards modules in which the module is embedded in. They are of
two types corresponding to solid and dashed arrows, respectively.
The solid arrow indicates that the highest weight vector of the embedded
module $v_{\hla_e}$ may be written as $v_{\hla_e}=(f_i)^{\ga_i}v_{\hla}$
where $v_{\hla}$ is the highest weight of the module which contains
the embedded module.
The dashed lines corresponds to more complicated expressions.

The discussion above refers to the case when $k>0$ For the opposite case
when $k<0$ the analysis is analogous with one essential difference, the
corresponding embedding diagram terminates as is shown in fig.(5.3).

\begin{figure}[H]
\begin{picture}(400,40)(0,0)

\put(130,20){.....}

\put(162,1){\circle*{2}}
\put(162,39){\circle*{2}}
\put(198,2){\vector(-1,1){34}}
\put(198,38){\vector(-1,-1){34}}
\put(168,1){\vector(-1,0){4}}
\put(171,1){\line(1,0){4}}
\put(178,1){\line(1,0){4}}
\put(185,1){\line(1,0){4}}
\put(192,1){\line(1,0){4}}

\put(168,39){\vector(-1,0){4}}
\put(171,39){\line(1,0){4}}
\put(178,39){\line(1,0){4}}
\put(185,39){\line(1,0){4}}
\put(192,39){\line(1,0){4}}

\put(200,1){\circle*{2}}
\put(200,39){\circle*{2}}
\put(238,2){\vector(-1,1){34}}
\put(208,1){\vector(-1,0){4}}
\put(211,1){\line(1,0){4}}
\put(218,1){\line(1,0){4}}
\put(225,1){\line(1,0){4}}
\put(232,1){\line(1,0){4}}

\put(238,38){\vector(-1,-1){34}}
\put(208,39){\vector(-1,0){4}}
\put(211,39){\line(1,0){4}}
\put(218,39){\line(1,0){4}}
\put(225,39){\line(1,0){4}}
\put(232,39){\line(1,0){4}}

\put(240,1){\circle*{2}}
\put(240,39){\circle*{2}}
\put(278,2){\vector(-1,1){34}}
\put(248,1){\vector(-1,0){4}}
\put(251,1){\line(1,0){4}}
\put(258,1){\line(1,0){4}}
\put(265,1){\line(1,0){4}}
\put(272,1){\line(1,0){4}}

\put(278,38){\vector(-1,-1){34}}
\put(248,39){\vector(-1,0){4}}
\put(251,39){\line(1,0){4}}
\put(258,39){\line(1,0){4}}
\put(265,39){\line(1,0){4}}
\put(272,39){\line(1,0){4}}


\put(280,1){\circle*{2}}
\put(280,39){\circle*{2}}

\put(298,22){\vector(-1,1){15}}
\put(298,18){\vector(-1,-1){15}}
\put(300,20){\circle*{2}}

\end{picture}
\caption{Embedding of Vermamodules for $\widehat{su}(2)$ II}
\end{figure}

This in not the whole story. If we begin at a weight which is an integral power
of $k+c_g$ we get the following embeddings depicted in fig.(5.4). The
top one refers to $k>0$ while the bottom one to $k<0$.

\begin{figure}[H]
\begin{picture}(400,40)(0,0)

\put(110,5){....}
\put(130,5){\circle*{2}}
\put(162,5){\vector(-1,0){30}}
\put(164,5){\circle*{2}}
\put(196,5){\vector(-1,0){30}}
\put(198,5){\circle*{2}}
\put(230,5){\vector(-1,0){30}}
\put(232,5){\circle*{2}}

\put(100,30){\circle*{2}}
\put(132,30){\vector(-1,0){30}}
\put(134,30){\circle*{2}}
\put(166,30){\vector(-1,0){30}}
\put(168,30){\circle*{2}}
\put(200,30){\vector(-1,0){30}}
\put(208,30){....}

\end{picture}
\caption{Embedding of Vermamodules for $\widehat{su}(2)$ III}
\end{figure}
We hence see that modules over affine algebras with central extensions
$k>0$  will contain infinitely many singular vectors, while in the
opposite case $k<0$ there is always a finite number of singular vectors.

For the case of $k<0$ the module on which the diagram is terminating is
obviously irreducible and the weights are known as anti-dominant weights.
The corresponding representations play an important r$\hat{\rm o}$le in
paper I. The structure of the singular vectors for arbitrary
representations will be one of the fundamental ingrediences in paper IV.

\begin{itemize}

\item V.G. Kac, {\sc Infinite dimensional Lie algebras},
Cambridge University Press 3'd ed. 1990

\item J. Fuchs, {\sc Affine Lie algebras and quantum groups},
Cambridge University Press 1992

\item P. Goddard and D. Olive, {\it Kac-Moody and
Virasoro algebras in relation to quantum physics}
Int. J. Mod. Phys. 1 (1986) 303-414

\end{itemize}

\chapter{Becchi-Rouet-Stora-Tyutin Quantization}

Many, if not all, of the most important physical theories
are known as gauge theories. The best known example is ordinary
electrodynamics governed by Maxwells equations. A gauge theory is a
theory which is described by more variables than there are physically
independent degrees of freedom. The physical degrees of freedom are
invariant under gauge transformations.
One gets constraint equations in addition to the ordinary
equations of motion which reduces the number of degrees
of freedom to the appropriate physical ones. \\
\ind History provides us with several ways of quantizing constrained
systems. In fig.(6.1) \cite{Hwang-Marnelius90-91} we have tried to indicate
the main steps and
differences of three different ways towards a quantum theory.
The one passing the lower left corner, BRST quantization, will be the
subject of this chapter. The diagonal way is Dirac and Gupta-Bleuler
kind of quantization, which will be briefly mentioned. Considering route
three, canonical quantization, we will here only comment that the
"Standard theory" is often a non-linear theory.

We will make explicit and extensive use of the BRST quantization
technique in papers I, IV and V. It proves to be a most convenient tool
for analyzing the issues of the papers contained in this thesis. Other
quantization techniques, such as canonical quantization, prove to be much
less powerful at least for the known example of the gauged WZNW model. This
model has been analyzed using canonical quantization in \cite{Bowcock89}, and
using BRST in a number of publications and pionered in
\cite{Karabali-Park-Schnitzer-Yang89,Karabali-Schnitzer90}.

\begin{figure}[H]
\begin{picture}(390,350)(5,90)

\put(20,350){\framebox(50,40){\shortstack{Gauge\\theory}}}
\put(20,210){\framebox(50,40){\shortstack{Enlarged\\theory}}}
\put(20,70){\framebox(50,40){\shortstack{Quantized\\unreduced\\theory}}}
\put(320,70){\framebox(50,40){\shortstack{Physical\\model}}}
\put(320,350){\framebox(50,40){\shortstack{"Standard\\theory"}}}
\put(170,210){\framebox(50,40){\shortstack{Quantized\\unreduced\\theory}}}

\put(45,210){\vector(0,-1){100}}
\put(45,350){\vector(0,-1){100}}
\put(345,350){\vector(0,-1){240}}

\put(70,370){\vector(1,0){250}}
\put(70,90){\vector(1,0){250}}

\put(70,350){\vector(1,-1){100}}
\put(220,210){\vector(1,-1){100}}

\put(130,360){\shortstack{Reduction of phase space\\by imposing
constraints}}
\put(-15,300){\shortstack{Enlarging\\of\\phase space}}
\put(-18,160){Quantization}
\put(120,300){\shortstack{Quantization\\without\\constraints}}
\put(350,230){\shortstack{"Ordinary"\\quantization}}
\put(130,80){\shortstack{Reduction of phase space\\by BRST condition}}
\put(210,150){\shortstack{Imposing\\constraints}}

\end{picture} \\
\caption{Quantization procedures.}
\end{figure}
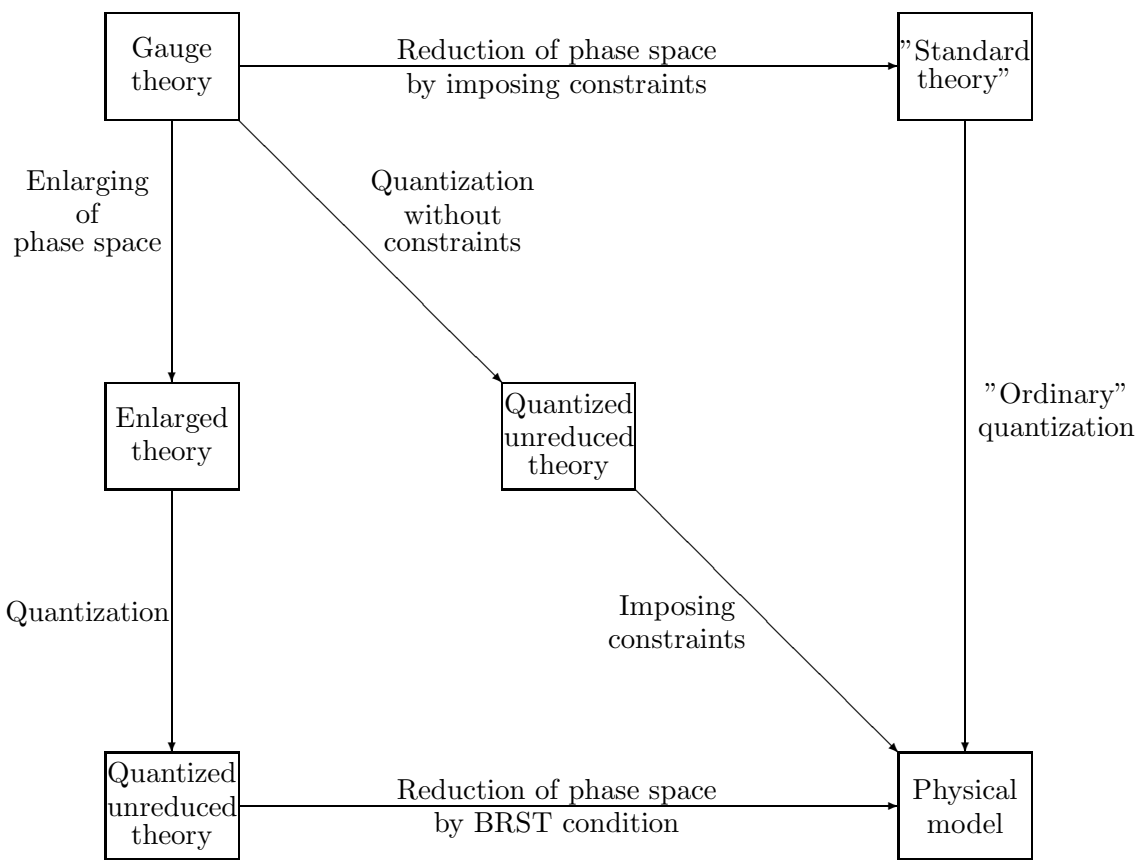

\section{Constrained Hamiltonian Systems}

We will start by reviewing the basic ingrediences of the global
formulation of dynamics and the action principle. The starting point is the
action functional
\be
S_L=\int_{t_1}^{t_2}L(q,\qd)dt,
\ee
where $L$ is the Lagrangian of the system under consideration and $q(t),
\qd(t)$ are the coordinates and velocities of the space considered,
(dot denotes as usual derivative with respect to the parameter $t$). The
classical
equations of motion are derived by requiring that the variation of
$S_L$ with respect to $\del q^i(t)$ vanishes. Performing this calculation
renders
us the Euler-Lagrange equations of motion
\be
\frac{d}{dt}\lef(\frac{\pr L}{\pr\qd^i}\rig)-\frac{\pr L}{\pr q^i}=0
\hs{5mm} \IFF \hs{5mm} \ddot{q}^j\frac{\pr^2L}{\pr\qd^i\pr\qd^j}=
\frac{\pr L}{\pr q^i}-\qd^j\frac{\pr^2L}{\pr\qd^i\pr q^j}
\hs{5mm} i=1,...,n.
\ee
We see that the accelerations $\ddot{q}^i$ are uniquely determined as
functions of $q$ and $\qd$ if and only if we can invert the matrix
\be
M_{ij}\equiv\frac{\pr^2L}{\pr\qd^i\pr\qd^j} \label{matrix}
\ee
i.e. if
its determinant does not vanish.

In order to make the transition to the
Hamiltonian formulation we substitute each velocity $\qd$ by its
conjugate momenta
\be
 p_i=\frac{\pr L}{\pr\qd^i},
\ee
and define the Hamiltonian of the system as the Legendre transform
\be
H=L-\qd^ip_i. \label{ham}
\ee
If the matrix in eq.(\ref{matrix}) is not invertible, i.e. we have a
singular Lagrangian, it is not possible to invert all velocities as
functions of coordinates and momenta. This means that we have some
constraints
\be
\Phi_{\al}(q,p)=0 \hs{20mm} \al=1,...,k. \label{constr}
\ee
that will relate the momenta that are not independent.
These equations define a submanifold of phase space which is assumed to
be smoothly embedded.

The Hamiltonian is a function of only $p$ and $q$ as can be seen from the
variation of (\ref{ham}) which yields
\be
\del H=\dot{q}^i\del p_i-\del q^i\frac{\pr L}{\pr q^i} \label{hamvary}.
\ee
$\del p$ is not an independent variation but (\ref{hamvary}) tells us that
the dependence of $H$ on $\dot{q}$ enters only via the dependence of $p$.

\ind A simple, yet illuminating, example of a system with singular Lagranian
is depicted in fig.(6.2)
\cite{Henneaux-Teitelboim92}.
We here have the Lagrangian $L=\frac{1}{2}(\qd^1-\qd^2)^2$
with momenta
$p_1=\qd^1-\qd^2\ \ p_2=\qd^2-\qd^1$. The primary constraint is $\Phi
=p_1+p_2=0$. All of $\qd$-space is mapped onto a line in $p$-space
and, furthermore, all lines for which $\qd^2-\qd^1$ are held constant are
mapped onto a single point of the
line $p_1+p_2=0$ in $p$-space. This means that the transformation
$q\rightarrow p$ is neither one-to-one nor onto. \\

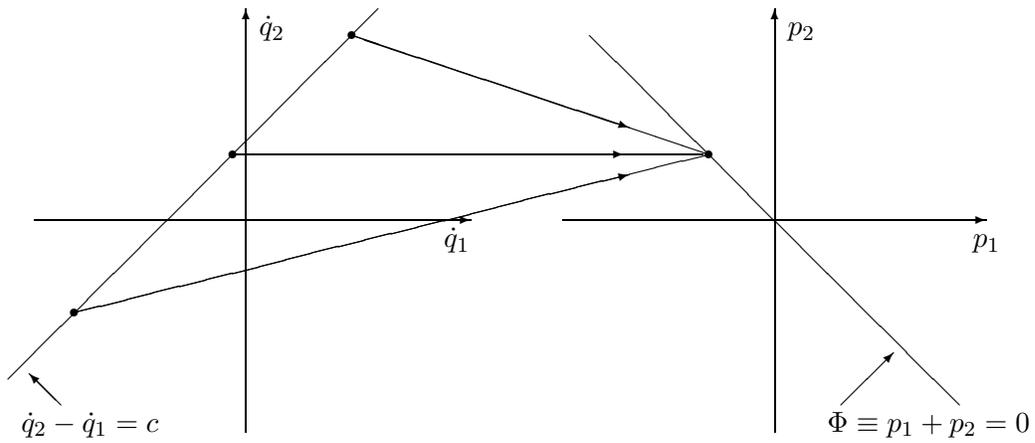
\begin{figure}[H]
\begin{picture}(400,200)(1,1)

\put(20,100){\vector(1,0){165}}
\put(100,20){\vector(0,1){160}}

\put(10,40){\line(1,1){140}}

\put(95,125){\line(1,0){178}}
\put(275,125){\line(-3,1){135}}
\put(275,125){\line(-4,-1){240}}

\put(95,125){\vector(1,0){148}}
\put(140,170){\vector(3,-1){105}}
\put(35,65){\vector(4,1){210}}

\put(105,170){$\dot{q}_2$}
\put(175,90){$\dot{q}_1$}
\put(15,20){$\dot{q}_2-\dot{q}_1=c$}

\put(95,125){\circle*{3}}
\put(140,170){\circle*{3}}
\put(35,65){\circle*{3}}

\put(30,30){\vector(-1,1){12}}

\put(220,100){\vector(1,0){160}}
\put(300,20){\vector(0,1){160}}

\put(300,100){\line(1,-1){70}}
\put(300,100){\line(-1,1){70}}

\put(275,125){\circle*{3}}

\put(305,170){$p_2$}
\put(375,90){$p_1$}
\put(320,20){$\Phi\equiv p_1+p_2=0$}

\put(325,30){\vector(1,1){20}}


\end{picture} \\
\caption{An example where the transformation
$\dot{q}\rightarrow p$ is neither one-to-one nor onto.}
\end{figure}
\ind When we have systems with singular Lagrangians, which will be the ones we
consider
in the following, the Hamiltonian (\ref{ham}) is not uniquely determined off
the submanifold defined by the constraints. We could thus replace the
Hamiltonian with the total Hamiltonian
\be
H^{tot}=H+v^{\al}\Phi_{\al},
\ee
where $v^{\al}$ are arbitrary functions of $q,\ \qd$, referred to as Lagrange
multipliers. \\
\ind  The constraints of eq.(\ref{constr}) must of course be zero for all
times. The time evolution of those primary constraints,
\be
\{\Phi_{\al},H^{tot}\}_{PB}=0,
\ee
may yield other constraints, known as secondary constraints, which may
be
independent of the primary constraints. This process must be repeated
until no more independent constraints are obtained. \\
\ind The distinction between primary-, secondary-, tertiary- etc.
constraints
are of little importance. It is, however, important to classify constraints
according to whether their Poisson bracket algebra closes or not. First
class constraints $\phi$ satisfy
\be
\{\phi,\Phi_{\al}\}_{PB}=f_{\al}^{\ \bet}\Phi_{\bet}
\ee
with all constraints, and second class constraints (in the following denoted
$\psi$) do not. We thus have two sets of constraints. The first class
constraints that satisfy a closed Poisson bracket algebra
\be
\{\phi_a,\phi_b\}_{PB}=f_{ab}^{\ \ c}\phi_c
\ee
among themselves, and the second class constraints whose most important
property, proven by Dirac \cite{Dirac50,Dirac64}, is
\be
det\{\psi_i,\psi_j\}_{PB}\neq0.
\ee
The presence of all constraints means that there are some degrees of
freedom which are physically redundant. For the case of second class
constraints we have one redundant degree of freedom for each constraint,
while for first class constraints there are two redundant degree of
freedom for each constraint. (In field theory those statements
are taken to be valid at each point in the space.) \\
\ind In order to quantize a system one may first try to get rid of the
second class constraints. One method was invented by the ubiquitous Dirac.
One introduces a new bracket, the Dirac bracket, to replace the Poisson
bracket, defined as
\be
\{A,B\}^\ast\equiv\{A,B\}_{PB}-\{A,\psi_i\}_{PB}(C^{-1})^{ij}\{\psi_j,B\}_{PB}
\ee
where $C^{ij}=\{\psi_i,\psi_j\}_{PB}$. It can be shown that the Dirac bracket
has the essential features of the Poisson bracket. Furthermore, the Dirac
bracket fulfills the important identity $\{A,\psi_i\}^{\ast}=0$.
Using the Dirac bracket, and
imposing $\psi_i=0$, we now only have first class constraints that satisfy
a closed Dirac bracket algebra. \\
\ind The quantization now proceeds following one route or the other indicated
in fig.(6.1). The canonical way usually means that one introduces new
constraints,
known as "gauge choices", one for each first class constraint. These pairs
now constitute second class constraints and using these a Dirac bracket is
constructed in order to eliminate them. We thus end up with a true physical
system, which is quantized by letting dynamical variables turn into operators
and letting Dirac brackets become commutators. This is, however, not a
straightforward task, for a general system. Furthermore, quantization
is usually not performed in a covariant way.
\\
\ind The other two ways of fig.(6.2) are both covariant in the sense that
one quantizes before imposing the constraint. One difference is if
Grassmann-odd variables, ghosts, are introduced or not.
In the diagonal way,
the one without ghosts, one quantizes the original Hamiltonian system. Then one
projects out the physical quantum theory, both operators and states. There
are two different prescriptions how to project out states, one by Dirac
\be
\phi_a|phys\ran=0,
\ee
and one by Gupta-Bleuler concerning quantum electrodynamics,
\be
\lan phys|\phi_a|phys\ran=0
\ee
which is somewhat weaker than Dirac's. One may note that the consistency
condition of the Dirac projection allows no quantum anomalies to appear in the
quantum constraint algebra, since this would ruin the projection
condition.

In favour of BRST quantization one must mention that it has so far have
always coincided with other quantization techniques except for the cases
where other techniques have proven to be insufficient. For
those cases the BRST quantization have turned out to be consistent.

With these sketchy remarks we now turn to the main point of interest,
the BRST quantization.

\section{The BRST Quantization Method}

As mentioned above, one essential feature of BRST
quantization is the introduction of ghosts. Ghosts are scalar objects of
opposite Grassmann parity to the constraints, where Grassmann even means
bosonic and Grassmann odd fermionic. The basic difference between odd and
even Grassmann parity is the utilization of anti-commutators between two
odd objects and commutators otherwise. \\
\ind Ghosts appears in for example the concept of
path-integral quantization of non-Abelian gauge theories. The originators
were Fadeev and Popov \cite{Fadeev-Popov67}. Heuristic derivations of
the Fadeev-Popov trick applied to for example electrodynamics can be
found in most books on quantum field theory e.g. \cite{Ryder85},
which also include a more rigorous
treatment of the non-Abelian case. \\
\ind We here state the result for the electromagnetic case. We have the
partition
function (or more correctly the generating functional of Greens functions),
given by
\be
Z=\int {\cal D}A_{\mu}e^{iS}
\ee
where $S$ is the action of the theory, and $A_{\mu}$ is the ordinary
vector potential in electromagnetism. The gauge invariance of $S$ causes
a divergence in $Z$. In order to remove this divergence we
gauge-fix, and use the well-known formula
\be
detM=\int{\cal D}\eta{\cal D}\bar{\eta}e^{\int\bar{\eta}M\eta}
\ee
to take care of the determinant that will appear in this procedure.
Here $\eta$ and $\bar{\eta}$ are objects of odd Grassmann parity.
We then find
\be
Z&\sim&\int{\cal D}A_{\mu}{\cal D}\eta{\cal
D}\bar{\eta}e^{i\int({\cal L}-\frac{1}{2\al}\chi^2-\bar{\eta}M\eta)} \nn \\
&=&\int{\cal D}A_{\mu}{\cal D}\eta{\cal D}\bar{\eta}e^{i\int{\cal
L}_{eff}}
\ee
${\cal L}_{eff}={\cal L}+{\cal L}_{GF}+{\cal L}_{FP}$ where
${\cal L}_{GF}=
-\frac{1}{2\al}\chi^2$ is the gauge fixing term, a convenient
choice is $\chi=\pr_{\mu}A^{\mu}$, and ${\cal L}_{FP}=-\bar{\eta}M\eta$, the
Fadeev-Popov term. \\
\ind Turning to the case of pure Yang-Mills theory, we instead get
\be
{\cal L}_{eff}=-\frac{1}{4}F_a^{\mu\nu}F^a_{\mu\nu}-\frac{1}{2\al}
\pr_{\mu}A^{\mu}_a\pr_{\mu}A^{\mu a}+\pr^{\mu}\bar{\eta}_a(
\pr_{\mu}\eta^a-gf_{\ b}^{a\ c}A_{\mu c}\eta^b). \label{leff}
\ee
As noted independently by Becchi, Rouet and Stora \cite{Becchi-Rouet-Stora76}
and
Tyutin
\cite{Tyutin75} the related action is invariant under the global
BRST transformation
\be
\del A_{\mu a}(x)&=&\la(\pr_{\mu}\eta_a+gf_a^{\ bc}A_{\mu b}\eta_c) \nn \\
\del\eta^a(x)&=&\la \frac{g}{2}f^a_{\ bc}\eta^b\eta^c \nn \\
\del\bar{\eta}_a(x)&=&-\la\frac{1}{\al}\pr_{\mu}A^{\mu}_a . \label{trans}
\ee
$\la$ is a constant Grassmann-odd parameter. The BRST transformation
(\ref{trans}) has the important property that it is nilpotent i.e. two
subsequent transformations yield zero. In the above form of
eq.(\ref{trans}) this is only true on-shell, but one may restate off-shell
nilpotency be reformulating eq.(\ref{leff}) in a more general form. \\
\ind From these important discoveries, a rapid development of BRST
quantization techniques has emerged, which are valid on a much more
general basis. \\
\ind Consider a gauge theory, which on the classical level, has $m$
independent first class constraints $\phi_a\ a=1,...,m$. We assume that
$\phi_a$ is hermitean and bosonic. More general situations may also be
considered in an analogous manner. In the corresponding quantum theory
$\phi_a$ could obey an algebra
\be
[\phi_a,\phi_b]=iU_{ab}^{\ \ c}\phi_c. \label{alg}
\ee
In general $U_{ab}^{\ \ c}$ may be complicated functions and one
distinguishes
between different cases by introducing the concept of rank of a theory.
We will soon illuminate this point further. \\
\ind In order to BRST quantize the system we introduce ghosts $\eta^a$,
one for each independent constraint, and its conjugate momenta
${\cal P}_a$, such that
\be
\{\eta^a,\cP_b\}=\del^a_{\ b}.
\ee
As was proven in reference \cite{Fradkin-Fradkina78}, a nilpotent BRST charge
$Q$ can always be found\footnote{The proof is for the case of finite number
of degrees of freedom}, and the
explicit form is
\be
Q=\phi_a\eta^a+\sum_{i=1}^NC^{a_1...a_i}\cP_{a_1}...\cP_{a_i}
\label{fvbvop}
\ee
Here the factors $C^{a_1...a_i}$ include ghosts, and $N$ is the rank of
the theory. Rank 0 is an Abelian theory. Considering the case of rank 1
we may for example have constant functions
$U_{ab}^c=f_{ab}^c$ i.e. eq.(\ref{alg}) is a Lie algebra, and we find
\be
Q=\phi_a\eta^a-\frac{i}{2}f_{ab}^{\ \ c}\cP_c\eta^a\eta^b+
\frac{i}{2}f_{ab}^{\ \ b}\eta^a.
\ee
This BRST operator differs from the general form (\ref{fvbvop}). The
difference is due to a reordering of the ghost term.
One may note, however,
that a given theory may have several different BRST charges of different
ranks related by canonical transformations. G. F\"ul\"op provides
us with an example of this situation in \cite{Fulop92}. \\
\ind $Q$ is nilpotent by construction, self-adjoint and assumed to be
conserved. Since we have assumed hermitean constraints we have both
hermitean ghosts and ghost-momenta. We now require that physical states
are annihilated by the BRST charge \cite{Curci-Ferrari76,Kugo-Ojima78}
\be
Q\phr=0. \label{proj}
\ee
The nilpotency of the BRST charge tells us that any state of the form
$Q|...\ran$ obeys (\ref{proj}), but those states also decouple from all
other physical states, i.e. they are null states. The true physical states
are thus given by the cohomology of $Q$. Define Ker$Q=\{|A\ran\ ;\
Q|A\ran=0\}$ and Im$Q=\{|A\ran\ ;\ |A\ran=Q|...\ran\}$. The cohomology is
given by Ker$Q$ modulo Im$Q$. This means that we have equivalence classes
of solutions such that $\phr$ and $\phr'=\phr+Q|...\ran$ are identified. \\
\ind It usually convenient to introduce the ghost number operator $N_g$
given by
\be
N_g=\frac{1}{2}(\eta^a\cP_a-\cP_a\eta^a)\ee
This operator counts the difference between the number of ghosts and
the number of ghost-momenta, as can be seen from the relations
\be
[N_g,\eta^a]=\eta^a \hs{10mm} [N_g,\cP_a]=-\cP_a .
\ee
Also
\be
[N_g,Q]=Q.
\ee
It is therefore possible, as first noticed in \cite{Kugo-Ojima79}, to classify
states
according to their ghost number
\be
N_g|A,n\ran=n|A,n\ran .
\ee

If we assume that the state space is irreducible we may classify states
following Kugo and Ojima \cite{Kugo-Ojima79}. The irreducibility means that
we may find a basis for the state space, consisting of states $|S_q\ran$
such that $\lan S_{-q}|S_q\ran \neq 0$.
The index $q$ is the ghost number of the state.

Take now an arbitrary state $|A_q\ran$. Obviously $|A_q\ran$ is either BRST
invariant or not and if BRST invariant it is either BRST exact or not. We
start by considering
the physical states, i.e. states $|A_q\ran$ such that $Q|A_q\ran=0$,
$|A_q\ran\neq Q|B_{q-1}\ran$. Take also $\lan A_{-q}|A_q\ran=1$. We first
assume that $\lan A_{-q}|=\lan C_{-q-1}|Q$ but this leads immediately to
a contradiction since then $1=\lan A_{-q}|A_q\ran=\lan C_{-q-1}|Q|A_q\ran=0$.
Thus a physical state cannot couple to a BRST exact state. The other
possibility would be that $\lan A_{-q}|Q=\lan D_{-q+1}|\neq 0$ which
we assume to hold. We have that there exists $|D_{q-1}\ran$ such that
$\lan D_{-q+1}|D_{q-1}\ran=1$. This gives
$1=\lan D_{-q+1}|D_{q-1}\ran=\lan A_{-q}|Q|D_{q-1}\ran=\lan A_{-q}|E_q\ran$.
If we chose our basis such that $|A_q\ran$ is in the basis, we must have
$|E_q\ran=|A_q\ran+|F_q\ran$ where $|F_q\ran$ decouples from $\lan A_{-q}|$.
Thus $|A_q\ran+|F_q\ran=Q|D_{q-1}\ran$. Since $|A_q\ran$ is BRST invariant
by assumption, $|F_q\ran$ must also be BRST invariant. Also $|F_q\ran$ can
obviously not be physical and hence $|F_q\ran=Q|G_{q-1}\ran$ and
$|A_q\ran=Q(|D_{q-1}\ran-|G_{q-1}\ran)$ in contradiction to the starting
point. Thus physical states couples to physical states.

Consider now
the case when $|A_q\ran$ is not BRST invariant i.e. $Q|A_q\ran = |B_{q+1}\ran$.
Consider then $1=\lan B_{-q-1}|B_{q+1}\ran=\lan B_{-q-1}|Q|A_q\ran$ thus
$|B_{-q-1}\ran$ cannot be BRST invariant rather $Q|B_{-q-1}\ran=|A_{-q}\ran$
where $\lan A_{-q}|A_q\ran=1$. Thus states that are not BRST invariant couples
to state which are BRST exact.

If we had started with an state $|A_q\ran$ which is BRST exact we would in
an analogous manner have found the reverse statement as we must. It is clear
from this argument that all unphysical states fall into quartets related as
the states $|A_q\ran,\ |A_{-q}\ran,\ |B_{q+1}\ran,\ |B_{-q-1}\ran$  above.

We have made use of this classification of states in paper I and especially
paper IV.

\ind The essential steps to solve the cohomology are to introduce a
gradation, a degree, such that we may isolate a small part $d_0$ of the
full BRST
charge $Q$ and then solve the cohomology for $d_0$. Finally, one relates the
cohomology of $d_0$ to the full cohomology. This can be done order by order
in the degree, when commutation relations only preserve the degree to
highest order, or to all orders, when commutation relations preserve the
degree exactly. The order by order procedure is developed by Hwang and
Marnelius in \cite{Hwang-Marnelius89} and used to prove a ghost decoupling
theorem
in a general setting. It is also used in paper I of this thesis and
we refer to this for more detailed accounts. An example of the exact
procedure may be found in \cite{Bouwknegt-McCarthy-Pilch92}, where it is
applied to two-dimensional
gravity coupled to $c\leq1$ matter. This is actually built on known
mathematical techniques of spectral sequences of filtered complexes,
see for example \cite{Bott-Tu82}. \\

\begin{itemize}

\item P. Dirac, {\sc Lectures on quantum mechanics}, Belfer Graduate
School of Science, Yeshiva University, New York 1964

\item M. Henneaux and C. Teitelboim, {\sc Quantization of gauge systems},
Princeton University Press 1992

\item S. Hwang and R. Marnelius, {\it Unpublished lecture notes on
BRST quantization}, Graduate course, Chalmers University of Technology and
G\"oteborg University, 1990-91

\item R. Marnelius, {\it Unpublished lecture notes on Classical field theory},
Graduate course, Chalmers University of Technology and G$\ddot{\rm o}$teborg
University

\item U. M\aa rtensson, {\sc Derivation and BRST quantization of models for
spinning relativistic particles}, PhD thesis, G$\ddot{\rm o}$teborg 1993

\end{itemize}

\chapter{Wess-Zumino-Novikov-Witten Model}

The WZNW theories are, as mentioned above, a conformal
field theory
which symmetry algebra is a semidirect sum of the Virasoro algebra and
an affine
Lie algebra. The generators of the holomorphic symmetry algebra are
at least the stress tensor $T(z)$ and the affine symmetry generator $J(z)$.
Furthermore, it is an example of a so-called rational conformal
field theory i.e. a conformal field theory which has a finite number of
primary fields with respect to the symmetry algebra. This is actually
only the case for compact groups and finite dimensional representations.

We also require that the stress-energy tensor of a WZNW theory is of
Sugawara type, namely
\be
T(z)=a\ka_{ab}\!:\!J^aJ^b\!:(z). \label{sugawara}
\ee
Here $a$ is some normalization constant $\ka_{ab}$ is the Killing form
of the Lie algebra g that appears in the decomposition $J(z)=t_aJ^a(z)$
fore some Lie algebra generators $t^a$.

The WZNW models have a Lagrangian description in terms of an action
which with properly chosen couplings, is conformally invariant. In addition
to the conformal symmetry there is an affine symmetry of the action.

The WZNW models and various versions of what is known as gauged WZNW models
have attracted a considerable interest from string theorists.
{}From the string point of view the WZNW models is a closed bosonic string
propagating on a group manifold. Exact results for correlation functions
have been presented \cite{Knizhnik-Zamolodchikov84} and critical dimensions
are calculated in references \cite{Jain-Shankar-Wadia85,Redlich-Schnitzer86}.
These models are solvable
in the sense that one may find the spectrum, vertex operators,
and analyze modular invariance. In fact, a closed bosonic string
propagating on a group manifold always give modular invariant theories
\cite{Gepner-Witten86} provided one chooses integrable representations for the
affine Lie algebra. These models are unphysical since they include a
tachyonic ground state.

The gauged WZNW models, which from the conformal point of view correspond
to so-called Goddard-Kent-Olive coset construction \cite{Goddard-Kent-Olive85}
in the algebraic formulation \cite{Karabali-Schnitzer90},
provide with a wider range of possible conformal field theories.
They hence equip the string theory with a plethora of choices for
backgrounds. There are various ways of gaugeing the WZNW model. We will below
concentrate on what is known as vector gaugeing.

A possibility to get rid of the tachyon is to include fermionic degrees of
freedom in the WZNW model yielding a supersymmetric WZNW model
\cite{DiVecchia-Knizhnik-Petersen-Rossi85,Abdalla-Abdalla85}. This model
may serve as backgrounds for the heterotic string if gauged in a
so-called chiral way, see \cite{Giveon-Rabinovici-Tseytlin93} and
references therein. The vector gauged supersymmetric WZNW model is related to
the algebraic superconformal coset construction
\cite{Kazama-Suzuki89,Kazama-Suzuki89:2} in much the
same fashion as for the non-supersymmetric way.

\section{General Properties}

Consider a WZNW model defined on a Riemann surface ${\cal M}$
with Lie group $G$ valued fields g. The action looks like
\cite{Witten84,Gepner-Witten86,Knizhnik-Zamolodchikov84}
\be
S_k=\frac{k}{16\pi}\int_{\cal M}{\rm d}\si{\rm d}\tau Tr(\pr_{\mu}{\rm g}
\pr^{\mu}{\rm g}^{-1})+
\frac{k}{24\pi}\int_{\cal B}{\rm d}^3y\ep^{abc}Tr({\rm g}^{-1}\pr_a
{\rm g}{\rm g}^{-1}\pr_b{\rm g}
{\rm g}^{-1}\pr_c{\rm g}) \label{wznwaction}
\ee
where ${\cal M}$ is the boundary of ${\cal B}$ on which g is assumed to
be well defined. (\ref{wznwaction}) is conformally invariant \cite{Witten84},
and we impose the closed string boundary condition
${\rm g}(0,\tau)={\rm g}(2\pi,\tau)$. A general variation of (\ref{wznwaction})
gives the result
\be
\del S=\frac{k}{8\pi}\int_{\cal M}{\rm d}\si{\rm d}\tau Tr({\rm g}^{-1}
\del{\rm g}\eta^{\mu\nu}\pr_{\mu}(
{\rm g}^{-1}\pr_{\nu}{\rm g})-
{\rm g}^{-1}\del{\rm g}\ep^{\mu\nu}\pr_{\mu}({\rm g}^{-1}\pr_{\nu}{\rm g})).
\label{varywzw}
\ee
{}From this we can obtain the equations of motion which in light-cone
coordinates $x_{\pm}=(\tau\pm\si)/\sqrt{2}$ become
\be
\pr_-({\rm g}^{-1}\pr_+{\rm g})=0 \hs{10mm} \pr_+(\pr_-{\rm g}{\rm g}^{-1})=0.
\ee
Note that the first of these equations follow directly from (\ref{varywzw})
and the second by noting $0=\pr_-({\rm g}^{-1}\pr_+{\rm g})=
-{\rm g}^{-1}\pr_-{\rm g}{\rm g}^{-1}\pr_+{\rm g}+{\rm g}^{-1}\pr_-\pr_+{\rm
g}=
{\rm g}^{-1}\pr_+(\pr_-{\rm g}{\rm g}^{-1}){\rm g}$.
\\
\ind Actually, the action (\ref{wznwaction}) possesses a much larger
symmetry than the conformal one known as the affine symmetry for reasons
that will become clear below. It is invariant under
\be
{\rm g}(x_+,x_-)\lra\bar{\Om}^{-1}(x_-){\rm g}(x_+,x_-)\Om(x_+),
\label{wznwinv}
\ee
$\Om$ and $\bar{\Om}$ are arbitrary group-valued
functions. A convenient tool for verifying the affine symmetry
is to use an identity known as the
Polyakov-Wiegmann identity \cite{Polyakov-Wiegmann83,Polyakov-Wiegmann84},
\be
S_k({\rm g}{\rm h})=S_k({\rm g})+\!S_k({\rm h})-
\frac{k}{8\pi}\!\int {\rm d}\si{\rm d}\tau Tr\lef(\eta^{\mu\nu}{\rm g}^{-1}
\pr_{\mu}{\rm g}\pr_{\nu}
{\rm h}{\rm h}^{-1}\!+\ep^{\mu\nu}{\rm g}^{-1}\pr_{\mu}{\rm g}\pr_{\nu}
{\rm h}{\rm h}^{-1}\rig)
\label{pwid}
\ee
This is found by a straightforward calculation starting from the action
(\ref{wznwaction}). In light-cone coordinates the last piece becomes
${\rm g}^{-1}\pr_+{\rm g}\pr_-{\rm h}{\rm h}^{-1}$.
Using the Polyakov-Wiegmann identity (\ref{pwid}) it is know
simple to verify the affine invariance (\ref{wznwinv}).

The symmetry (\ref{wznwinv}) gives rise to an infinite
number of conserved currents obeying the relations
\be
\pr_-J=0 \hs{10mm} \pr_+\bar{J}=0.
\ee
Here $J(x_+)=J^a(x_+)t_a=(k/2){\rm g}^{-1}\pr_+{\rm g},\
\bar{J}(x_-)=\bar{J}^a(x_-)t_a
=-(k/2)\pr_-{\rm g}{\rm g}^{-1}$. $t_a$ are antihermitean matrices representing
the Lie
algebra $g$ of $G$, i.e. $[t_a,t_b]=f_{ab}^{\ \ c}t_c$. \\
\ind The variation of the currents $J$ and $\bJ$  under the infinitesimal
version of (\ref{wznwinv}), $\Om(x_+)=I+\om(x_+),\
\bar{\Om}(x_-)=I+\bar{\om}(x_-)$,
yields the result
\be
\del_{\om}J(x_+)=[J(x_+),\om(x_+)]+\frac{k}{2}\pr_+\om(x_+) \nn \\
\del_{\bar{\om}}\bJ(x_-)=[\bJ(x_-),\bar{\om}(x_-)]-
\frac{k}{2}\pr_-\bar{\om}(x_-) \label{wznwtrans}
\ee
This shows, introducing the holomorphic and anti-holomorphic coordinates
$z=e^{i(\tau+\si)}$, $\bz=e^{-i(\tau+\si)}$, and using
\be
\del_{\om}A(z,\bz)=\oint_z\frac{d\xi}{2\pi i}J^a(\xi)\om_a(\xi)A(z,\bz),
\ee
that $J$ and $\bJ$ obey affine Lie algebras of level $k$,
i.e. we have, using Laurent expansions of $J^a(z)=\sum_{m\in{\Bbb Z}}
J^a_mz^{-m-1}$,
\be [J_m^a,J_n^b]=if^{ab}_{\ \ c}J^c_{m+n}+
\frac{k}{2}m\del^{ab}\del_{m+n,0} \label{kacmoody}
\ee
as well as for $\bJ$. Also $\del_{\om}\bJ(\bz)=0\
\del_{\bar{\om}}J(z)=0$, which means that $J$ and $\bJ$ commutes. In what
follows we will only bother to display the holomorphic sector. \\
\ind As shown for conformal field theories in two dimensions the
energy-momentum tensor splits into a holomorphic part, and an
anti-holomorphic one. The holomorphic part of the
energy-momentum tensor, for the WZNW theory, is given by the
Sugawara construction as
\be
T(z)&=&\frac{1}{k+c_g}:J^a(z)J_a(z): \nn \\
&=&\frac{1}{k+c_g}\sum_{\stackrel{\scriptstyle{m\in{\Bbb Z}}}{n\in{\Bbb Z}}
}:J^a_{n-m}J_{a,m}:z^{-n-2} .
\ee
$c_g$ is the second Casimir of the adjoint representation of $g$
defined as $f_{ab}^{\ \ c}f_{dc}^{\ \ a}=\del_{bd}c_g$. From this we may
identify the Laurent coefficients of the energy-momentum tensor to be
\be
L_n=\frac{1}{k+c_g}\sum_{m\in{\Bbb Z}}:J^a_mJ_{a,n-m}: \label{viraskm}
\ee
{}From eq.(\ref{kacmoody}) we may verify that $L_n$ indeed satisfies a
Virasoro algebra, c.f. (\ref{virasoro}), with $D$ replaced by $c$, and
furthermore we have the important relation
\be
[L_n,J_m^a]=-mJ^a_{m+n} \label{ljcom}.
\ee
The conformal charge of the Virasoro algebra is found to be
\be
c=\frac{kd_g}{k+c_g} \label{confcha}
\ee
where $d_g$ is the dimension of the group $G$. Eq.(\ref{ljcom}) means that
we have the operator product expansion
\be
T(z)J^a(w)=\frac{1}{(z-w)^2}J^a(w)+\frac{1}{z-w}\pr_zJ^a(w)+{\rm r.t.}.
\ee
$J^a(z)$ is thus a weight (1,0) Virasoro-primary field, i.e. primary
with respect to the Virasoro algebra. The current is not, however,
primary with respect to
affine Lie algebra  as can be seen from
eq.(\ref{wznwtrans}), or alternatively from (\ref{kacmoody}). This fact
is rather obvious since the currents are the generators of the symmetry. \\
\ind A primary field with respect to the current algebra transforms as
\be
\phi(z,\bz)\lra\lef(\bar{\Om}^{(\bar R)}(\bz)\rig)^{-1}\phi(z,\bz)\Om^{(R)}(z),
\ee
$\Om^{(R)}(z), \ \bar{\Om}^{(\bar R)}(\bz)$ belongs to some
representation $R$ and $\bar{R}$ of the Lie algebra $g$. Infinitesimally
we may write them as
\be
\Om^{(R)}(z)=I+\om^a(z)t_a^{(R)},
\ee
and likewise for $\bar{\Om}^{(\bar R)}(\bz)$. $t^{(R)}_a$ is as usual an
antihermitean matrix in the representation $R$. Since the currents are
the symmetry generators we should have the following operator product
expansion
\be
J^a(z)\phi^{\al}_{\ \bet}(w,\bar{w})=\frac{(t^a)^{\al}_{\ \ga}}{z-w}
\phi^{\ga}_{\ \bet}(w,\bar{w})+{\rm r.t.}, \label{affprimope}
\ee
for primary fields $\phi(w,\bar{w})$ of the affine symmetry.
We will in what follows often ignore the indices $\al,\bet,\ga,...$
of the representation.
Using (\ref{affprimope})
we may write down Ward identities of the affine symmetry, compare to
eq.(\ref{virward}) in the Virasoro case,
\be
\lan J^a(z)\phi_1(w_1,\bar{w}_1)...\phi_N(w_N,\bar{w}_N)\ran=
\sum_{j=1}^N\frac{t^a_j}{z-w_j}\lan\phi_1(w_1,\bar{w}_1)...
\phi_N(w_N,\bar{w}_N)\ran. \label{kmward}
\ee
It is not difficult to verify that primary fields of the affine algebra
are also primary with respect to the Virasoro algebra with conformal weight
\be
h=\frac{c^{(R)}}{k+c_g} \hs{10mm} c^{(R)}I=t^at_a.
\ee
\ind In analogy to Lie algebra analysis we introduce highest
weight representations of the affine Lie algebra see chapter 5.
We build
the state space on ground states $|R,\al,0\ran$ that satisfy
\be
J^a_n|R,\al,0\ran=0 \hs{5mm} n>0 \hs{10mm}
J^a_0|R,\al,0\ran=\sum_{\beta}(t^a)_{\al\beta}|R,\beta,0\ran
\ee
i.e. $|R,\al,0\ran$ is a highest weight state with respect to the currents,
and transforms in the representation $R$ of the Lie algebra $g$.
As for the Virasoro case we
have the correspondence between primary fields and the ground state
\be
|R,0\ran=\phi^{(R)}(0,0)|0\ran .
\ee
General states are constructed by acting on $|R,0\ran$ with currents of
negative mode number yielding some state $|R,N\ran$, where $N$ indicates
the sum of all the different mode numbers, that is the grade. \\
\ind We have the necessary and sufficient condition, for
the representations to be unitary,
\be
\al\cdot\la\geq 0 \label{unitarrep1}
\ee
for all simple roots $\al$, and
\be
\th\cdot\la\leq\frac{k}{2} \label{unitarrep2}
\ee
where $\th$ is the highest root and $\la$ the highest
weight of the vacuum representation.

To see that the condition (\ref{unitarrep1}) is
necessary, we consider the the $su(2)$ subalgebra
generated by $J^{\pm\al}_0$ and $\frac{2}{\al^2}\al_iJ^i_0$.
Take a highest weight state $|\la\ran$. The norm of the descendant
$J_0^{-\al}|\la\ran$ becomes, using the algebra,
\be
\lan\la|J^{\al}_0J_0^{-\al}|\la\ran=\frac{2}{\al^2}\al\cdot\la
\lan\la|\la\ran.
\ee
Hence, we must have $\al\cdot\la\geq0$. It is sufficient to demand this for
the simple roots since it will then follow for all positive roots. The
condition (\ref{unitarrep2}) follows from identical considerations of the
$\widehat{su}(2)$ subalgebra generated by
$J^{\mp\al}_{\pm1}$ and $\frac{2}{\al^2}(-\al_iJ^j_0+K/2)$,
and the descendant
$J_{-1}^{\al}|\la\ran$. We now find
\be
\lan\la|J^{-\al}_1J_{-1}^{\al}|\la\ran=\frac{2}{\al^2}(-\al\cdot\la+k/2)
\lan\la|\la\ran.
\ee
Hence we must have $\al\cdot\la\leq k/2$. The most restrictive condition comes
from the highest root $\th$. This can be seen as follows. Take for simplicity
the case of a simply-laced algebra and the square of the root to be two.
Then we have $\al^{(i)}\cdot\la^{(j)}=
\del^{ij}$, where $\al^{(i)}$ are the simple roots and $\la^{(j)}$ are the
fundamental weights. Any positive root may be expanded as $\al=
\sum m_i\al^{(i)}$ with non-negative integers $m_i$. Furthermore any weight
on the weight lattice is a sum of fundamental weights with integer coefficients
$\la=\sum n_j\la^{(j)}$. From the
the condition $\al\cdot\la\geq0$ we see that for unitary weights all the
coefficients $n_j$ are non-negative. The highest root may be defined as
the root $\th$ for which $\th-\al=\sum l_k\al^{(k)}$ with non-negative
integers $l_k$ valid for any positive root $\al$. It then follows
that $(\th-\al)\cdot\la=\sum l_kn_k\geq0$

It follows from representation theory of $su(2)$ that the
eigenvalues of $\frac{2}{\al^2}(-\al_iJ^j_0+K/2)$ must be integers. Since
$\frac{2}{\al^2}\al\cdot\la$ are integers for any weight $\la$ on the weight
lattice we also find
that $k$ must be an integer.


\section{Supersymmetric WZNW model}

The supersymmetric generalization of the WZNW model
at the conformal point is given by the
action \cite{DiVecchia-Knizhnik-Petersen-Rossi85,Abdalla-Abdalla85}
\be
S_k({\cal G})=\frac{k}{16\pi}\int{\rm d}^2x{\rm d}^2\th\bD{\cal G}^{-1} D{\cal
G}+
\frac{k}{16\pi}\int{\rm d}^2x{\rm d}^2\th{\rm d}t
{\cal G}^{-1}\frac{{\rm d}{\cal G}}{{\rm d}t}\bD{\cal G}^{-1}\ga_5D{\cal G}
\label{swznwaction}
\ee
where ${\cal G}$ is a superfield, $D$ and $\bD$ are superderivatives and $k$
is known as the level of the supersymmetric WZNW model.

We adopt
the conventions in what follows. Gamma matrices
\be
\ga^0=\left(\begin{array}{cc} 0 & 1 \\ -1 & 0 \end{array}\right) \hs{5mm}
\ga^1=\left(\begin{array}{cc} 0 & 1 \\ 1 & 0 \end{array}\right) \hs{5mm}
\ga^5=\left(\begin{array}{cc} 1 & 0 \\ 0 & -1 \end{array}\right)
\ee
Conjugation of spinors
\be
\bar{\th}=(-\ga^0\th)^T=(\ga_0\th)^T
\ee
where we have indicated the metric $\eta^{\al\bet}=(-1,1)$.
Superderivative
\be
D_{\al}=\frac{\pr}{\pr\bth^{\al}}+i(\ga^{\mu}\th)_{\al}\pr_{\mu}
\ee
which, using light-cone coordinates
\be
x^{\pm}=x^0\pm x^1 \hs{5mm} \frac{\pr}{\pr x^{\pm}}=
\frac{\pr}{\pr x^0}\pm\frac{\pr}{\pr x^1} \hs{5mm} \th^{\pm}=
\ga_{\pm}\th \hs{5mm} \ga_{\pm}=\frac{1}{2}(1\pm \ga_5)
\ee
has the components
\be
D_{\al}=\left(\begin{array}{c} -D^+ \\ D^- \end{array}\right)=
\left(\begin{array}{c} -\left(\frac{\pr}{\pr \th^-}-i\th^-
\frac{\pr}{\pr x^{+}}\right) \\ \left(\frac{\pr}{\pr \th^+}-i\th^+
\frac{\pr}{\pr x^{-}}\right) \end{array}\right).
\ee

The equations of motion of (\ref{swznwaction}) take the form
\be
\bD({\cal G}^{-1}\ga_+D{\cal G})=D^-({\cal G}^{-1} D^+
{\cal G})=0 & &  \nn \\
D^+(D^-{\cal G}{\cal G}^{-1})=0. & &
\ee
It is convenient to introduce the component form for the superfield ${\cal G}$
\be
& & {\cal G}={\rm g}+i\bth\psi+\frac{1}{2}i\bth\th F \nn \\
& & {\cal G}^{-1}={\rm g}^{-1}+i\bth\psi^{\dagg}+\frac{1}{2}i\bth\th F^{\dagg}.
\ee
Here g is a group $G$ valued bosonic field, $\psi$ a spinor transforming
in an appropriate representation of $G$ and $F$ is an auxiliary field.
Using the definition $\cG\cGi=1$ we get
\be
& & \psi^{\dagg}=-\rgi\psi\rgi \nn \\
& & F^{\dagg}=-\rgi F\rgi-i\rgi\bar{\psi}\rgi\psi\rgi. \label{inversesusyg}
\ee

We may now rewrite the equations of motion in components to find
\be
F=i\psi_2\rgi\psi_1 \hs{10mm} \pr_-(\rgi\psi_1)=0
\hs{10mm} \pr_-\lef(\rgi\pr_+\rmg+i\rgi\psi_1\rgi\psi_1\rig)=0
\ee
and finally using these equations and (\ref{inversesusyg}) the last one
become
\be
\rgi\pr_+(\psi_2\rgi)\rmg=0
\ee

In components, and using the equations of motion to eliminate the
auxiliary field $F$,
we may find the action (\ref{swznwaction})
\be
S_k(\cG)=S_k(\rmg)+
\frac{ik}{8\pi}\int{\rm d}^2x\left(\rgi\psi_1\pr_-(\rgi\psi_1)+
\psi_2\rgi\pr_+(\psi_2\rgi)\right) \label{swznwcomp}
\ee
where $S(\rmg)$ is the ordinary bosonic WZNW action (\ref{wznwaction}).
This shows
that $\rgi\psi_1$ and $\psi_2\rgi$ are components of a free fermion $\chi$
which may be written as \cite{DiVecchia-Knizhnik-Petersen-Rossi85}
\be
\chi=\rgi\ga_+\psi+\ga_-\psi\rgi. \label{decouple}
\ee
Furthermore, if we change variables to $\chi$, bosons and fermions will
decouple.
In the path-integral approach this will, however, introduce a shift in the
level of the WZNW model $k\lra k-c_g$ \cite{Schnitzer89}, where $c_g$ is the
quadratic Casimir in the adjoint representation of G.

The appearance of the anomaly responsible for the shift of levels
becomes intuitively clear in the algebraic approach. The superaffine Lie
algebra couples fermions and bosons in the way indicated in
(\ref{superaffine}). One may introduce another affine Lie algebra
which decouples from the fermions. As is easily verified,
the level of this decoupled
affine Lie algebra will be shifted by $-c_g$ compared to the affine
Lia algebra one started with. The action of this decoupled system is then a
bosonic WZNW
model with shifted level and free fremions.
In the action, however, the change of variables
(\ref{decouple}) does not change anything. Hence, in the path-integral
approach, the shift must appear as an anomaly, i.e. from the measure.

The action (\ref{swznwaction}) is invariant under superconformal
transformations as well as superaffine transformations. The latter
are given by
\be
\cG\lra\Om_-^{-1}\cG\Om_+ \label{saffinv}
\ee
for superfields $\Om_+$ and $\Om_-$ satisfying
$D^-\Om_+=D^+\Om_-=0$. The verification of the invariance (\ref{saffinv}) is
easiest in components.
Introducing components for the fields
$\Om_{\pm}=\om_{\pm}\mp i\th^{\mp}\mu_{\pm}$ where
$\pr_{\pm}\om_{\mp}=\pr_{\pm}\mu_{\mp}=0$, we may identify the
affine transformation and its superpartner
\be
\rmg\lra\om_-^{-1}\rmg\om_+ \hs{5mm} \psi_1\lra\om_-^{-1}\psi_1\om_++\bet_1
\hs{5mm} \psi_2\lra\om_-^{-1}\psi_2\om_++\bet_2 \label{safftrans}
\ee
where we have $\bet_1=\om_-^{-1}\rmg\mu_+$ and $\bet_2=\mu_-^{\dagg}\rmg\om_+$.
$\mu^{\dagg}$ is found in a similar way as $\psi^{\dagg}$ above. The
verification
of the invariance (\ref{saffinv}) is now a straightforward matter using the
component form of the action (\ref{swznwcomp}), the Polyakov-Wiegmann identity
(\ref{pwid}) and noting that
\be
\rgi\psi_1\lra\om_+^{-1}\rgi\psi_1\om_++\om_+^{-1}\mu_+
\hs{10mm} \psi_2\rgi\lra\om_-^{-1}\psi_2\rgi\om_-+\mu_-^{\dagg}\om_-
\ee
under the affine transformation (\ref{saffinv}).

The
conserved quantities that are related to the invariance (\ref{saffinv}) may be
constructed in a standard fashion, compare with the bosonic case in
for example \cite{Knizhnik-Zamolodchikov84}. They are found to be
\cite{DiVecchia-Knizhnik-Petersen-Rossi85}
\be
\cGi D^+\cG=-i\rgi\psi_1+\th^-(\rgi\psi_1\rgi\psi_1-i\rgi\frac{\pr \rmg}{\pr
x^+})-
\th^+\th^-\rgi\frac{\pr(\psi_2\rgi)}{\pr x^+}\rmg, \label{currents}
\ee
and similarly for $D^-\cG\cGi$. On-shell the $\th^+\th^-$ term vanishes and
we may identify the generators of the superaffine transformation
(\ref{safftrans}). \\

A large class of conformal field theories can be classified
using Goddard, Kent and Olives (GKO) coset construction
\cite{Goddard-Kent-Olive85,Goddard-Kent-Olive86}.
Actually this method first appeared as an example several years earlier
\cite{Halpern71,Halpern75}. The gauged WZNW model provides us with a Lagrangian
formulation of this construction \cite{Karabali-Schnitzer90}.

There also exists a similar construction for the $N=1$ superconformal field
theories known as the Kazama-Suzuki coset construction
\cite{Kazama-Suzuki89,Kazama-Suzuki89:2}. Here the gauged supersymmetric
WZNW model provides with the Lagrangian formulation, see paper V.

\section{Goddard-Kent-Olive and Kazama-Suzuki Coset Constructions}

The GKO and Kazama-Suzuki coset constructions
uses purely algebra methods.
It enables us to increase the set of consistent
models substantially. In fact without coset formulations the Sugawara
stress-energy tensor gives only conformal central charges which are
larger than one.

Consider a Virasoro algebra c.f. (\ref{virasoro}). As usual,
we assume that
we have highest weight states (\ref{hw}) on which we build our
state space by applying products of $L_{-m}, \ m>0$. We represent the
Virasoro algebra in terms of an affine Lie algebra $\hat{ g}$ via the
Sugawara prescription (\ref{viraskm}). It is also assumed that the underlying
Lie algebra $g$ corresponds to a compact group $G$. The central charge of
the Virasoro algebra is given by eq.(\ref{confcha}). \\
\ind Take now $h$ to be a subalgebra of $g$. We define
\be
L^G_m=\frac{1}{k+c_g}\sum^{d_g}_{\stackrel{\scriptstyle{A=1}}{n\in{\Bbb
Z}}}:J^A_{m-n}J^A_n: \hs{5mm} L^H_m=\frac{1}{k+c_h}
\sum^{d_h}_{\stackrel{\scriptstyle{a=1}}{n\in{\Bbb Z}}}:J^a_{m-n}J^a_n:.
\ee
They satisfy Virasoro algebras of central charges $kd_g/(k+c_g)$ and
$kd_h/(k+c_h)$, respectively. (We adopt in this section the convention
that capital group indices take values in $g$ and lower case indices in
$h$.) This corresponds to energy-momentum tensors of the Sugawara form
\be
T^G=\frac{1}{k+c_g}:J^A(z)J_A(z): \hs{5mm}
T^H=\frac{1}{k+c_h}:J^a(z)J_a(z):,
\ee
and we define
\be
T^{GKO}\equiv T^G-T^H \label{gkoenmom}
\ee
\ind Due to the feature (\ref{ljcom}) we find that
\be
[L^G_m-L^H_m,J^a_n]=0
\ee
which has the obvious consequence that
\be
[L^G_m-L^H_m,L^H_n]=0
\ee
and hence
\be
[L^G_m-L^H_m,L^G_n-L^H_n]=[L^G_m,L^G_n]-[L^H_m,L^H_n].
\ee
This shows that not only $L^G_m$ and $L^H_m$ satisfy Virasoro algebras,
but so does also their difference $K_m=L^G_m-L^H_m$. The corresponding
central charge is
\be
c=\frac{kd_g}{k+c_g}-\frac{kd_h}{k+c_h}. \label{confanom}
\ee
If $h$ is not simple but, for example, consists of two simple pieces, we
will subtract one piece for each simple part. \\
\ind If we take the coset $SU(2)_k\times SU(2)_1/SU(2)_{k+1}$ we will
reproduce exactly the discrete series of central charges of
eq.(\ref{discr}) with $m=k+2$,
as was found in reference \cite{Goddard-Kent-Olive85}. \\
\ind Furthermore, we will get as condition on the state space of the coset
that
\be
J^a_n|\phi\ran=0 \hs{5mm} n>0 \hs{15mm}
J^{\al}_0|\phi\ran=0 \hs{5mm} \forall\ \al\in\De^+. \label{convcoset}
\ee
We will refer to this as the GKO coset condition in the following.

The Kazama-Suzuki coset construction is a generalization of the GKO
construction for an N=1 superconformal algebra.
We are here using a different notation compared to the one in paper V.
In order to compare one should exchange $J$ and $\hJ$ as well as
$k$ and $\hk$. We take an N=1 super
conformal algebra and represent its generators in terms of a superaffine
Lie algebra
\be
& & L_m=\frac{1}{k}\ka_{AB}\lef(\sum_l
:\hJ^A_{m-l}\hJ^B_l+\sum_r(r-\frac{1}{2}n)
:\chi^A_{n-r}\chi^B_r:+a\del_{n,0}\rig) \nn \\
& & G_r=\frac{2}{k}\ka_{AB}\sum_s:J^A_{r-s}\chi^B_s:+\frac{4i}{3k^2}f_{ABC}
\sum_{s,t}:\chi^A_{r-s-t}\chi^B_s\chi^C_t: \label{supervir2}
\ee
where $a=3/16$ for the Ramond and $a=0$ for Neveu-Schwartz boundary
conditions. We have here used the affine generator
\be
\hJ_m^A\equiv J^A_m+\frac{i}{k}f^A_{\ \ BC}\sum_r:\chi^B_{n-r}\chi^C_r:
\label{affdecop}
\ee
which obeys an affine Lie algebra of level $k-c_g$ and decouples from the
fermions $\chi^A$. Note that this is then just an affine algebra
plus a free fermion.

We now want to make a decomposition of $L_m$ and $G_r$ in terms of a
subalgebra
$h$ and and a coset $g/h$.
We thus want an orthogonal decomposition of the generators
\be
L^{G}_m=L^{G/H}_m+L^H_m \hs{15mm}
G^{G}_r=G^{G/H}_r+G^H_r \label{decompvir}
\ee
such that
\be
[L^{G/H}_m,L^H_n]=0 \hs{5mm} [G^{G/H}_r,L^H_m]=0  \hs{5mm}
[L^{G/H}_m,G^H_r]=0 \hs{5mm} \{G^{G/H}_r,G^H_s\}=0
\ee
The decomposition is given by using the $N=1$ generators (\ref{supervir2})
for the algebra $g$ as well as the subalgebra $h$ and using
(\ref{decompvir}) we may obtain expressions for $L^{G/H}_m$ and $G^{G/H}_r$.
It is of crucial importance to note that the generator $\hJ^a$ to be used in
(\ref{supervir2}) for $L^H$ and $G^H$ is not the restriction of $\hJ^A$ to
the subalgebra but rather the generator given by substituting uppercase indices
by lowercase indices in (\ref{affdecop}). In this way $\hJ^a$ will have
level $k-c_h$ while the restriction of $\hJ^A$ to the subalgebra would
still have level $k-c_g$.
The central charge of the coset will be
\be
c_{G/H}=c_G-c_H=\frac{1}{2}d_g+\frac{(k-c_g)d_g}{k}-
\frac{1}{2}d_h-\frac{(k-c_h)d_h}{k}
\ee
As usual, $d_g$ and $d_h$ denote the dimensions of the underlying
Lie algebras.

Goddard, Kent and Olive were able to construct a superconformal series
for the special choice of $(g\oplus g)/g$ where the generators of
the affine Lie algebra were all represented by bilinears in fermions.
They then obtained a superconformal algebra of central charge
\be
c=d_g\frac{k(k+3c_g)}{2(k+c_g)(k+2c_g)} \label{specialsuper}
\ee
This central charge is found to be a special case of the Kazama-Suzuki
construction. If we introduce the shifted levels $\hat{k}=k-c_g$, take
the diagonal coset $(g_1\oplus g_2)/g_{1+2}$ and choose $\hk_2=0$ we obtain
(\ref{specialsuper}) but in terms of $\hat{k}_1$ instead of $k$.

\section{The Gauged Wess-Zumino-Novikov-Witten Model}

The WZNW model possesses, as pointed out above,
an affine invariance given
by (\ref{wznwinv}). In addition there is also a global invariance
\be
{\rm g}(x_+,x_-)\lra {\rm h}^{-1}_L{\rm g}(x_+,x_-){\rm h}_R
\ee
where ${\rm h}_L,\ {\rm h}_R$ take values in some sub-group $H$ of $G$.
In order to
promote this global invariance to a local one
\be
{\rm g}(x_+,x_-)\lra {\rm h}^{-1}_L(x_+){\rm g}(x_+,x_-){\rm h}_R(x_-)
\ee
we introduce non-propagating gauge fields $A_+$ and $A_-$. We require
$A_+$ and $A_-$ to transform as
\be
A_+\lra {\rm h}^{-1}_L(x_+)(A_++\pr_+){\rm h}_L(x_+), \hs{10 mm}
A_-\lra {\rm h}^{-1}_R(x_-)(A_-+\pr_-){\rm h}_R(x_-)
\ee
in order to have the covariant derivatives $D_+{\rm g}\equiv
\pr_+ {\rm g}+A_+{\rm g}$ and
$D_-{\rm g}\equiv\pr_-{\rm g}-{\rm g}A_-$ to transform in the desired
fashion $D_+{\rm g}\lra {\rm h}^{-1}_LD_+{\rm g}{\rm h}_R$ and $D_-{\rm g}
\lra {\rm h}^{-1}_LD_-{\rm g}{\rm h}_R$.
We now choose ${\rm h}_L$ and ${\rm h}_R$ to take values in the same
sub-group and
substitute the derivatives for covariant derivatives in the action.
This gives the action for the {\it chirally} gauged WZNW model
\be
S_k({\rm g},A)=S_k({\rm g})-\frac{k}{4\pi}\int{\rm d}^2x Tr\lef(A_+\pr_-
{\rm g}{\rm g}^{-1}-
A_-{\rm g}^{-1}\pr_+{\rm g}-A_+{\rm g}A_-{\rm g}^{-1}\rig). \hspace{3mm}
\ee

For our purposes, we instead of this left right invariance,
desire a gauged WZNW
model with so-called vector invariance i.e. invariance under the vector
transformation
\be
{\rm g}(x_+,x_-)\lra{\rm h}^{-1}(x_+,x_-){\rm g}(x_+,x_-){\rm h}(x_+,x_-).
\label{vectorinv}
\ee
We may now use the Polyakov-Wiegmann identity (\ref{pwid}) to convince
ourselves that it is sufficient to add the term $A_+A_-$ to the
chirally gauged WZNW action to get an invariant action under (\ref{vectorinv}).
The resulting action becomes
\be
S_k({\rm g},A)=S_k-\frac{k}{4\pi}\!\int\! {\rm d}^2x Tr\lef(A_+\pr_-
{\rm g}{\rm g}^{-1}-
A_-{\rm g}^{-1}\pr_+{\rm g}-A_+{\rm g}A_-{\rm g}^{-1}+A_-A_+\rig)\!.
\hspace{3mm}
\ee
$A_{\pm}$ transforms in the adjoint representation of $H$ which is taken
to be an anomaly-free vector subgroup of $G$. Following
\cite{Karabali-Schnitzer90},
we may parametrize the
gauge fields as
\be
A_+=-\pr_+{\rm h}{\rm h}^{-1} \hs{10mm} A_-=-\pr_-\tilde{{\rm h}}
\tilde{{\rm h}}^{-1} \label{gaugeparam}
\ee
where h and $\tilde{{\rm h}}$ are group elements of $H$. \\
\ind The partition function is given as
\be
Z=\int{\cal D}{\rm g}{\cal D}A_+{\cal D}A_-e^{(-S_k({\rm g},A))}.
\ee
Using the parametrization (\ref{gaugeparam}), and the Polyakov-Wiegmann
identity (\ref{pwid})
we can write the action as
\be
S_k({\rm g},A)=S_k\lef({\rm h}^{-1}{\rm g}\tilde{{\rm h}}\rig)
-S_k\lef({\rm h}^{-1}\tilde{{\rm h}}\rig).
\ee
This change of variables will render us Jacobians ${\rm det}D_+{\rm det}D_-$,
where
$D_{\pm}$ are covariant derivatives. Following Polyakov-Wiegmann
\cite{Polyakov-Wiegmann83,Polyakov-Wiegmann84} we rewrite this as
\be
{\rm det}D_+{\rm det}D_-=e^{S_{2c_h}({\rm h}^{-1}\tilde{{\rm h}})}{\rm det}
\pr_+{\rm det}\pr_-.
\ee
$c_h$ denotes, as usual, the quadratic Casimir of the adjoint
representation of $H$. The two remaining determinants are represented by a
$bc-$ghost system. The $b-$ghost has
conformal weight one and the $c-$ghost weight zero. Changing variables
${\rm h}^{-1}{\rm g}\tilde{{\rm h}}\lra {\rm g}$, choosing the gauge
$\tilde{{\rm h}}=1$ i.e. $A_-=0$
and finally changing variables again into ${\rm h}^{-1}\lra {\rm h}$ we may
write
down the final form of the partition function
\be
Z&=&\int{\cal D}{\rm g}{\cal D}{\rm h}{\cal D}b_+{\cal D}b_-
{\cal D}c_+{\cal D}c_-
exp\lef(-S_k({\rm g})\rig)exp\lef(S_{k+2c_h}({\rm h})\rig) \nonumber \\
&\ &exp\lef(-\int {\rm d}^2xTr\lef(b_+\pr_-c_++b_-\pr_+c_-\rig)\rig).
\label{partfunc}
\ee
\ind The total action (\ref{partfunc}) gives us a total conserved
current consisting of three parts, (now in holomorphic coordinates as
above),
\be
J^{a,tot}(z)=J^a(z)+\jt^a(z)+J^a_{gh}(z).
\ee
The three parts of the current each obey affine Lie algebras of levels
$k$, $-k-2c_h$ and $2c_h$ respectively. This means that $J^{a,tot}(z)$
satisfy an affine Lie algebra of level zero. \\
\ind The energy-momentum tensor is found to be
\be
T(z)&=&\frac{1}{k+c_g}:J^A(z)J_A(z):-\frac{1}{k+c_h}:\jt^a(z)\jt_a(z):-
:b^a(z)\pr_zc_a(z): \nonumber \\
&\equiv& T^G+T^H+T^{gh}
\ee
where, as above, capital indices take values in $g$, and lower case indices
in $h$. The total conformal anomaly becomes
\be
c^{tot}=\frac{kd_g}{k+c_g}+\frac{(-k-2c_h)d_h}{(-k-2c_h)+c_h}-2d_h=
\frac{kd_g}{k+c_g}-\frac{kd_h}{k+c_h}
\ee
which, as noted in \cite{Karabali-Park-Schnitzer-Yang89}, coincides
with the conformal anomaly of
the GKO construction (\ref{confanom}).

\ind The total action of eq.(\ref{partfunc}) is invariant under the BRST
transformation \cite{Karabali-Schnitzer90,Bastianelli91}
\be
\delta_B{\rm g}&=&c_-{\rm g}-{\rm g}c_+ \nonumber \\
\delta_B{\rm h}&=&c_-{\rm h}-{\rm h}c_+ \nonumber \\
\delta_Bc_{\pm}&=&\frac{1}{2}\{c_{\pm},c_{\pm}\} \nonumber \\
\delta_Bb_+&=&\frac{k}{4\pi}{\rm g}^{-1}\pr_+{\rm g}
-\frac{k+2c_h}{4\pi}{\rm h}^{-1}\pr_
+{\rm h}+\{b_+,c_+\} \nonumber \\
\delta_Bb_-&=&-\frac{k}{4\pi}\pr_-{\rm g}{\rm g}^{-1}+
\frac{k+2c_h}{4\pi}\pr_-{\rm h}{\rm h}^{-1}
+\{b_-,c_-\}  .   \label{brst}
\ee
The BRST charge is then found to be
\ind
\be
Q=\oint\frac{dz}{2\pi
i}\left[:c_a(z)(J^a(z)+\jt^a(z)):-\frac{i}{2}f^{ad}_{\ \
 e}
:c_a(z)c_d(z)b^e(z):\right]\label{brstq}.
\ee
We note that the BRST transformation (\ref{brst}) is not nilpotent. The
transfomations on the $c$-ghosts and the currents are nilpotent but for
example
\be
\del^2_Bb_+=\frac{k}{4\pi}\rgi\pr_+c_-\rmg-\frac{k+2c_h}{4\pi}{\rm h}^{-1}
\pr_+c_-{\rm h}+\lef(-\frac{k}{4\pi}+\frac{k+2c_h}{4\pi}\rig)\pr_+c_+
\ee
A similar expression holds for $\del^2_Bb_-$. We thus se that nilpotency is
aquired on shell
for $c$-ghosts and {\em if} the level of the auxiliary sector $\tilde{k}$ had
been $\ti{k}=-k$ and not $\ti{k}=-k-2c_h$ as above. This is, however, not a
surprise since the contribution of $-2c_h$ to the level of the auxiliary sector
is a quantum anomaly and
(\ref{brst}) is just a classical invariance. On the quantum level, like in the
BRST charge, the ghosts will contribute by the necessary amount to aquire
nilpotency.

\ind This may also be seen from another point of view. Classically we
have a constraint, $J^a(z)\approx 0$, which is of second class if $k\neq0$. In
order to make it first class we introduce a new set of variables
$\jt^a(z)$ which are affine currents of level $\ti{k}$. The new constraits
will be first class if we take $\ti{k}=-k$. This will give us the BRST
operator (\ref{brstq}). The
requirement of nilpotency of the BRST operator gives that the current
$\jt^a(z)$ must have level $-k-2c_H$, where the additional $-2c_h$ is a
quantum correction originating from the ghosts.

Using the BRST charge we find that \cite{Karabali-Schnitzer90}
\be
T(z)=T^{GKO}+\frac{1}{k+c_h}\lef[Q,:b_a(z)\lef(J^a(z)+\jt^a(z)\rig):\rig]
\ee
where $T^{GKO}$ is given by eq.(\ref{gkoenmom}). This means that the
energy-momentum tensors will coincide when inserted between physical
states since $Q|phys\ran=0$. \\
\ind When we have an abelian $H$ it was shown, \cite{Karabali-Schnitzer90},
that the two
constructions are
identical, i.e. that the GKO coset condition eq.(\ref{convcoset}) and the
BRST projection $Q|phys\ran=0$ yield the same spectrum of physical
states. \\
\ind The generalization to non-abelian groups $H$ is, as mentioned before,
the subject of paper I. We approach the problem using the BRST technique.
The state space will be given by the BRST cohomology, which is
solved by a standard technique known as spectral sequences of filtered
complexes.
The essential step is to introduce a grading of the operators such
that one may start by analyzing a smaller and hopefully easier part of the
BRST operator. In the generic case the degree may either be exact or not.
Exact means that the degree is preserved by the commutation relations. If the
degree
is exact one relates the cohomology of the part of the BRST operator to the
full
BRST operator. In the
case where the degree is not exact, which is the one at hand in paper I,
one must solve the BRST cohomology
order by order in the degree.

It is found that the states in the cohomology will be ghost
free if one restricts the representations to integrable representations of the
G sector and to representations that excludes null-vectors in the H sector.
We will refer in what follows to the latter type of representations as
anti-dominant highest weight
representations. The states in the cohomology will be
restricted to highest weight primary states in both the G and the H sectors,
and to a ghost vacuum for the ghosts. We take the level of the affine Lie
algebra $k$ to be positive for the G sector and, hence, it follows that it
is negative in the H sector. The restrictions of representations thus
confines us to the modules furthest to the left in figure (5.2) for
the G sector and the module furthest to the right in figure (5.3) for
the H sector. This means that there exists infinitely many null-vectors
in the G module. If we, however, restrict ourselves to the
irreducible sub-module in the G sector we have a unique solution. The BRST
condition (\ref{proj}) will then coincide with the conventional coset condition
(\ref{convcoset}).

The cohomology for arbitrary representations will in the general case include
states at ghost numbers different from zero. Those states arise due to the
presence of null-vectors in the Verma modules over affine Lie algebras. In
paper IV
we provide with a technique by which one may construct non-trivial states
in the cohomology for arbitrary representations. The basic idea is to use
the fact that in the irreducible $L$ sub-module the null-vectors are
identified with the zero element. Thus if a state $|S\ran$ in $L$ is BRST
exact with a null-vector
$|N\ran$ i.e. $Q|N\ran=|S\ran$ then $|S\ran$ may be a non-trivial state
in the cohomology over $L$. Conversely, if a state $|S\ran\in L$ obeys
$Q|S\ran=|N\ran$
where $|N\ran$ is null, then in the irreducible sub-module
$|S\ran$ will be BRST invariant. All states
of those two forms are not in the cohomology, we refer, however, to paper
IV for an explicit account for the restrictions for non-triviality of such
states.
The rigorous proof of the cohomology is provided by Stephen Hwang
\cite{Hwang95}. \\

We now wish to gauge the supersymmetric WZNW model in a manifestly
supersymmetric
way. The result becomes \cite{Tseytlin94}, similarly to the non-supersymmetric
WZNW
model, a sum of two supersymmetric WZNW models the original and an auxiliary.
We denote their levels $\hat{k}$ and $\hat{\ti{k}}$, respectively.
The determinant from the gauge fixing may be represented by a superghost system
of conformal weight $(0,\frac{1}{2})$ which in components is equivalent
to a (0,1) fermionic ghost system, and a $(\frac{1}{2},\frac{1}{2})$ bosonic
ghost
system. The entire ghost system contributes to the conformal charge by $-3c_h$.

In paper V we show that the effective action possesses a BRST symmetry which
is nilpotent if and only if $\hat{k}+\hat{\ti{k}}=0$. For this choice of
level for the auxiliary sector the total conformal anomaly becomes
\be
C_{tot}&=&C({\rm G},k-\frac{1}{2}c_g)+C({\rm H},-k-\frac{1}{2}c_h)+
\frac{1}{2}d_g+\frac{1}{2}d_h-3d_h \nn \\
&=&\lef(C({\rm G},k-\frac{1}{2}c_g)+\frac{1}{2}d_g\rig)
-\lef(C({\rm H},-k-\frac{1}{2}c_h)+\frac{1}{2}d_h\rig) \\
& & {\rm with} \hs{3mm} C(G,k)\equiv \frac{kd_g}{k+\frac{1}{2}c_g}. \nn
\ee
This may be recognized as the conformal anomaly of the algebraic N=1 coset
construction of Kazama and Suzuki \cite{Kazama-Suzuki89:2}.

In complete analogy to the bosonic case we were able to show that the
total stress-energy tensor of the gauged supersymmetric WZNW model
coincides with the Kazama-Suzuki coset stress-energy tensor plus a
term which is BRST exact. \\

In paper I the BRST invariant branching function is also defined which leads us
to the last subject of this thesis, characters. This is also the
subject of paper II and III. \\

\begin{itemize}

\item J. Fuchs, {\sc Affine Lie algebras and quantum groups},
Cambridge University Press 1992

\item V.G. Knizhnik and A.B. Zamolodchikov, {\it Current algebra and
Wess-Zumino model in two dimensions}, Nucl. Phys. B247 (1984) 83

\item M. Henningsson, {\sc Out of flatland: Applications of
Wess-Zumino-Witten models to string theory}, PhD thesis,
G$\ddot{\rm o}$teborg 1992

\end{itemize}

\np
\thispagestyle{plain}


\chapter{Characters And Modular Invariance}

Consistent string theories are required to be completely
reparametrization invariant. We have above discussed local reparametrization
invariance which for example manifests itself in the vanishing of the
stress-energy tensor. There may, however, be global reparametrization
invariances left and requiring invariance under these is
equivalent to requiring
modular invariance.

If modular invariance is not met we should be able to find this at
the one loop level in string perturbation theory. We hence consider
this case.

The one loop contribution in the string perturbation series comes
from the path-integral on the torus. We hence require the
contribution to be invariant under global reparametrizations
of the torus. This is also intuitively clear. A conformal field
theory on the torus must be independent of how we choose to parametrize
the torus. Reparametrizations on the torus are generated by the modular group.

To define a torus we use the complex coordinate $w$ of
a conformal field theory on a cylinder, and identify two periods.
We can take one
period to be $w\equiv w+1$, and the other $w\equiv w+\tau$,
where $\tau=\tau_1+i\tau_2$ is the modular parameter. $\tau_1$ and $\tau_2$ are
taken
to be real.
(Usually one takes $w\equiv w+2\pi$ and $w\equiv w+2\pi\tau$ but we have
normalized by $2\pi$.) By this
we of course mean that conformal fields $\phi(w)$ that are defined on the
torus should satisfy $\phi(w+1)=\phi(w)$, and
$\phi(w+\tau)=\phi(w)$. In coordinates this means that we
identify the opposite sides of the parallelogram as depicted in fig.(8.1).
\begin{figure}[H]
\begin{picture}(250,100)(5,10)

\put(160,15){\vector(0,1){90}}
\put(160,15){\vector(1,0){115}}

\put(160,15){\line(1,2){30}}
\put(240,15){\line(1,2){30}}
\put(190,75){\line(1,0){80}}

\put(179,45){\line(1,0){8}}
\put(195,45){\line(1,0){8}}
\put(211,45){\line(1,0){8}}
\put(227,45){\line(1,0){8}}
\put(243,45){\line(1,0){8}}

\put(202,18){\line(1,2){5}}
\put(209,33){\line(1,2){5}}
\put(216,48){\line(1,2){5}}
\put(223,63){\line(1,2){5}}

\put(125,100){$Im(w)$}
\put(280,10){$Re(w)$}
\put(186,78){$\tau$}
\put(260,78){$\tau+1$}
\put(238,5){$1$}

\put(190,75){\circle*{2}}
\put(270,75){\circle*{2}}

\put(240,13){\line(0,1){4}}

\end{picture} \\
\caption{A torus with its two nontrivial cycles (dashed lines).}
\end{figure}
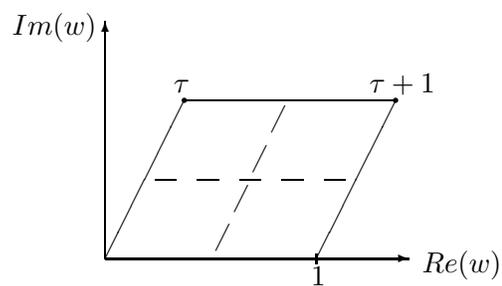

\ind The action of the group of modular transformations may be
visualized by cutting
along one of the non-trivial cycles, and the reglue after a
twist of 1 (read $2\pi$). If we take $w=\si_1+\tau\si_0$ for real $\si_0$
and $\si_1$ we may cut along
the lines of constant $\si_0$ and $\si_1$.
The cycle corresponding to $\si_0$ constant is the one parallel
to the real axis. Cuts
along $\si_0$ and $\si_1$ corresponds to the transformations $T:\tau\lra\tau+1$
and $U:\tau\lra\tau/(\tau+1)$, respectively. We may represent
these transformations in diagrams as in fig.(8.2).
Note that $\tau+1$ and $\tau/(\tau+1)$ are the new ratio of periods.
\begin{figure}[H]
\begin{picture}(450,70)(5,5)



\put(10,10){\line(1,2){15}}
\put(50,10){\line(1,2){15}}
\put(25,40){\line(1,0){40}}
\put(10,10){\line(1,0){40}}

\put(25,40){\circle*{2}}
\put(65,40){\circle*{2}}

\put(21,43){$\tau$}
\put(55,43){$\tau+1$}
\put(80,28){$T$}

\put(75,25){\vector(1,0){20}}


\put(100,10){\line(2,1){60}}
\put(140,10){\line(2,1){60}}
\put(160,40){\line(1,0){40}}
\put(100,10){\line(1,0){40}}

\put(120,40){\circle*{2}}
\put(160,40){\circle*{2}}

\put(116,43){$\tau$}
\put(150,43){$\tau+1$}



\put(240,10){\line(1,2){15}}
\put(280,10){\line(1,2){15}}
\put(255,40){\line(1,0){40}}
\put(240,10){\line(1,0){40}}

\put(255,40){\circle*{2}}
\put(295,40){\circle*{2}}

\put(251,43){$\tau$}
\put(285,43){$\tau+1$}
\put(310,28){$U$}

\put(305,25){\vector(1,0){20}}


\put(345,10){\line(1,2){15}}
\put(345,10){\line(2,1){60}}
\put(405,40){\line(1,2){15}}
\put(360,40){\line(2,1){60}}

\put(360,40){\circle*{2}}
\put(405,40){\circle*{2}}

\put(356,43){$\tau$}
\put(400,28){$\tau+1$}

\end{picture} \\
\caption{Modular transformations on the torus.}
\end{figure}
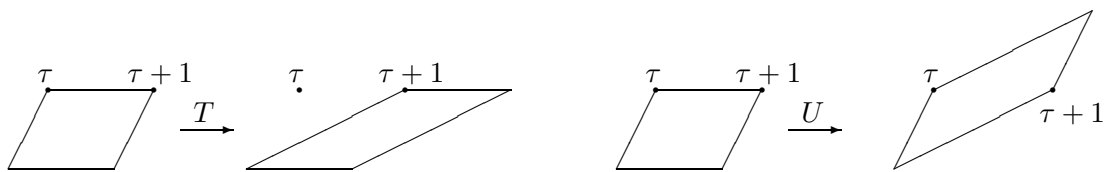

\ind The transformations $T$ and $U$ generate the group of transformations
\be
\tau \lra \frac{a\tau+b}{c\tau+d}, \hs{10mm} \lef(\begin{array}{cc}
a\ \ \ b \\ c\ \ \ d \end{array} \rig) \in Sl(2,{\Bbb Z}).
\ee
The modular group is then $Sl(2,{\Bbb Z})/{\Bbb Z}_2$, since we can
reverse the signs of $a$, $b$, $c$ and $d$ without changing the
the physical content.
One conventionally chooses the transformation $T$ and $S$,
$S\equiv T^{-1}UT^{-1}
:\tau\lra -1/\tau$ instead of $U$, to generate the modular group.

When, in string theory, one wishes to calculate
loop amplitudes $\grave{\rm a}$ la Polyakov \cite{Polyakov81,Polyakov81:2}, one
integrates over all inequivalent metrics of the world sheet. If one considers a
one-loop zero point diagram, one finds the (gauge-fixed) expression
\be
\int\frac{d^2\tau}{(Im\tau)^2}Z(\tau) \label{loopamp}
\ee
where $Z(\tau)$ is the partition function on the torus,
and $d^2\tau/(Im\tau)^2$ is
the modular invariant measure.

Under $T$ and $S$ we have the transformations
\be
T&:&d\tau d\bar{\tau}\lra d(\tau+1)d(\bar{\tau}+1)=d\tau d\bar{\tau} \nn \\
T&:&\tau_2^2\lra\tau_2^2 \nn \\
S&:&d\tau d\bar{\tau}\lra d(-\frac{1}{\tau})d(-\frac{1}{\bar{\tau}})
=|\tau|^{-4}d\tau d\bar{\tau} \nn \\
S&:&\tau_2^2\lra \lef(Im(-\frac{1}{\tau})\rig)^2=(|\tau|^{-2}\tau_2)^2.
\ee
It thus follows that $d^2\tau/\tau_2^2$ is invariant under the
modular group. \\
\ind We must now also demand that the partition function is
modular invariant, i.e.
\be
Z(\tau)=Z(\tau+1)=Z(-\frac{1}{\tau}).
\ee
{}From the requirement of modular invariance it follows that we
must then restrict the region of integration in the parameter space of $\tau$.
We find that the fundamental region ${\cal F}$ is sufficient
to integrate over.
\be
& &{\cal F}=
\{\tau|\tau\in{\cal H}, |{\rm Re}\tau|<\frac{1}{2}, |\tau|>1\}\nn \\
& &\cup
\{i\infty\}\cup\{\tau|{\rm Re}\tau =-\frac{1}{2},|\tau|\geq1\}\cup
\{\tau|\ |\tau|=1,-\frac{1}{2}\leq{\rm Re}\tau\leq0\}
\ee
Here ${\cal H}$ denotes the upper half of the complex plane.
The rest of the upper-half of the complex plane is
covered by transformations of the modular group of points in
the fundamental region. Thus every point
outside this region is in the same equivalence class as some point
inside the strip.

In order to find a consistent string theory we thus have to investigate
the modular properties of the partition function. In the theories of our
interest the partition function factorizes into a holomorphic and an
anti-holomorphic part such that
\be
Z(\tau,\bar\tau)=\sum_{a,b}M_{ab}\bar{\chi}^a(\bar\tau)\chi^b(\tau).
\ee
The functions $\chi(\tau)$ and $\bar{\chi}(\bar{\tau})$ will be denoted
characters. The factorization of the partition function is related to
the fact that the symmetry algebras are direct sums of holomorphic and
anti-holomorphic parts. The modular properties of the partition function
is hence encoded in the modular properties of the characters. \\

We first consider the critical closed bosonic string for the flat case
and in the light-cone gauge. The \pfu\ is given by the trace over the Hilbert
space of the zero mode of the Virasoro algebra
\be
Z(\tau)={\rm Tr}e^{2\pi i\tau(L_0-1)}e^{2\pi i\bar\tau(\bar{L}_0-1)}.
\ee
The Hilbert space
is the Fock space of oscillators $\al_{-n}^{\mu}$ in the 24 transverse
directions acting on a tachyonic vacuum. $L_0$ is given by $k^2/8+N$ where
$k$ is the tachyon momentum and $N$, the total mode number operator,
satisfies $[N,\al_{-n}]=n\al_{-n}$.

For each transverse direction we may consider the infinite tower of states
\be
|1,k\ran &\iff& \al_{-1}|0,k\ran \nn \\
|2,k\ran &\iff& \al_{-2}|0,k\ran,\ \ (\al_{-1})^2|0,k\ran \nn \\
|3,k\ran &\iff& \al_{-3}|0,k\ran,\ \ \al_{-2}\al_{-1}|0,k\ran \ \
(\al_{-1})^3|0,k\ran \nn \\
&...& \nn
\ee
The character will get a contribution of $e^{2\pi i\tau mn}\equiv q^{mn}$
from each $(\al_{-n})^m|k,0\ran$. It is not difficult to convince oneself
that the total mode number operator contributes to the {\pfu} by a factor
\be
\prod_{n=1}^{\infty}(1-q^n)^{-d}
\ee
where $d$ is the number of transverse directions, and we have used the
identity $1+q^n+q^{2n}+...=(1-q^n)^{-1}$. If we introduce the well-known
Dedekind $\eta$-function
\be
\eta(q)=q^{1/24}\prod_{n=1}^{\infty}(1-q^n)
\ee
we find the left character
\be
\chi_L(q)=q^{k^2/8}q^{(d-24)/24}(\eta(q))^{-d}.
\ee
$\eta(q)$ has very nice modular properties, it is a particular example of what
is known as a modular form.

For the closed bosonic string propagating on a group manifold, i.e. a
WZNW model, the analysis is more complicated. It is, however, greatly
simplified by the fact that the complicated part of the {\pfu} can be found
from the affine Lie algebra characters. The relevant characters have nice
modular
properties since we may formulate them in terms of modular forms.

\section{Characters}

We take an irreducible highest weight $\hg$ module $L(\hLa)$
over a dominant
weight $\hLa$ (\ref{dominant}). There exists a weight space decomposition
of $L(\hLa)$ into finite dimensional subspaces $L(\hLa)_{\hla}$. The dimensions
of
those subspaces will be denoted mult$_{\hLa}(\hla)$.

The character of the $\hg$ module $L(\hLa)$ is defined by
\be
{\rm ch}L(\hLa)=\sum_{\hla}{\rm mult}_{\hLa}(\hla)e^{\hla}
\ee
where the formal exponential $e^{\hla}$ operates as $(e^{\hla})h=e^{\hla(h)}$
for $h\in${\sc h} the Cartan sub-algebra.

We may write the character more explicitly if we introduce an arbitrary vector
in the root space. If we chose an orthonormal basis for the roots
of the finite Lie algebra $g$ to be $\nu_i$ $i=1,...,r$ then an arbitrary
vector in the affine root space is
\be
\hat{v}=-2\pi i(\sum_{i=1}^r\th_i\nu_i,\tau,u) \label{pararoot}
\ee
where $\th_i,\tau, u\in{\Bbb C}$. In what follows we will denote
$\sum_{i=1}^r\th_i\nu_i$ by $\th$.
Thus we may write the character
\be
{\rm ch}L(\hLa)=\sum_{\hla}{\rm mult}_{\hLa}(\hla)
e^{-2\pi i(\la\cdot\th+n\tau+uk/2)} \label{basisroot}
\ee
for a weight $\hla=(\la,k/2,n)$. For the special values $u=\th_i=0$ $i=1,...,r$
we have the contribution we desire for the string partition function.
Note that the states in the Verma modules all have negative grade so the
sign for the term proportional to $\tau$ in in (\ref{basisroot})
will turn out correctly for the
parametrization (\ref{pararoot}).
Also since the level is constant in the module we may take this part
outside the sum.

It is, however, worth while to consider the general point in root space
before taking $u=\th_i=0$ $i=1,...,l$ because the affine character may
be represented in terms of what is known as $\Th$-functions. $\Th$-functions
are,
like the Dedekind $\eta$-function, modular forms and have very nice and
well-studied modular properties.

For the case at hand, namely modules over dominant weights,
the character is given by the celebrated Kac-Weyl formula
\cite{Kac74}
\be
{\rm
ch}L(\hLa)=\frac{\sum_{\hat{w}\in\hat{W}}\ep(\hat{w})e^{(\hat{w}(\hLa+\hrho/2)
-\hrho/2)}}{\prod_{\hal\in\hDe_+}(1-e^{-\hal})^{{\rm mult}\hal}}.
\label{kacweyl}
\ee
Here $\ep(\hat{w})$ is plus or minus one corresponding to whether $\hat{w}$ can
be expressed in terms of even or odd number of simple reflexions. mult$\hal$
is the multiplicity of the affine roots, which is one for the real roots and
$r$ for the light-like roots.

The denominator of (\ref{kacweyl}) may be reformulated, using the character
for the trivial representation ch$L(0)=1$, into
\be
\prod_{\hal\in\hDe_+}(1-e^{-\hal})^{{\rm mult}\hal}=\sum_{\hat{w}\in\hat{W}}
\ep(\hat{w})e^{(\hat{w}(\hrho/2)
-\hrho/2)}. \label{idrepchar}
\ee
This reformulation is desirable in order to write the character in
terms of modular forms, and hence for examining modular properties.
In explicit calculations, however, like the ones presented in paper II and III,
the
first form is much more convenient. Note that for $\widehat{su}(2)$
(\ref{idrepchar})
corresponds to the famous Jacobi triple product identity.

The affine Weyl group $\hat{W}$ is a semidirect product of the finite Weyl
group
and translations on the co-root lattice see chapter 5.2.
We may use this to split the sum
over the affine Weyl group into sums over the finite Weyl elements and the
translations. The numerator of the character (\ref{kacweyl}) may then, in
the basis (\ref{pararoot}), be written as proportional to
\be
e^{-\pi iuk}\sum_{w\in W}\ep(w)
\sum_{t\in M}e^{-2\pi i\th\cdot(\si_{\al}(\La+\rho/2+t)-\rho/2)}
e^{2\pi i\tau(\La+\rho/2+t)^2/(k+c_g)}.
\ee
The constant of proportionality is $exp(-2\pi i\tau(\hLa+\hrho/2)^2/(k+c_g))$.
Here $M$ is the lattice spanned by the translation part of the
affine Weyl group $\hat{W}$.

In the basis for the root space (\ref{basisroot}) we define the $\Th$-function
\be
\Th_{\la}=e^{-2\pi iku}\sum_{\ga\in M+\la/k}e^{(\pi ik\tau|\ga|^2-2\pi
ik\ga\cdot\th)}
\ee
where $M$ is the lattice of the translation part of the Weyl group $\hat{W}$.
Using this we may rewrite the numerator and the denominator of the Kac-Weyl
formula
(\ref{kacweyl}) into
\be
& & e^{(-\hrho\cdot\hat{v}/2-2\pi i\tau\frac{(\hLa+\hrho/2)^2}{(k+c_g)})}
\sum_{w\in W}\ep(w)\Th_{w(\La+\rho/2)}\ee
and
\be
& & e^{(-\hrho\cdot\hat{v}/2-2\pi i\tau\frac{(\hrho/2)^2}{c_g})}
\sum_{w\in W}\ep(w)\Th_{w(\rho/2)}
\ee
respectively. We hence find the affine character expressed in therms of
$\Th$ functions to look like
\be
{\rm ch}L(\hLa)=e^{(-2\pi i\tau(\frac{(\hLa+\hrho/2)^2}{(k+c_g)}-
\frac{(\hrho/2)^2}{c_g}))}\frac{\sum_{w\in W}\ep(w)\Th_{w(\La+\rho/2)}}
{\sum_{w\in W}\ep(w)\Th_{w(\rho/2)}}. \label{affcharth}
\ee

The number $\frac{(\hLa+\hrho/2)^2}{k+c_g}-\frac{(\hrho/2)^2}{c_g}$ may be
rewritten using the very strange formula $\frac{(\rho)^2}{2c_g}=\frac{d_g}{12}$
valid for any simple Lie algebra $g$, into $\frac{c_R}{c_g+k}-
\frac{kd_g}{c_g+k}\frac{1}{24}$ where $c_R$ is the quadratic Casimir of the
representation with highest weight $\La$. This is then nothing but the
conformal dimension of the representation minus $\frac{1}{24}$ times the
conformal anomaly.

Finally we note that the appropriate extension of the action of the modular
group is
\be
\tau\rightarrow\frac{a\tau+b}{c\tau+d} \hs{10mm}
\th\rightarrow\frac{\th}{c\tau+d} \hs{10mm}
u\rightarrow u+\frac{c\th^2}{2(c\tau+d)}
\ee
under the transformation matrix
\be
\lef[\begin{array}{cc} a & b \\ c & d
\end{array}\rig]
\ee
Taking $u=\th=0$ gives the
ordinary action as it must. \\

The case of particular interest to us is the coset construction. For
dominant highest
weights the character of the full algebra may be decomposed into
characters of a sub-algebra with coefficients known as branching functions i.e.
\be
{\rm ch}L(\hLa)=\sum_{\hla}b_{\hLa,\hla}{\rm ch}L(\hla). \label{branch}
\ee
Here $\hla$ are the weights of the sub-algebra which ranges
over dominant weights. The generic branching function
is not easy to find using conventional methods. Conventional
calculations of branching functions have been restricted to
studies of special cases or by considering
general formulas obtained under some assumptions.\\

In paper I a branching function of the gauged WZNW model which respects the
BRST symmetry is introduced. This branching function
factorizes into three parts, one of which is the Kac-Weyl formula. The other
two are
also of simple forms. This indicates that this branching function should be
most suited for investigations of modular properties.

It is shown in paper I that the
BRST invariance of this branching function guarantees
that only the coset degrees of freedom are propagating. Furthermore,
this branching function respects the symmetries of the full group $G$, and does
not rely on a decomposition of $G$ in $H$. This is an essential and novel
feature compared to previous branching function calculations.

We also, in paper I, compute explicitly the branching functions for the
parafermion
theory
$SU(2)_k/U(1)$ as well as for minimal models $SU(2)_k\otimes SU(2)_1/
SU(2)_{k+1}$. We find that our result coincides with previously
obtained results using conventional methods.

In paper II we give an alternative proof of the correctness of the branching
function introduced in paper I. We prove that this branching function
coincides up to a possible normalization to the conventional definition
of a branching function (\ref{branch}). In the proof we make extensive
use of representation theory for affine Lie algebras. We also give the
general form for the type of branching functions known as string functions.

Paper III deals with the application of the branching function introduced
in paper I to a number of general coset constructions. Included in those
are for example $N=1$ superconformal minimal models $SU(2)_k\otimes
SU(2)_2/SU(2)_{k+2}$. Some of those cases has been studied before
using some assumption or for some specific choice of coset. We
partly verify previously obtained results.

\begin{itemize}

\item P. Ginsparg, {\sc Applied conformal field theory}, In: Fields,
Strings and critical phenomena, Les Houches, Session XLIX, 1988, Ed. by:
E. Brezin and J. Zinn-Justin, Elsevier Science Publishers 1989

\item W. Lerche and A. Schellekens, {\it The covariant lattice construction
of four-dimensional string theories}, Lectures delivered at the summer school
on
Strings and superstrings, Poiana Brasov, Romania, September 1987,
CERN-TH.4925, 1988

\item V.G. Kac, {\sc Infinite dimensional Lie algebras},
Cambridge University Press 3'd ed. 1990

\item V.G. Kac and D. Petersen, {\it Infinite-dimensional Lie algebras theta
functions and modular forms}, Adv. in Math. 53 (1984) 125

\end{itemize}



\begin{thebibliography}{10}

\bibitem{Scherk-Schwarz74}
J. Scherk and J.H. Schwarz, {\em Dual models for non-hadrons}, Nucl. Phys. B81
  (1974)  118.

\bibitem{Nambu70}
Y. Nambu, {\em Lectures at the Copenhagen symposium} (1970).

\bibitem{Goto71}
T. Goto, {\em Relativistic quantum mechanics of one-dimensional mechanical
  continuum and subsidiary condition of dual resonance model}, Prog. Theor.
  Phys. 46 (1971)  1560.

\bibitem{Brink-Di-Vecchia-Howe76}
L. Brink, P. Di Vecchia and P. Howe, {\em A locally supersymmetric and
  reparametrization invariant action for the spinning string}, Phys. Lett. 65B
  (1976)  471.

\bibitem{Banks-Dixon-Friedan-Martinec88}
T. Banks, L. Dixon, D. Friedan and E. Martinec, {\em Phenomenology and
  conformal field theory or can string theory predict the weak mixing angle?},
  Nucl. Phys. B299 (1988) pages 613--626.

\bibitem{Friedan-Qiu-Shenker84}
D. Friedan, Z. Qiu and S. Shenker, {\em Conformal invariance, unitarity, and
  critical exponents in two dimensions}, Phys. Rev. Lett. 52 (1984)  1575.

\bibitem{Goddard-Kent-Olive86}
P. Goddard, A. Kent and D. Olive, {\em Unitary representations of the Virasoro
  and super-Virasoro algebras}, Commun. Math. Phys. 103 (1986) pages 105--119.

\bibitem{Alvarez83}
O. Alvarez, {\em Theory of strings with boundaries: Fluctuations, topology and
  quantum geometry}, Nucl. Phys. B216 (1983) pages 125--184.

\bibitem{Gliozzi-Scherk-Olive77}
F. Gliozzi, J. Scherk and D. Olive, {\em Supersymmetry, supergravity theories
  and the dual spinor model}, Nucl. Phys. B122 (1977)  253.

\bibitem{Kac67}
V.G. Kac, {\em Simple graduated Lie algebras of finite growth}, Funct. Anal.
  Appl. 1 (1967)  328.

\bibitem{Kac68}
V.G. Kac, {\em Graduated Lie algebras and symmetric spaces}, Funct. Anal. Appl.
  2 (1968) pages 182--183.

\bibitem{Kac68:2}
V.G. Kac, {\em Simple irreducible graded Lie algebras of finite growth}, Math.
  USSR Izvestija 2 (1968) pages 1271--1311.

\bibitem{Moody67}
R.V. Moody, {\em Lie algebras associated with generalized Cartan matrices},
  Bull. Am. Math. Soc. 73 (1967)  217.

\bibitem{Moody68}
R.V. Moody, {\em A new class of Lie algebras}, J. of Alg. 10 (1968) pages
  211--230.

\bibitem{Kac90}
V.G. Kac, {\sc Infinite dimensional Lie algebras}, third edition, Cambridge
  Univ. Press (1990).

\bibitem{Gebert-Nicolai94}
R. Gebert and H. Nicolai, {\em $E_{10}$ for beginners}, hep-th/9411188 (1994).

\bibitem{Verma68}
D.-N. Verma, {\em Structure of certain induced representations of complex
  semisimple Lie algebras}, Bull. Am. Math. Soc. 74 (1968) pages 160--166.

\bibitem{Bernshtein-Gel'fand-Gel'fand71}
I.N. Bernshtein, I.M. Gel'fand and S.I. Gel'fand, {\em Structure of
  representations generated by vectors of highest weight}, Funct. Anal. Appl. 5
  (1971) pages 1--8.

\bibitem{Conze-Dixmier72}
N. Conze and J. Dixmier, {\em Id$\acute{e}$aux primitifs dans
  l'alg$\grave{e}$bre enveloppante d'une alg$\grave{e}$bre de Lie semi-simple},
  Bull. Sc. Math. $2^e$ s$\acute{e}$rie 96 (1972) pages 339--351.

\bibitem{Jantzen80}
J.C. Jantzen, {\sc Moduln mit einem h$\ddot{\rm o}$chsten Gewicht}, Lecture
  notes inmathematics, Ed. A. Dold and B. Eckman, Springer Verlag (1979).

\bibitem{Rocha-Caridi-Wallach83:2}
A. Rocha-Caridi and N. Wallach, {\em Highest weight modules over graded Lie
  algebras: Resolutions, filtrations and character formulas}, Trans. Am. Math.
  Soc. 277 (1983) pages 133--162.

\bibitem{Deodhar-Lepowsky77}
V. Deodhar and J. Lepowsky, {\em On multiplicity in the Jordan-H$\ddot{o}$lder
  series of Verma modules}, J. of Alg. 49 (1977) pages 512--524.

\bibitem{Kac-Kazhdan79}
V.G. Kac and D.A. Kazhdan, {\em Structure of representations with highst weight
  of infinite-dimensional Lie algebras}, Adv. in Math. 34 (1979) pages 97--108.

\bibitem{Malikov-Feigin-Fuks86}
F.G. Malikov, B.L. Feigin, D.B. Fuks, {\em Singular vectors in Verma modules
  over Kac-Moody algebras}, Funkt. Anal. Ego Prilozh. 20 (1986) ~25.

\bibitem{Hwang-Marnelius90-91}
S. Hwang and R. Marnelius, {\sc Unpublished lecture notes on BRST
  quantization}, Graduate course, Chalmers University of Technology and
  G$\ddot{\rm o}$tebor University (1990-91).

\bibitem{Bowcock89}
P. Bowcock, {\em Canonical quantization of the gauged Wess-Zumino model}, Nucl.
  Phys. B316 (1989) ~80.

\bibitem{Karabali-Park-Schnitzer-Yang89}
D. Karabali, Q-H Park, H. Schnitzer and Z. Yang, {\em A GKO construction based
  on a path integral formulation of gauged Wess-Zumino-Witten actions}, Phys.
  Lett. B 216 (1989)  307.

\bibitem{Karabali-Schnitzer90}
D. Karabali and H. Schnitzer, {\em BRST quantization of the gauged WZW action
  and coset conformal field theories}, Nucl. Phys. B329 (1990)  649.

\bibitem{Henneaux-Teitelboim92}
M. Henneaux and C. Teitelboim, {\sc Quantization of gauge systems}, Princeton
  University Press (1992).

\bibitem{Dirac50}
P. Dirac, {\em Generalized Hamiltonian dynamics}, Can. J. Math. 2 (1950)  129.

\bibitem{Dirac64}
P. Dirac, {\sc Lectures on quantum mechanics}, Belfer Graduate School of
  Science, Yeshiva University, New York (1964).

\bibitem{Fadeev-Popov67}
L.D. Fadeev and V.N. Popov, {\em Feynman diagrams for the Yang-Mills field},
  Phys. Lett. 25B (1967) ~29.

\bibitem{Ryder85}
L. Ryder, {\sc Quantum field theory}, Cambridge University Press (1985).

\bibitem{Becchi-Rouet-Stora76}
C. Becchi, A. Rouet and R. Stora, {\em Renormalization of gauge theories}, Ann.
  Phys. 98 (1976)  287.

\bibitem{Tyutin75}
I.V. Tyutin, {\em Gauge invariance in field theory and in statistical physics
  in the operator formalism}, Lebedev preprint, FIAN No. 39 (1975)
  unpublished.

\bibitem{Fradkin-Fradkina78}
E.S. Fradkin and T.E. Fradkina, {\em Quantization of relativistic systems with
  boson and fermion first- and second-class constraints}, Phys. Lett. 72B
  (1978)  343.

\bibitem{Fulop92}
G. F$\ddot{\rm u}$l$\ddot{\rm o}$p, {\em Transformations and BRST-charges in
  (2+1)-dimensional gravitation}, Mod. Phys. Lett. A7 (1992)  3495.

\bibitem{Curci-Ferrari76}
G. Curci and R. Ferrari, {\em An alternative approach to the proof of unitarity
  for gauge theories}, Nou. Cim. 35A (1976)  273.

\bibitem{Kugo-Ojima78}
T. Kugo and I. Ojima, {\em Manifestly covariant canonical formulation of
  Yang-Mills theories physical states subsidiary conditions and physical
  S-matrix unitarity}, Phys. Lett. 73B (1978)  459.

\bibitem{Kugo-Ojima79}
T. Kugo and I. Ojima, {\em Local covariant operator formalism of non-abelian
  gauge theories and quark confinement problem}, Suppl. Prog. Theor. Phys. 66
  (1979) ~1.

\bibitem{Hwang-Marnelius89}
S. Hwang and R. Marnelius, {\em BRST symmetry and a general ghost decoupling
  theorem}, Nucl. Phys. B320 (1989)  476.

\bibitem{Bouwknegt-McCarthy-Pilch92}
P. Bouwknegt, J. McCarthy and K. Pilch, {\em BRST analysis of physical states
  for 2D gravity coupled to c$\leq$1 matter}, Commun. Math. Phys. 145 (1992)
  541.

\bibitem{Bott-Tu82}
R. Bott and L. Tu, {\sc Differential forms in algebraic topology},
  Springer-Verlag (1982).

\bibitem{Knizhnik-Zamolodchikov84}
V.G. Knizhnik and A.B. Zamolodchikov, {\em Current algebra and Wess-Zumino
  model in two dimensions}, Nucl. Phys. B247 (1984) ~83.

\bibitem{Jain-Shankar-Wadia85}
S. Jain, R. Shankar and S. Wadia, {\em Conformal invariance and string theory
  in compact space: Bosons}, Phys. Rev. D32 (1985) pages 2713--2721.

\bibitem{Redlich-Schnitzer86}
A.N. Redlich and H. Schnitzer, {\em The Polyakov string in O(N) or SU(N) group
  space}, Phys. Lett. 167B (1986) pages 315--319.

\bibitem{Gepner-Witten86}
D. Gepner and E. Witten, {\em String theory on group manifolds}, Nucl. Phys.
  B278 (1986)  493.

\bibitem{Goddard-Kent-Olive85}
P. Goddard, A. Kent and D. Olive, {\em Virasoro algebras and coset space
  models}, Phys. Lett. 152 (1985) pages 88--92.

\bibitem{DiVecchia-Knizhnik-Petersen-Rossi85}
P. Di Vecchia, V.G. Knizhnik, J.L. Petersen and P. Rossi, {\em A supersymmetric
  Wess-Zumino lagrangian in two dimensions}, Nucl. Phys. B253 (1985)  701.

\bibitem{Abdalla-Abdalla85}
E. Abdalla and M.C.B. Abdalla, {\em Supersymmetric extension of the chiral
  model and the Wess-Zumino term in two dimensions}, Phys. Lett. 152 (1985)
  ~59.

\bibitem{Giveon-Rabinovici-Tseytlin93}
A. Giveon, E. Rabinovici and A. Tseytlin, {\em Heterotic string solutions and
  coset conformal field theories}, Nucl. Phys. B409 (1993) pages 339--362.

\bibitem{Kazama-Suzuki89}
Y. Kazama and H. Suzuki, {\em Characterization of N=2 superconformal models
  generated by the coset space method}, Phys. Lett. B 216 (1989)  112.

\bibitem{Kazama-Suzuki89:2}
Y. Kazama and H. Suzuki, {\em New N=2 superconformal field theories and
  superstring compactification}, Nucl. Phys. B321 (1989)  232.

\bibitem{Witten84}
E. Witten, {\em Non-abelian bosonization in two dimensions}, Commun. Math.
  Phys. 92 (1984)  455.

\bibitem{Polyakov-Wiegmann83}
A.M. Polyakov and P.B. Wiegmann, {\em Theory of nonabelian goldstone bosons in
  two dimensions}, Phys. Lett. 131B (1983)  121.

\bibitem{Polyakov-Wiegmann84}
A.M. Polyakov and P.B. Wiegmann, {\em Goldstone fields in two dimensions with
  multivalued actions}, Phys. Lett. 141B (1984)  223.

\bibitem{Schnitzer89}
H. Schnitzer, {\em A path integral construction of superconformal field
  theories from a gauged supersymmetric Wess-Zumino-Witten action}, Nucl. Phys.
  B324 (1989)  412.

\bibitem{Halpern71}
M.B. Halpern, {\em The two faces of a dual pion-quark model}, Phys. Rev. D4
  (1971) pages 2398--2401.

\bibitem{Halpern75}
M.B. Halpern, {\em Quantum `solitons' which are SU(N) fermions}, Phys. Rev. D12
  (1975)  1684.

\bibitem{Bastianelli91}
F. Bastianelli, {\em BRST symmetry from a change of variables and the gauged
  WZNW model}, Nucl. Phys. B361 (1991)  555.

\bibitem{Hwang95}
S. Hwang, {\em The BRST cohomology of affine Lie algebras}, G$\ddot{\rm
  o}$teborg ITP 95-04 (1995).

\bibitem{Tseytlin94}
A.A. Tseytlin, {\em Conformal sigma models corresponding to gauged
  Wess-Zumino-Witten theories}, Nucl. Phys. B411 (1994)  509.

\bibitem{Polyakov81}
A.M. Polyakov, {\em Quantum geometry of bosonic strings}, Phys. Lett. 103B
  (1981)  207.

\bibitem{Polyakov81:2}
A.M. Polyakov, {\em Quantum geometry of fermionic strings}, Phys. Lett. 103B
  (1981)  211.

\bibitem{Kac74}
V.G. Kac, {\em Infinite-dimensional Lie algebras and Dedekind's
  $\eta$-function}, Funct. Anal. Appl. 8 (1974) pages 68--70.

\end{thebibliography}
\end{document}